\documentclass[onecolumn,11pt,aps,pre,superscriptaddress]{revtex4-1}
\usepackage{graphicx}
\usepackage{mathptmx,amsmath,xcolor}
\newcommand{\beq}[0]{\begin{equation}}
\newcommand{\eeq}[0]{\end{equation}}

\newcommand{\bc}{\begin{center}}
\newcommand{\ec}{\end{center}}

\begin{document}

\title{Propagation of periodic wave trains along the magnetic
field in a collision-free plasma}
\author{G. Abbas}
%\email{gohar.abbas@gcu.edu.pk}
\affiliation{Department of Physics, Government College University Lahore
54000, Pakistan}
\author{P.\,G. Kevrekidis}
\affiliation{Department of Mathematics and Statistics, University of
  Massachusetts, Amherst MA 01003-4515, USA} 
\affiliation{Mathematical Institute, University of Oxford, Oxford, OX2
  6GG, UK}
\author{J. E. Allen}
\affiliation{Mathematical Institute, University of Oxford, Oxford, OX2
  6GG, UK}
\author{V. Koukouloyannis}
\affiliation{Department of Mathematics, Laboratory of Applied Mathematics
and Mathematical Modelling, University of the Aegean, Karlovassi, 83200
Samos, Greece}

\author{D.\,J. Frantzeskakis}
\affiliation{Department of Physics, National and Kapodistrian University of
Athens, Panepistimiopolis, Zografos, Athens 15784, Greece}

\author{N. Karachalios}
\affiliation{Department of Mathematics, Laboratory of Applied Mathematics
and Mathematical Modelling, University of the Aegean, Karlovassi, 83200
Samos, Greece}
\date{\today}

\begin{abstract}
In this work, a systematic study, examining the propagation of 
periodic and solitary wave along the magnetic field in a cold collision-free plasma, is presented. 
Employing the quasi-neutral approximation and the conservation of momentum flux and
energy flux in the frame co-traveling with the wave, the exact analytical
solution of the stationary solitary pulse is found analytically 
in terms of particle densities, parallel and transverse velocities, as well as 
transverse magnetic fields. Subsequently, this solution is generalized in
the form of periodic waveforms represented by cnoidal-type waves. These
considerations are fully analytical in the case where 
%
%%\textcolor{red}{
%
the total angular momentum flux $L$, due to the ion and electron motion together with the contribution due to the Maxwell stresses, vanishes. 
%
%
%the angular momentum flux $L$
%associated with the particles and transverse magnetic field vanishes.
% }.
%%%%%%%%%%%%%% PGK: Change re: L \neq 0.
%When this quantity is non-zero, only periodic solutions are found to be possible and are
%obtained numerically (as an exact analytical form is no longer evidently
%available).
%%%%%%%%%%%%%%%
%Graphical representations of the solutions and in
%particular, the 3D structures of the transverse magnetic and velocity field
%lines are presented.
A graphical representation of all associated fields %, namely the
%velocity fields of the electrons and ions, the magnetic and electric
%fields
%and charge density
is also provided.
\end{abstract}

\maketitle

%\preprint{AIP/123-QED}

% Force line breaks with \\


\section{Introduction}

The study of solitary waves, their dynamics and interactions, as well as the
associated notions of integrability and soliton theory have been long
thought to originate with the famous Fermi-Pasta-Ulam problem~\cite{fpu}.
Arguably, though, the explosion of interest in the field came a decade later
and is chiefly credited to the numerical study of the FPU problem, and
associated remarkable properties of the Korteweg-de Vries (KdV) equation, 
reported by Kruskal and Zabusky~\cite{kz}. What is perhaps far less
well-known is that a major part of the associated motivation (even cited as
such in the work of~\cite{kz}) stemmed from work at around the same time in
the study of traveling waves in plasmas. More well-known in this regard is
the seminal work of Washimi and Taniuti \cite{tan} connecting the dynamics
of ion-acoustic waves in plasmas to the KdV equation. Yet, a ``well-kept
secret'' since its inception already many years earlier (i.e., in
1958-1960) has been the work of Adlam and Allen~\cite{Allen1958,aa2}. These
authors developed a model for the propagation of a solitary wave in a
collisionless plasma along the $x-$ and $y-$directions coupled with dynamics in
the $x-$ and $y-$directions for the electric field and a transverse
magnetic field $B_z$.

%One of the outstanding problems in plasma physics research involves the
%understanding of the non-linear structure of travelling solitary waves. The
Recently, the propagation of finite amplitude waves across a magnetic field in cold plasmas 
%
%in a cold collision-free plasma 
%was first investigated by Adlam and Allen in 1958 \cite%
%{Allen1958}. The model was applied to large amplitude waves that travel into
%a compressed plasma containing a magnetic field. Recently, the study of such
%
has been revisited for the demonstration of the $\mathbf{j\times B}$ force
in a collisionless plasma \cite{Allen2017}. On the other hand, some time ago, Montgomery 
\cite{Montgomery1959} and Saffman \cite{Saffman1961} discussed the large
amplitude waves propagating along the magnetic field. Such parallel
propagating solitary waves have been further examined with the core filled
by oscillatory structures in more recent works. For example, Sauer {\it et al.} 
\cite{Sauer2002} and Dubinin {\it et al.} \cite{Dubinin2003} discussed stationary
nonlinear solutions, including localized waveforms 
%
%solitary amplitude 
with  oscillating phase, looking like envelope solitons in the wave frame. 
Keeping in view this framework, Cattaert and Verheest \cite{Cattaert2005} analyzed large
amplitude weakly nonlinear (KdV type) oscillatory solitary structures in the
plasma frame. Such a treatment has further been employed in solitary
waves in dusty plasmas~\cite{Gibson2017,Cramer2018}. 
%In the
%cited literature, these are also referred a non-linear version of
%whistler waves. This may not be a good analogy because the Poynting vector
%has no component in the direction of propagation. The whistler mode is
%basically an electromagnetic wave in which the propagation of energy is
%described by the Poynting vector.
In all the above cases, the core purpose is to obtain a one-dimensional
solution which describes the motion of a pulse or solitary wave in the
direction perpendicular or parallel to the magnetic field. Over the past
year, both of these directions have been further considered. More
specifically, in Ref.~\cite{abbas}, analytical expressions in
the co-traveling frame, for the form of the longitudinally propagating
solitary wave, were obtained. Concurrently, in the transverse field case, 
the work of~\cite{allen2020} examined various generalizations; these include 
the interaction of two solitary waves (identifying their repulsion), 
as well as periodic (cnoidal) solutions, that were identified numerically, 
which generalize the solitary waveform (and possess the latter as a special limit).

%The present investigation explores a more detailed study than that followed
%by other workers. For example, for large amplitudes, a useful but rather
%complicated numerical study for finding the oscillatory structures inside
%the core of solitary wave amplitude has been done by Dubinin et al \cite%
%{Dubinin2003}. An analytical treatment of such oscillatory solitary
%amplitudes has been done by Cattaert and Verheest \cite{Cattaert2005} but
%for weakly nonlinear case.

The present work studies analytically --and complements the analytics with
numerical identification when needed-- the system of equations for fully
nonlinear electromagnetic waves propagating along a uniform magnetic field in
a cold collisionless plasma. The oscillatory solutions of transverse
particle velocities and transverse magnetic fields are derived without
linearization. We shall solve the present problem using the concept of quasineutrality. In particular, 
it is assumed that an 
electrostatic field is produced by an infinitesimal difference between the electron and ion 
densities; the Gauss' law for the electric field is not employed. The concept 
is valid in the case where the electron plasma frequency is much
greater than the electron gyrofrequency. However, in a
  strongly nonlinear case, the possible deviation from quasineutrality
  condition in the longitudinal direction can be significant. In
  Ref. \cite{Trukh2016}, the properties of ion- and electron-acoustic
  (longitudinal) solitons in a plasma without a magnetic field are
  considered with the understanding that solitons carry out one-way
  transfer of charged particles at a distance of several Debye
  radii. In particular, the electric currents with a DC component were
  induced. Subsequently \cite{Trukh2018}, it was shown that these
  currents can be significant in the case of large amplitudes. These
  considerations were further analyzed in the experimental study of
  dust-acoustic soliton currents \cite{Trukh2019}. This is, as
  can be observed from the above works, a particularly
  active direction of research, yet here we will restrict
considerations
to the regime where quasineutrality is a valid approximation.

The problem of interest herein is exposed in its full generality by involving nine
space-time dependent fields, namely five velocity fields, two (transverse)
magnetic fields, an electric field and a charge density field. Expressions
for all of them are provided. The starting point is a reduction of the
problem to an effective two- and eventually (through a polar decomposition)
one degree-of-freedom problem for the transverse magnetic field components.
In these variables, the solitary wave case of~\cite{abbas} is initially
retrieved and is subsequently generalized to periodic
solutions in the form of cnoidal waves. 
%More interestingly, we present a three dimensional structure
%of the oscillatory pulse for the case in which the plasma and magnetic field
%are uniform at the boundaries.
The derivation of the solitary wave is found to be analytically possible 
in the case of a vanishing angular momentum flux $L$ (which is due to the 
ion and electron motion and the contribution due to the Maxwell stresses) 
in the plane of the transverse magnetic field.
%%%%%%%%%%%%% PGK: change re: L \neq 0
In the Appendix, we discuss the case of
non-zero $L$ which is of mathematical interest as it
yields solely a possibility for periodic
waves which we numerically reconstruct.
%%%%%%%%%%%%%
 In the astrophysical context, the present study may
  have applications in the Earth foreshock region and Jupiter’s bow shock
  region where  oscillatory solitary structures are expected
  naturally under many different conditions~\cite{Sauer2001}.
  Nevertheless, we are not presently aware of an experiment that has
observed the relevant structures.

Our presentation is structured as follows. In Section~II, we revisit the
analytical formulation of the problem in dimensionless units and reduce it
to the two degree-of-freedom system for the transverse magnetic fields. In
Section~III, we discuss the analytical solutions and connect the problem to
a Duffing oscillator, obtaining also the cnoidal wave solutions for the
case of vanishing $L$.
%We also discuss the differences present in
%the case with non-vanishing $L$ and construct the solely periodic
%orbits of the latter.
In Section~IV, we summarize our findings and
present a number of conclusions, as well as directions for future
research.
%%%%%%%%%%%%% PGK: change re: L \neq 0
The Appendix contains the case of  non-vanishing $L$ and a
mathematical
discussion of the reconstruction of the (solely) periodic  orbits of
the latter.

\section{Basic equations of motion}

The framework of physical interest, similarly to the case of~\cite{abbas}, is
as follows. We wish to describe ions and electrons propagating under a 
\textit{common} velocity field and electric field along the $x$-direction.
In the quasi-neutral setting of interest (where the charge carriers satisfy 
$n_{e}\approx n_{i}=n$), the opposite charges bear unequal velocities along
the $y-$ and $z-$directions. In these directions, there exists a nontrivial
(non-constant) magnetic field $B_{y}$ and $B_{z}$. For a cold collisionless
plasma, first we write the equations in SI units governing the motion of a
cold gas consisting of electrons and one type of positive ion, as follows: 
\begin{equation}
\left( \frac{\partial }{\partial t}+\mathbf{v}_{s}\mathbf{\cdot \nabla }\right) 
\mathbf{v}_{s}=\frac{q_{s}}{m_{s}}\left[ \mathbf{E}+\mathbf{v}_{s}\mathbf{%
\times B}\right] ,  \label{eq1}
\end{equation}%
and%
\begin{equation}
\frac{\partial n_{s}}{\partial t}+\mathbf{\nabla \cdot}\left( n_{s}\mathbf{v}%
_{s}\right) =0.  \label{conti 2}
\end{equation}%
The index s can either stand for e, i.e., electrons or for i, i.e., ions.
In response, the electromagnetic fields generated by particle motion are:
\begin{align}
\mathbf{\nabla \times E}& \mathbf{=-}\frac{\mathbf{\partial B}}{\partial t},
\notag \\
\mathbf{\nabla \times B}& \mathbf{=}\mu _{0}\sum_{s} q_{s}n_{s}\mathbf{v}%
_{s}+\epsilon _{0}\mu _{0}\frac{\mathbf{\partial E}}{\partial t},
\label{Maxwell3}
\end{align}
where $m_{s}=m_{e,i}$ are the respective masses, $\epsilon_0$ and $\mu _{0}$ denote 
vacuum's electric permittivity and magnetic permeability, respectively, $q_{s}=\pm e$ 
is the charge, and $n_{s}$ is the density of the charges. 
We add Eq.~(\ref{eq1}) for ions and electrons and obtain:
\begin{equation}
nD\left( m_{i}\mathbf{v}_{i}+m_{e}\mathbf{v}_{e}\right) =\frac{1}{\mu _{0}}%
\left[ \left( \mathbf{B \cdot\nabla }\right) \mathbf{B}-\frac{1}{2}\mathbf{\nabla 
}\left( \mathbf{B}^{2}\right) \right].  
\label{7}
\end{equation}
Here, we have used the total derivative
notation $D=\frac{\partial }{\partial t}+v_{x}\frac{\partial
}{\partial x}$ for compactness. 
The electric stresses are negligible, since quasi-neutrality has been assumed. 
The above vector equation only refers to change in momentum flux. 
From Eq.~(\ref{eq1}), the energy equation can also be written down immediately as follows:
\begin{equation}
\frac{m_{t}n}{2}Dv_{x}^{2}+\frac{m_{i}n}{2}Dv_{i\bot }^{2}+\frac{m_{e}n}{2}%
Dv_{e\bot }^{2}=0.  
\label{17}
\end{equation}

\subsection{Normalized equations}

In what follows, and although it is possible to work with Eqs.~(\ref{eq1})-(\ref{17}) 
as expressed in dimensional units, it is convenient --for notational simplicity and more
straightforward connection with the numerical results-- to use  
%we will show everything in 
dimensionless equations. Before that, recall that the full set of
fields involves, in addition to the density, the magnetic and electric fields, 
$\mathbf{B}=(B_{1},B_{y}(x),B_{z}(x))$ and $\mathbf{E}=(E(x),0,0)$ respectively, as well as 
the velocity field $(v_{x},v_{ey,iy},v_{ez,yz})$. Notice the common
velocity along the x-axis $v_x$ of the electrons and the ions. 
More specifically, we start by rewriting the equations in normalized form.
The unit for the distance is $d=\sqrt{%
m_{e}m_{i}/e^{2}\mu _{0}n_{1}(m_{e}+m_{i})}$.  The corresponding
characteristic speed $v_{A}=B_{1}/\sqrt{\mu _{0}N_{1}\text{ }(m_{e}+m_{i})}$
is the Alfv\'{e}n velocity, $\Omega _{e}=eB_{1}/m_{e}$ is the electron
angular frequency and $\alpha =m_{e}/m_{i}$ is the mass ratio. The unit for
the velocity employed in this paper was obtained by multiplying the
distance $d$ by $\Omega _{e}$ (i.e., $v^{\ast }=\frac{v_{A}}{%
\sqrt{\alpha }})$. We then define the normalized variables: $t\mapsto\Omega
_{e}\,t,E_{x}\mapsto E_{x}/v^{\ast }B_{1}$, $x\mapsto x/d,$ $\mathbf{%
v\mapsto v/}v^{\ast }$, $\mathbf{B\mapsto B/}B_{1}.$ In these units,
the equations become dimensionless in the form that we now discuss.

\paragraph{Dimensionless Equations for Electrons and Ions:}

By implementing the above mentioned rescaling, we acquire the dimensionless form of the
Newtonian equations of motion of the particles (electrons and ions with
respective subscripts), involving the force from the electric and the
magnetic field ($\mathbf{v \times B}$). 
%in the dimensionless units. 

\begin{equation}
\left( \frac{\partial }{\partial t}+v_{x}\frac{\partial }{\partial x}\right)
\left( 
\begin{array}{c}
v_{x} \\ 
v_{ey} \\ 
v_{ez}%
\end{array}%
\right) =-\left[ \left( 
\begin{array}{c}
E_{x} \\ 
0 \\ 
0%
\end{array}%
\right) +\left( 
\begin{array}{c}
v_{ey}B_{z}-v_{ez}B_{y} \\ 
-v_{x}B_{z}+v_{ez} \\ 
v_{x}B_{y}-v_{ey}%
\end{array}%
\right) \right],  \label{ElecEqM58}
\end{equation}

%\paragraph{Dimensionless Equations for Ions:}

\begin{equation}
\left( \frac{\partial }{\partial t}+v_{x}\frac{\partial }{\partial x}\right)
\left( 
\begin{array}{c}
v_{x} \\ 
v_{iy} \\ 
v_{iz}%
\end{array}%
\right) =\alpha \left[ \left( 
\begin{array}{c}
E_{x} \\ 
0 \\ 
0%
\end{array}%
\right) +\left( 
\begin{array}{c}
v_{iy}B_{z}-v_{iz}B_{y} \\ 
-v_{x}B_{z}+v_{iz} \\ 
v_{x}B_{y}-v_{iy}%
\end{array}%
\right) \right].  \label{IonEqM59}
\end{equation}

Notice once again that in the above equations the electric
field lies along the $x$-direction, while the magnetic field $B_1$ is
constant along the same direction and has now been normalized to unity. In equations \eqref{ElecEqM58}-\eqref{IonEqM59}, the derivatives are the total derivatives 
involving traveling along the $x-$direction, hence a potential traveling 
configuration will simply mean that we set the partial
derivatives with respect to the (dimensionless) time to zero, allowing the
waves to move along the $x$ direction without changing shape.

\paragraph{Additional Field Equations.}

In addition to equations \eqref{ElecEqM58}-\eqref{IonEqM59}, we have the continuity equation for
the density (of both ions and electrons in our quasi-neutral setting): 
\begin{eqnarray}
\left( \frac{\partial }{\partial t}+v_{x}\frac{\partial }{\partial x}%
\right)n + n \frac{\partial v_x}{\partial x}=0,  \label{contin}
\end{eqnarray}

while the full set of $9$ equations for the $9$ fields is completed by the two
components of Amp{\'e}re's Law along the $y-$ and $z-$direction: 
\begin{equation}
\frac{\partial B_{y}}{\partial x}=\frac{n}{\alpha +1}\left(
v_{iz}-v_{ez}\right),  \label{AmpY60}
\end{equation}%
\begin{equation}
\frac{\partial B_{z}}{\partial x}=-\frac{n}{\alpha +1}\left( v_{iy}-v_{ey}\right).
\label{AmpZ61}
\end{equation}

It is worthwhile to note that algebraic manipulations of the $6$ Newtonian
equations can lead to a reformulation of $3$ of them as momentum flux
equations~\cite{abbas} in the following form: 
\begin{eqnarray}
Dv_{x}+\frac{\alpha }{n}\frac{\partial B_{\bot }^{2}}{\partial x} &=&0, 
\notag \\
D\left( \frac{1}{\alpha }v_{iy}+v_{ey}\right) -\frac{\alpha +1}{n}\frac{%
\partial B_{y}}{\partial x} &=&0,  \label{MFlux62} \\
D\left( \frac{1}{\alpha }v_{iz}+v_{ez}\right) -\frac{\alpha +1}{n}\frac{%
\partial B_{z}}{\partial x} &=&0,  \notag
\end{eqnarray}
where again we use $D=\frac{\partial }{\partial t}+v_{x}\frac{\partial }{\partial x}$,  and $B_\perp^2=B_x^2+B_y^2$. 
%\textcolor{red}{
Here, it is interesting to note that the electric stresses are negligible 
compared to the magnetic stresses when the concept of quasineutrality is employed.
%}.

\subsection{Equations in co-travelling frame}

Remarkably, and despite their complexity for a multitude of fields, it is
possible to tackle the above $9$ equations when looking for a traveling
wave. We thus now turn to the setting of solutions that do not depend on
time explicitly. There are two potential approaches towards reducing
the problem. One is to attempt to solve the equations involving the
transverse magnetic components as a function of the speeds, and formulate
ordinary differential equations (ODEs) for the latter. The second is to
express the velocity fields in terms of $B_y$ and $B_z$; here, we follow 
this latter approach. As explained before, in the co-traveling frame, the partial derivative with respect to time vanishes. This way, the relevant Newtonian ODEs read:

\paragraph{For Electrons:}
%The relevant Newtonian ODEs now read:
%
\begin{equation}
v_{x}\frac{\partial }{\partial x}\left( 
\begin{array}{c}
v_{x} \\ 
v_{ey} \\ 
v_{ez}%
\end{array}%
\right) =-\left[ \left( 
\begin{array}{c}
E_{x} \\ 
0 \\ 
0%
\end{array}%
\right) +\left( 
\begin{array}{c}
v_{ey}B_{z}-v_{ez}B_{y} \\ 
-v_{x}B_{z}+v_{ez} \\ 
v_{x}B_{y}-v_{ey}%
\end{array}%
\right) \right],  \label{ElecEq63}
\end{equation}

\paragraph{For Ions:}

\begin{equation}
v_{x}\frac{\partial }{\partial x}\left( 
\begin{array}{c}
v_{x} \\ 
v_{iy} \\ 
v_{iz}%
\end{array}%
\right) =\alpha \left[ \left( 
\begin{array}{c}
E_{x} \\ 
0 \\ 
0%
\end{array}%
\right) +\left( 
\begin{array}{c}
v_{iy}B_{z}-v_{iz}B_{y} \\ 
-v_{x}B_{z}+v_{iz} \\ 
v_{x}B_{y}-v_{iy}%
\end{array}%
\right) \right].  
\label{IonEq64}
\end{equation}
The continuity equation is now reduced to a simple algebraic equation: 
\begin{equation}
n\text{ }v_{x}=C  \label{EqCont67}
\end{equation}

%\subsubsection{Momentum Flux equations}

The momentum flux equations (\ref{MFlux62}) can now be integrated to yield: 
\begin{eqnarray}
\frac{Cv_{x}}{\alpha }+\frac{1}{2}B_{\bot }^{2} &=&E_{1},  \notag \\
\frac{C}{\alpha +1}\left( \frac{1}{\alpha }v_{iy}+v_{ey}\right) -B_{y}
&=&E_{2},  \notag \\
\frac{C}{\alpha +1}\left( \frac{1}{\alpha }v_{iz}+v_{ez}\right) -B_{z}
&=&E_{3},  \label{MFlux68}
\end{eqnarray}
where $E_{1,2,3}$ are the corresponding integration constants. 
%\textcolor{red}{
Equations~(\ref{MFlux68}) describe the momentum flow in the three directions $x$, $y$ and $z$. 
The first term on the left-hand side (in all three equations) is clearly the rate of flow 
of momentum of the particles in the directions $x$, $y$ and $z$. The second term is the rate 
of flow of momentum associated with the magnetic field.
%}

\paragraph{Equations for the two-dimensional potential well.}

One can now combine the momentum flux equations for the $y-$ and $z-$%
components of the velocity field with the Amp{\'e}re's law components, and express 
$v_{iy,ey}$ and $v_{iz,ez}$ in terms of the transverse magnetic
field components $B_y$ and $B_z$. The resulting equations read:
\begin{equation}
v_{iy,ey}=\frac{\alpha B_{y}}{C}\mathbf{+}\frac{\left( -\alpha ,1\right) }{n}%
\frac{\partial B_{z}}{\partial x}+\frac{\alpha }{C}E_{2},  \label{Vy69}
\end{equation}%
\begin{equation}
v_{iz,ez}=\frac{\alpha B_{z}}{C}\mathbf{+}\frac{\left( \alpha ,-1\right) }{n}%
\frac{\partial B_{y}}{\partial x}+\frac{\alpha }{C}E_{3}.  \label{Vz70}
\end{equation}

We now differentiate the components of Ampere's law with respect to $x$, 
multiply by $v_x$, and substitute the resulting velocity derivatives from Newton's 
equations, as well as the velocity components in terms of the magnetic fields $B_y, B_z$ 
[see Eqs.~(\ref{Vy69})-(\ref{Vz70})]. This way, we can obtain the following 
pair of equations involving \textit{solely} the transverse magnetic field components: 
%This algebraic manipulation leads to: 
\begin{equation}
v_{x}\frac{\partial }{\partial x}\left( v_{x}\frac{\partial B_{y}}{\partial x%
}\right) \mathbf{=-}\frac{\partial \Phi }{\partial B_{y}}-\frac{\left(
1-\alpha \right) }{n}C\frac{\partial B_{z}}{\partial x},  \label{Eq73}
\end{equation}
\begin{equation}
v_{x}\frac{\partial }{\partial x}\left( v_{x}\frac{\partial B_{z}}{\partial x%
}\right) \mathbf{=-}\frac{\partial \Phi }{\partial B_{z}}+\frac{\left(
1-\alpha \right) }{n}C\frac{\partial B_{y}(x)}{\partial x},  \label{Eq74}
\end{equation}
where the effective potential $\Phi$ can be defined as: 
\begin{eqnarray}
\Phi &\equiv &\Phi \left( B_{y},B_{z}\right) =\frac{1}{8}\alpha B_{\bot }^{4}%
-\frac{1}{2}\alpha \left( E_{1}-1\right) B_{\bot }^{2}+\alpha \left(
E_{2}B_{y}+E_{3}B_{z}\right).  \label{PseuPot71} \\
&&  \notag
\end{eqnarray}

%\begin{eqnarray}
%\frac{\partial \Phi }{\partial B_{y}} &=&\frac{\alpha B_{\bot }^{2}B_{y}}{2}%
%-\alpha \left( E_{1}-1\right) B_{y}+\alpha E_{2}  \nonumber \\
%\frac{\partial \Phi }{\partial B_{z}} &=&\frac{\alpha B_{\bot }^{2}B_{z}}{2}%
%-\alpha \left( E_{1}-1\right) B_{z}+\alpha E_{3}  \label{PseudoDiff72}
%\end{eqnarray}

\subsubsection{Conservation Laws for the Two-Dimensional System}

By multiplying the first of Eqs.~(\ref{Eq73}) 
by $\frac{\partial B_y}{\partial x}$ and similarly Eq.~(\ref{Eq74}) 
by $\frac{\partial B_z}{\partial x}$, and adding the two we get 
%It is now immediate to infer from the above system %of equations of motion
%that they possess a conserved energy in the form: 
\begin{equation}
\frac{1}{2}\left[ \left( v_{x}\frac{\partial B_{y}}{\partial x}\right)
^{2}+\left( v_{x}\frac{\partial B_{z}}{\partial x}\right) ^{2}\right] +\Phi
\left( B_{y},B_{z}\right) =W={\rm const}.  \label{EnEq75}
\end{equation}
%\textcolor{red}{
The above equation implies that $W$ is a conserved quantity for the system. Here, $W$ stands for 
%known to be 
the difference between the flux of total kinetic energy and the 
kinetic energy carried by the plasma particles along $x-$direction.
%} 
%Note that Eq.~(\ref{EnEq75}) can be obtained upon .

%\subsubsection{Angular Momentum Flux Equation}

One can similarly manipulate the two equations (\ref{Eq73})-(\ref{Eq74}) (multiplying the first by $B_z $ 
and the second by $B_y$ and subtracting) to obtain another conserved quantity, namely:
\begin{equation}
B_{z}\left( v_{x}\frac{\partial B_{y}}{\partial x}\right) -B_{y}\left( v_{x}%
\frac{\partial B_{z}}{\partial x}\right) +\frac{\left( 1-\alpha \right) }{2}%
B_{\bot }^{2}-\alpha \int \left( E_{2}B_{z}-E_{3}B_{y}\right) dx=L={\rm const},
\label{AngMoment76}
\end{equation}
%\textcolor{red}{
where $L$ represents the total angular momentum flux.
%}. 
The explicit presence of an integral in Eq.~(\ref{AngMoment76}) suggests
that it is far more amenable to analytical (or semi-analytical)
manipulations in the case of $E_2=E_3=0$. We will return to this point in
what follows.

\paragraph{Polar Coordinate Analysis of the ODEs.}
Using a polar decomposition in the form: 
\begin{equation}
B_{y}=B_{\bot }\cos \theta,  \quad B_{z}=B_{\bot }\sin \theta,  \notag
\end{equation}
we obtain%
\begin{equation}
\left( \frac{\partial B_{y}}{\partial x}\right) ^{2}+\left( \frac{\partial
B_{z}}{\partial x}\right) ^{2}=\left( \frac{\partial B_{\bot }}{\partial x}%
\right) ^{2}+\left( B_{\bot }\frac{\partial \theta }{\partial x}\right)^{2}.
\label{Eq77}
\end{equation}

Upon substituting these in Eq.~(\ref{EnEq75}), we obtain:%
\begin{equation}
\frac{1}{2}\left[ \left( v_{x}\frac{\partial B_{\bot }}{\partial x}\right)
^{2}+B_{\bot }^{2}\left( v_{x}\frac{\partial \theta }{\partial x}\right) ^{2}%
\right] +\Phi \left( B_{\bot }\right) =W.  \label{Eq78}
\end{equation}
Furthermore, the left hand side of equation Eq.~(\ref{Eq78}) is reshaped as: 
\begin{equation}
B_{z}\left( \frac{\partial B_{y}}{\partial x}\right) -B_{y}\left( \frac{%
\partial B_{z}}{\partial x}\right) =-B_{\bot }^{2}\frac{\partial \theta }{%
\partial x}.  \label{Eq79}
\end{equation}%
%
%
%Eq. (76) therefore becomes 
Then, reformulation of the corresponding conservation law of Eq.~(\ref%
{AngMoment76}) leads to: 
\begin{equation}
\left[ \frac{\left( 1-\alpha \right) }{2}-v_{x}\frac{\partial \theta }{%
\partial x}\right] B_{\bot }^{2}+\alpha \int B_{\bot }\left( E_{2}\sin
\theta -E_{3}\cos \theta \right) dx=L.  \label{Eq80}
\end{equation}
%
%\subsubsection{Elimination of $v_{x}\frac{\partial \protect\theta }{\partial
%x}$}
%
Next, using Eq.~(\ref{Eq80}), we find 
%substituting $v_{x}\frac{\partial \theta }{\partial x}$ from Eq.~(\ref{Eq80}), namely
%
%$v_{x}\frac{\partial \theta }{\partial x}$ is given
%by 
\begin{equation}
v_{x}\frac{\partial \theta }{\partial x}=\frac{\left( 1-\alpha \right) }{2}-%
\frac{L}{B_{\bot }^{2}}+\frac{\alpha }{B_{\bot }^{2}}\int B_{\bot }\left(
E_{2}\sin \theta -E_{3}\cos \theta \right) dx,  \label{Eq81}
\end{equation}
%Using this substitution 
%
which is substituted into Eq.~(\ref{Eq78}) to give: 
\begin{eqnarray}
&&\frac{1}{2}\left( v_{x}\frac{\partial B_{\bot }}{\partial x}\right) ^{2}+%
\frac{B_{\bot }^{2}}{2}\left( \frac{\left( 1-\alpha \right) }{2}-\frac{L}{%
B_{\bot }^{2}}+\frac{\alpha }{B_{\bot }^{2}}\int B_{\bot }\left( E_{2}\sin
\theta -E_{3}\cos \theta \right) dx\right) ^{2}+  \label{Eq82} \\
&&+\frac{\alpha B_{\bot }^{4}}{8}-\alpha \left( E_{1}-1\right) \frac{B_{\bot
}^{2}}{2}+\alpha B_{\bot }\left( E_{2}\cos \theta +E_{3}\sin \theta \right)
-W=0  \notag
\end{eqnarray}

%\subsubsection{E$_{2}$=E$_{3}$=0 and L=0}
It is now evident that the integral involving $E_{2,3}$ precludes us from
further continuing the analysis. Hence, we will hereafter assume that $%
E_2=E_3=0$. We now distinguish the following two possibilities. Either the 
constant $L$ can be selected to vanish (in line with what
was also done in Ref.~\cite{abbas}), or it can be selected to be
$L\neq0$.
%We
%will examine both scenarios  %what follows 
%in the next Section. However,
%%%%%%%%%%% PGK: Change re: L \neq 0
The former scenario is of direct physical relevance given the
vanishing of the angular momentum. The latter ($L \neq 0$) case  is a topic of
mathematical
interest even though it does not appear to us to have a direct
physical interpretation in the setting at hand. For this reason,
here we focus our analytical considerations to the case of $L=0$,
while we relegate the topic of $L \neq 0$ to an Appendix given the
mathematical interest in the latter case in its own right.
%%%%%%%%%%%%
Then, the
energy conservation of Eq.~(\ref{Eq82}) reads: 
%Considering E$_{2}$=E$_{3}$=0 and L=0, we have%
\begin{equation}
\frac{1}{2}\left( v_{x}\frac{\partial B_{\bot }}{\partial x}\right) ^{2}+%
\frac{B_{\bot }^{2}}{8}\left( 1-\alpha \right) ^{2}+\frac{\alpha B_{\bot
}^{4}}{8}-\alpha \left( E_{1}-1\right) \frac{B_{\bot }^{2}}{2}-W=0,
\label{Eq83}
\end{equation}%
%
%
%or 
%\begin{equation}
%\frac{1}{2}\left( \frac{\partial B_{\bot }}{\partial x}\right) ^{2}+\frac{1}{%
%v_{x}^{2}}\left[ \frac{B_{\bot }^{2}}{8}\left( 1-\alpha \right) ^{2}+\frac{%
%\alpha B_{\bot }^{4}}{8}-\alpha \left( E_{1}-1\right) \frac{B_{\bot }^{2}}{2}%
%-W\right] =0  \nonumber
%\end{equation}%
where the first of the momentum flux equations allows us to express: 
\begin{equation*}
v_{x}=\frac{\alpha }{C}\left( E_{1}-\frac{1}{2}B_{\bot }^{2}\right).
\end{equation*}%
%
%
%therefore 
%\begin{equation}
%\frac{1}{2}\left( \frac{\partial B_{\bot }}{\partial x}\right) ^{2}+\frac{%
%C^{2}}{\alpha ^{2}\left( E_{1}-\frac{1}{2}B_{\bot }^{2}\right) ^{2}}\left[ 
%\frac{B_{\bot }^{2}}{8}\left( 1-\alpha \right) ^{2}+\frac{\alpha B_{\bot
%}^{4}}{8}-\alpha \left( E_{1}-1\right) \frac{B_{\bot }^{2}}{2}-W\right] =0 
%\label{Eq84}
%\end{equation}%
%or 
%\begin{equation}
%\left( \frac{\partial B_{\bot }}{\partial x}\right) ^{2}+\frac{C^{2}}{\alpha
%^{2}\left( E_{1}-\frac{1}{2}B_{\bot }^{2}\right) ^{2}}\left[ B_{\bot
%}^{2}\left\{ \frac{\left( 1-\alpha \right) ^{2}}{4}-\alpha \left(
%E_{1}-1\right) \right\} +\frac{\alpha B_{\bot }^{4}}{4}-W\right] =0
%\label{Eq84}
%\end{equation}

%In comparison with the results reported in Abbas2020 (Accepted), the
%constants can be estimated as 
In line with the discussion of~\cite{abbas}, we can express the
dimensionless constants of Eq.~(\ref{Eq83}) as $E_{1}=M^{\ast 2}/\alpha $
and C=$M^{\ast },$ where $M^{\ast }=v_{1}/v_{Ae}$ is the electron Alfven
Mach number $0.5<M^{\ast }<0.707.$ 
Substituting into Eq.~(\ref{Eq83}), and 
%
%Using these in the equation, we get, 
%\begin{equation}
%\frac{1}{2}\left( v_{x}\frac{\partial B_{\bot }}{\partial x}\right) ^{2}+%
%\frac{B_{\bot }^{2}}{8}\left( 1-\alpha \right) ^{2}+\frac{\alpha B_{\bot
%}^{4}}{8}-\left( M^{\ast 2}-\alpha \right) \frac{B_{\bot }^{2}}{2}-W=0
%\label{Eq85}
%\end{equation}%
introducing the dimensionless parameter $\lambda ^{2}=M^{\ast 2}-\frac{%
\left( 1+\alpha \right) ^{2}}{4}$, we obtain: 
\begin{equation}
\frac{1}{2}\left( v_{x}\frac{\partial B_{\bot }}{\partial x}\right) ^{2}+%
\frac{B_{\bot }^{2}}{8}\left[ \alpha B_{\bot }^{2}-4\lambda ^{2}\right] -W= 
\frac{1}{2}\left( v_{x}\frac{\partial B_{\bot }}{\partial x}\right) ^{2}+V_{%
\mathrm{eff}}-W=0.  \label{Eq86}
\end{equation}%
%
%
%or 
%\begin{equation}
%\frac{1}{2}\left( v_{x}\frac{\partial B_{\bot }}{\partial x}\right)
%^{2}+V_{\mathrm{eff}}-W=0  \label{Eq87}
%\end{equation}
The effective potential now naturally assumes the form of the quartic one
arising in a Duffing oscillator. Hence it is amenable to analytical
considerations as discussed in the following Section.

\section{Analytical and Numerical Solutions}

%%%%%%%%%%% PGK: Change re: L \neq 0
We now tackle the solutions of Eq.~(\ref{Eq82}), focusing as discussed above
on the scenario where $E_2=E_3=0$. We separate the analytically
tractable
and physically relevant
case of $L=0$ which we present here from the numerically examined one
of
$L \neq 0$ that is relegated to Appendix \ref{appA}.
%\subsection{$L=0$ Case}
We notice that in all the relevant quantities that we have examined so far,
we do not simply find derivatives such as $\frac{\partial}{\partial x}$, but
rather these derivatives always appear multiplied by $v_x$. This renders it
rather natural to consider a transformation of coordlnates from $x$ to a new
spatial variable $x^{\prime}$ that ``absorbs'' this factor of $v_x$. This is
done by considering the $x\mapsto x^{\prime}$ transformation defined through 
$\frac{\partial}{\partial x^{\prime}}=v_x\frac{\partial}{\partial x}$. Then
Eq.~\eqref{Eq83} becomes:
\begin{equation}
\frac{1}{2}\left(\frac{\partial B_{\bot }}{\partial x^{\prime}}\right)
^{2}+V_{\mathrm{eff}}=W,  \label{eq:energy}
\end{equation}
with the effective potential $V_{\mathrm{eff}}$ given by:
\begin{equation}
V_{\mathrm{eff}}=-pB_\bot^2+qB_\bot^4,  \label{eq:veff}
\end{equation}
where $p=\lambda^2/2$ and $q=\alpha/8$. Since $V_{\mathrm{eff}}$ is of the
form of a well-known double-well potential or Duffing oscillator, we can
seek solutions in the form of Jacobi elliptic functions \cite{abra}. A key advantage of this
approach is that then the special solitonic solutions of the earlier work of~%
\cite{abbas} merely become special cases of the more general elliptic
function waveforms. In particular, we choose: 
\begin{equation}
B_\bot(x^{\prime})=A~ \text{dn}(bx^{\prime},k),  \label{eq:bperp}
\end{equation}
where $\text{dn}$ is a Jacobi elliptic function and $k$ is the elliptic modulus. 
Bearing in mind that 
\begin{equation*}
\frac{\partial B_\bot}{\partial x^{\prime}}=-bk^2A\mathrm{cn}%
^2(bx^{\prime},k)\mathrm{sn}^2(bx^{\prime},k),
\end{equation*}
and by also using the identities involving the Jacobi elliptic functions: 
\begin{equation*}
k^2\mathrm{cn}^2(bx^{\prime},k)=k^2-1+\mathrm{dn}^2(bx^{\prime},k),
\end{equation*}
\begin{equation*}
k^2\mathrm{sn}^2(bx^{\prime},k)=1-\mathrm{dn}^2(bx^{\prime},k),
\end{equation*}
we get: 
\begin{equation}
\frac{1}{2}\left(\frac{\partial B_\bot}{\partial x^{\prime}}\right)^2=\frac{%
A^2b^2(k^2-1)}{2}+\frac{(2-k^2)b^2}{2}B_\bot^2-\frac{b^2}{2A^2}B_\bot^4.
\label{eq:kineticenergy}
\end{equation}
We substitute \eqref{eq:kineticenergy} to \eqref{eq:energy} and, by
comparing with \eqref{eq:veff}, we infer the solvability conditions: 
\begin{equation}
W=\frac{A^2b^2(k^2-1)}{2}\, , \quad A=\sqrt{\frac{p}{q(2-k^2)}}=\frac{%
2\lambda}{\sqrt{\alpha(2-k^2)}}\quad \mathrm{and}\quad b=\sqrt{\frac{2p}{%
2-k^2}}=\frac{\lambda}{\sqrt{2-k^2}}.  \label{eq:ab}
\end{equation}
Thus, there exists a solution of Eq.~\eqref{eq:energy} of the form of Eq.~\eqref{eq:bperp} 
with $A$ and $b$ given by \eqref{eq:ab}. The inverse
transformation $x^{\prime}\mapsto x$ is defined through the direct $\frac{%
\partial}{\partial x^{\prime}}=v_x\frac{\partial}{\partial x}$
transformation which implies $\frac{{\mathrm{d}} x}{{\mathrm{d}} x^{\prime}}%
=v_x$. Thus, using Eq.~\eqref{MFlux68}, it holds that: 
\begin{equation}
x=\int v_x(x^{\prime}){\mathrm{d}} x^{\prime}=\frac{\alpha}{C}\int \left(
E_1-\frac{1}{2}B_\bot^2\right) {\mathrm{d}} x^{\prime}.
\end{equation}
%where $v_x$ is taken from \eqref{MFlux68}. 
For the specific form of $B_\bot$, the above becomes 
\begin{equation}
x=\frac{\alpha E_1}{C}x^{\prime}-\frac{\alpha A^2}{2bC}E({\rm am}(b~x^{%
\prime},k),k)=Cx^{\prime}-\frac{2\lambda}{C\sqrt{2-k^2}}~E({\rm am}(b~x^{%
\prime},k),k),
\end{equation}
where $E$ stands for the Jacobi integral of the second kind, and ${\rm am}$ for
the Jacobi amplitude function. It is worth noting that the only free
parameter in these solutions is the elliptic modulus $k$. As $k$ varies
between $0\leq k \leq 1$, we switch from a constant (equilibrium) solution
at the minimum of $V_\mathrm{eff}$ (for $k=0$) to a soliton solution of
vanishing energy (per Eq.~\eqref{eq:ab}) for $k=1$. Any intermediate value
gives rise to a periodic solution with periodicity $T=2 K(k)$ where $K(k)$
stands for the complete elliptic integral of the first kind.

%The kind of the possible solutions depend on the the value of $k$. We examine first the case $k=1$.

\subsubsection{The Soliton Solution for $k=1$}

\begin{figure}[tbp]
\includegraphics[height=5cm]{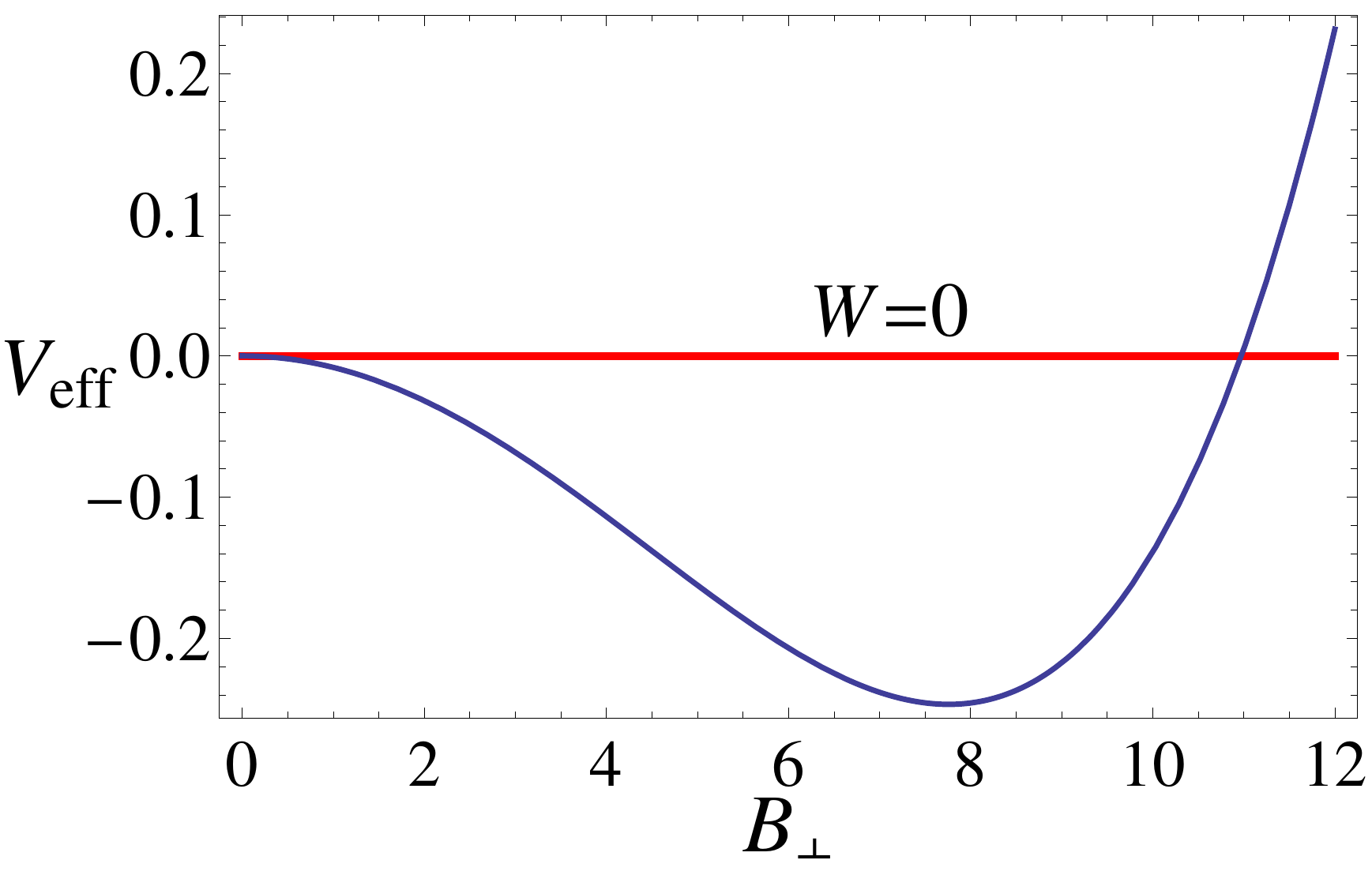}\hspace{0.5cm} %
\includegraphics[height=5cm]{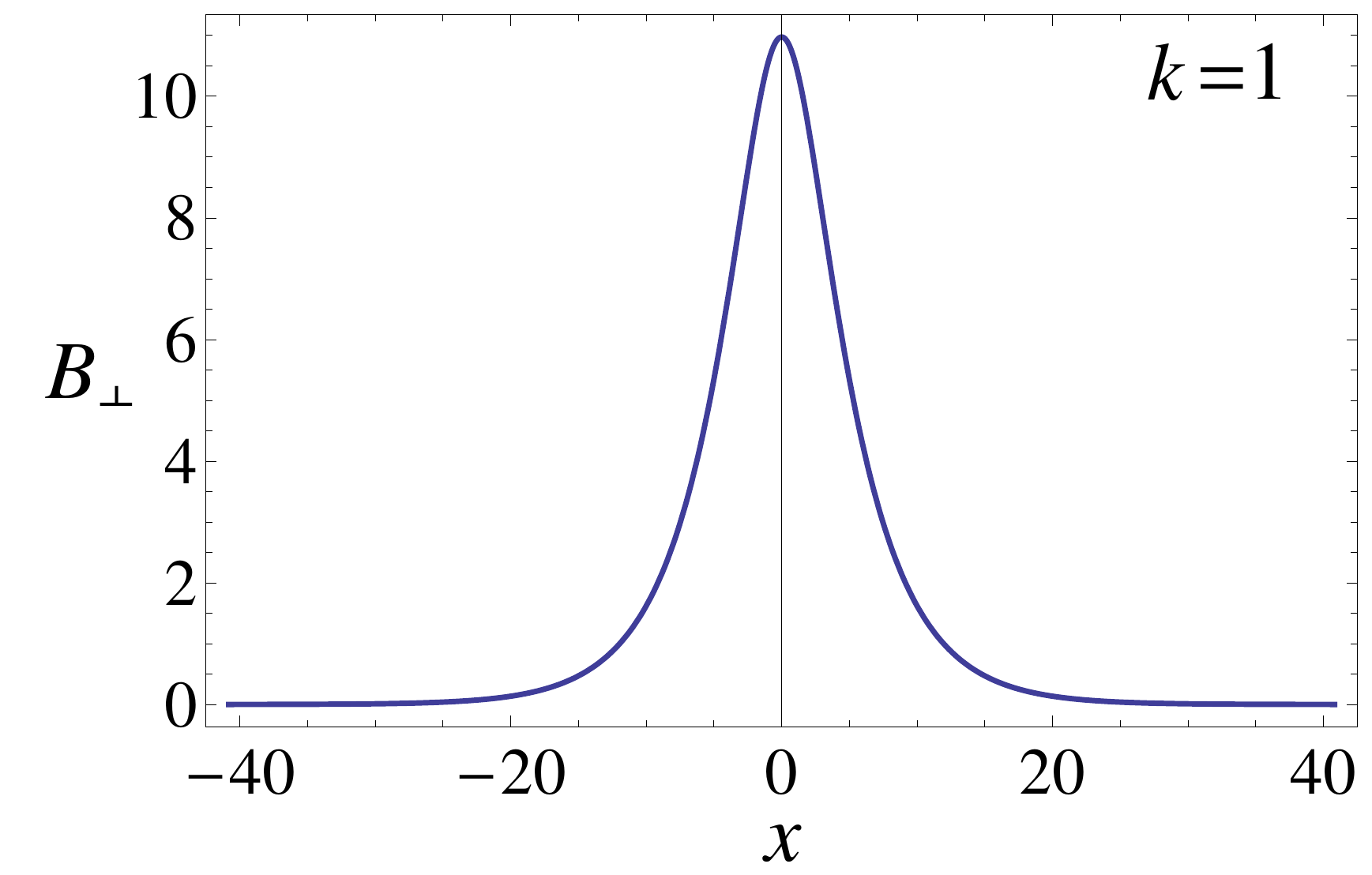}
\caption{Effective Potential (left panel) and soliton solution) 
for $B_\perp$ (right panel, in the case $L=0 $ and $k=1$ as given analytically by Eqs.~(\protect\ref{eq:veff}) 
and (\protect\ref{eq:bperp}).}
\label{fig:V_k_1}
\end{figure}

In this case, $W=0$ as can be seen also in Fig.~\ref{fig:V_k_1}. For all the
calculations in this work, the values of the parameters $C=M^*=0.5164%
\Rightarrow\lambda=0.128$, have been used. Since $W=0$
corresponds to the local maximum of the effective potential $V_\mathrm{eff}$%
, the corresponding solution is a homoclinic one. The form of the $B_\bot$
solution will be a solitonic one as the one depicted in the right panel of
Fig.~\ref{fig:V_k_1}. In particular, for $k=1$, solution \eqref{eq:bperp}
becomes 
\begin{equation}
B_\bot(x^{\prime})=A~\mathrm{sech}(b x^{\prime})=\frac{2\lambda}{\sqrt{\alpha%
}}~\mathrm{sech}(\lambda x^{\prime}),  \label{eq:bperpk1}
\end{equation}
while the $x^{\prime}\mapsto x$ transformation becomes in this case 
\begin{equation}
x=\int_0^{x^{\prime}} v_x(t){\mathrm{d}} t=\frac{1}{C}\int\left[\alpha
E_1-2\lambda^2~\mathrm{sech}^2(\lambda x^{\prime})\right]~{\mathrm{d}}
x^{\prime}=\frac{1}{C}\left[\alpha E_1x^{\prime}-2\lambda~\mathrm{tanh}%
(\lambda x^{\prime})\right]=Cx^{\prime}-\frac{2\lambda}{C}~\mathrm{tanh}%
(\lambda x^{\prime}).  \label{eq:xxpk1}
\end{equation}
The combined result of \eqref{eq:bperpk1} and \eqref{eq:xxpk1} provides the
exact form of $B_\bot=B_\bot(x)$ shown in the right panel of Fig.~\ref%
{fig:V_k_1}.

As mentioned above, the equation for $v_x$ is calculated by the first one
among Eqs.~\eqref{MFlux68}, rewritten here 
\begin{equation*}
v_x=\frac{\alpha}{C} \left( E_1-\frac{1}{2}B_\bot^2\right).
\end{equation*}
%By using the first of Eqs. \eqref{MFlux68}, 
On the other hand, $\theta$ can be calculated through \eqref{Eq81}, which
reads for $E_2=E_3=L=0$ 
\begin{eqnarray}
\frac{\partial \theta }{\partial x^{\prime}}=\frac{\left( 1-\alpha \right) }{%
2} \Rightarrow \theta=\frac{\left( 1-\alpha \right) }{2}x^{\prime},
\label{eq:theta}
\end{eqnarray}
by considering (without loss of generality) that the constant of
integration $\theta_0=0$. Using %
\eqref{eq:theta} and \eqref{eq:bperpk1}, we can calculate $B_y$ and $B_z$
through $B_{y}=B_{\bot }\cos \theta ,B_{z}=B_{\bot }\sin \theta$, while $n$
can be calculated through $n=\frac{C}{v_x}.$ The fields $v_{iy,ey}$ and $%
v_{iz,ez}$ are calculated by Eqs.~\eqref{Vy69} and \eqref{Vz70}, namely (for our
case of $E_2=E_3=0$):
\begin{equation}
v_{iy,ey}=\frac{\alpha B_{y}}{C}\mathbf{+}\frac{\left( -\alpha ,1\right) }{C}%
\frac{\partial B_{z}}{\partial x^{\prime}},
\end{equation}%
\begin{equation}
v_{iz,ez}=\frac{\alpha B_{z}}{C}\mathbf{+}\frac{\left( \alpha ,-1\right) }{C}%
\frac{\partial B_{y}}{\partial x^{\prime}},
\end{equation}
and finally $E_x$ is given by the first of \eqref{ElecEq63}, which in the
present case reads 
\begin{equation}
E_x=\frac{1}{\alpha}\frac{\partial v_x}{\partial x^{\prime}}%
-(v_{iy}B_z-v_{iz}B_y).  \label{elec}
\end{equation}
Thus, by recalling also \eqref{eq:bperpk1} and \eqref{eq:xxpk1}, we can
calculate all the fields that are needed for the full description of our
system. The calculation for the present case is shown in Fig.~\ref%
{fig:Fields_k_1}. These results are in line with the recent analysis of Ref.~\cite%
{abbas} and illustrate the ability of our formulation to not only retrieve
these earlier findings (in an arguably more intuitive fashion from a dynamical
systems point of view), but also the potential to extend them as will be
done below. It is worthwhile to also note that the $B_\bot$ field, as well
as ones that depend directly on that such as $v_x$ and $n$, have a ``regular''
solitonic form. However, quantities involving the components of the magnetic
field and the ones of the velocities (for both ions and electrons) in the
transverse directions feature oscillations whose origin is now more
transparent. These stem from the polar decomposition of the transverse
magnetic field, endowing one of its components with a cosinusoidal and
another with a sinusoidal variation, so that the relevant fields bear
substantial resemblance to the notions of envelope solitons~\cite{rem}.

\begin{figure}[tbp]
\begin{tabular}{ccc}
\includegraphics[scale=0.3]{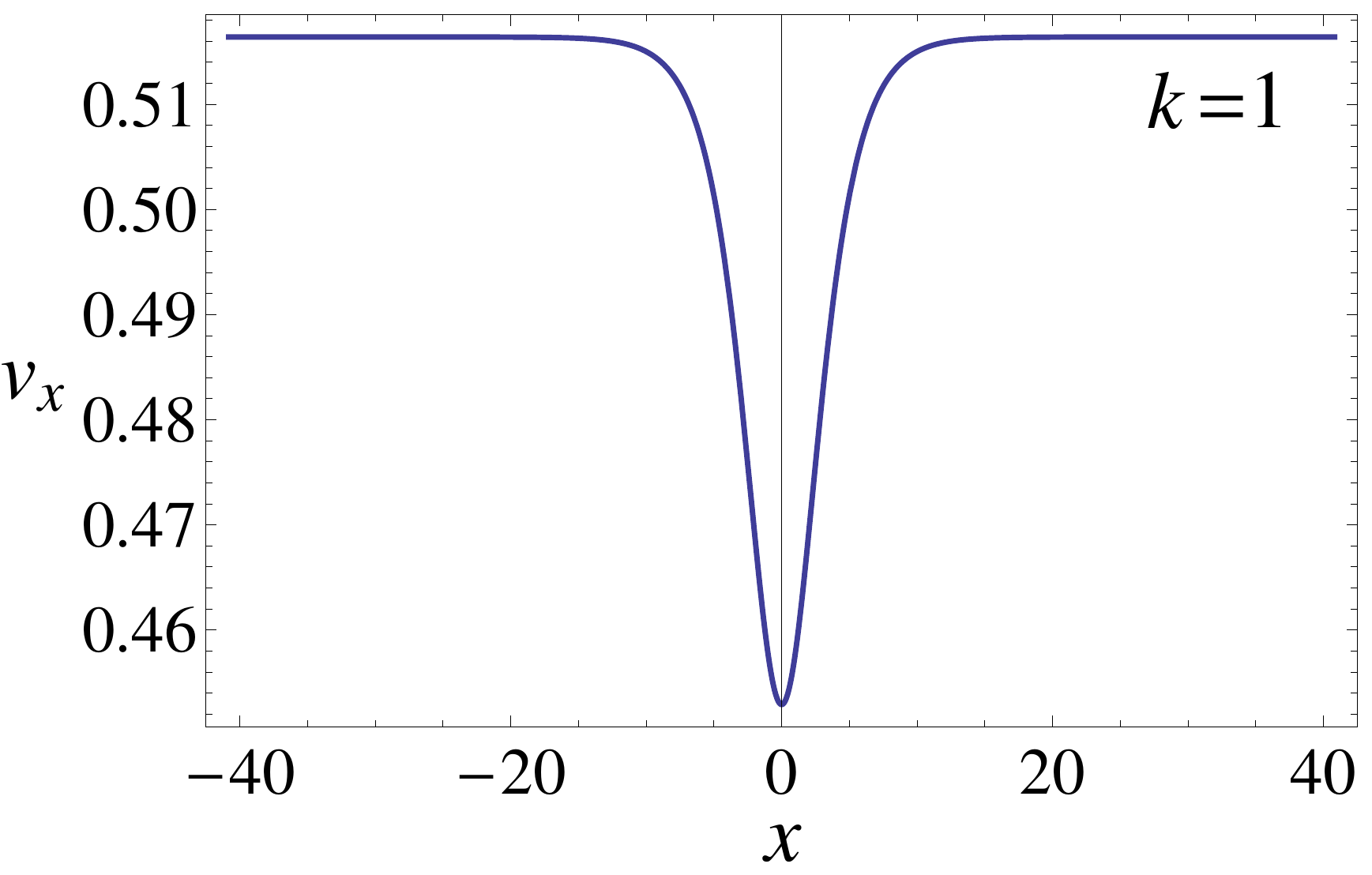} & %
\includegraphics[scale=0.3]{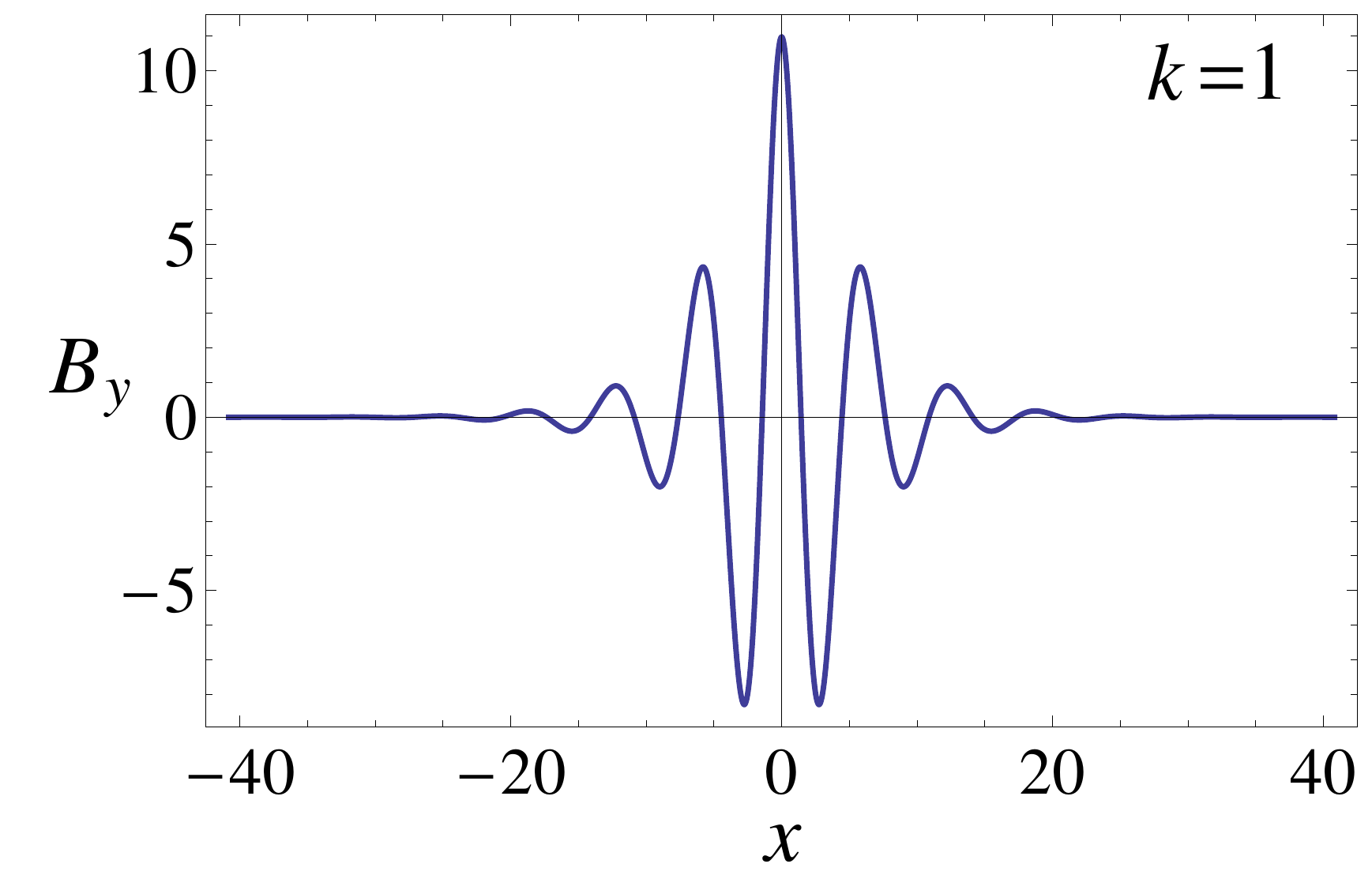} & %
\includegraphics[scale=0.3]{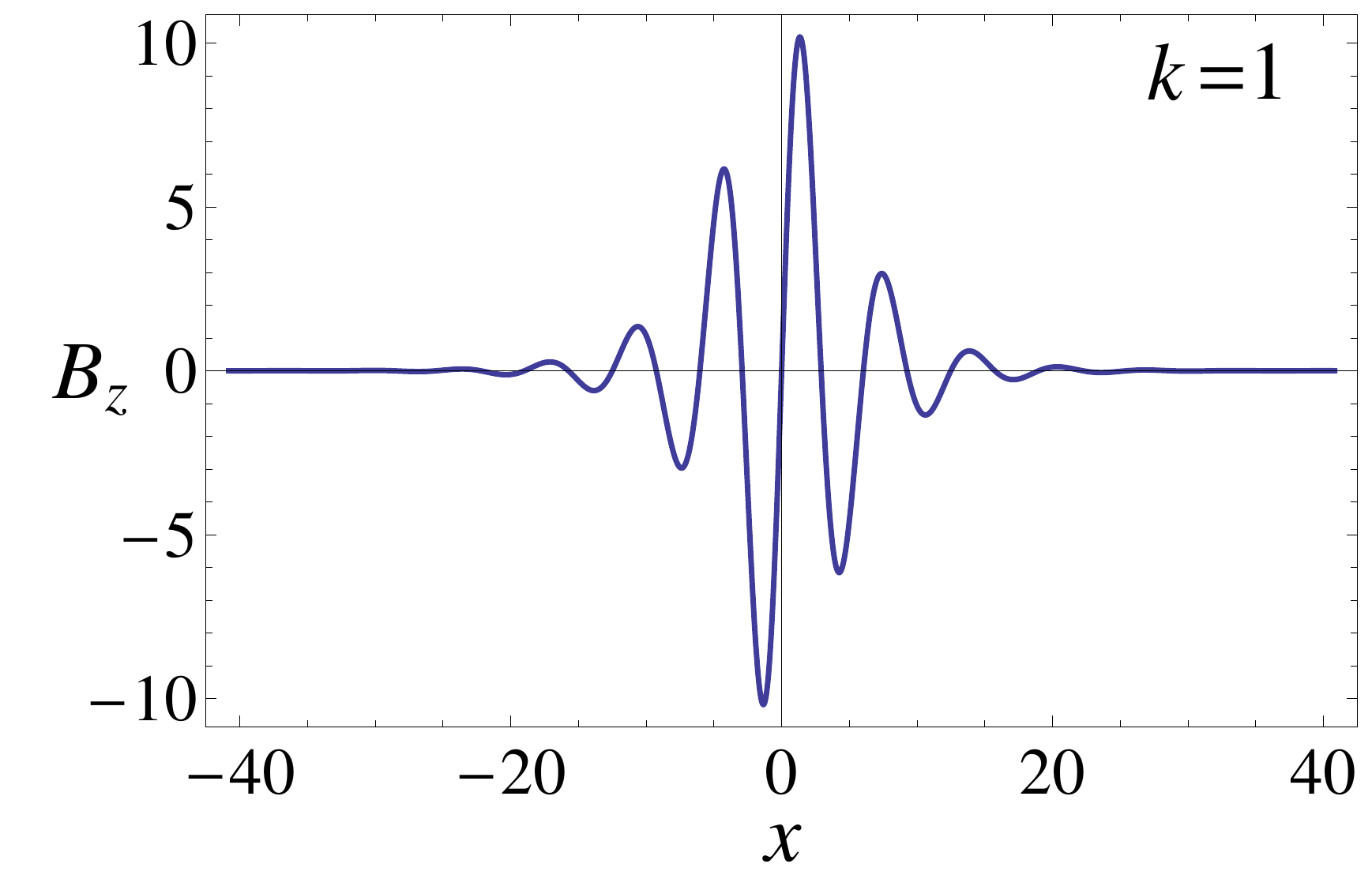} \\ 
\includegraphics[scale=0.3]{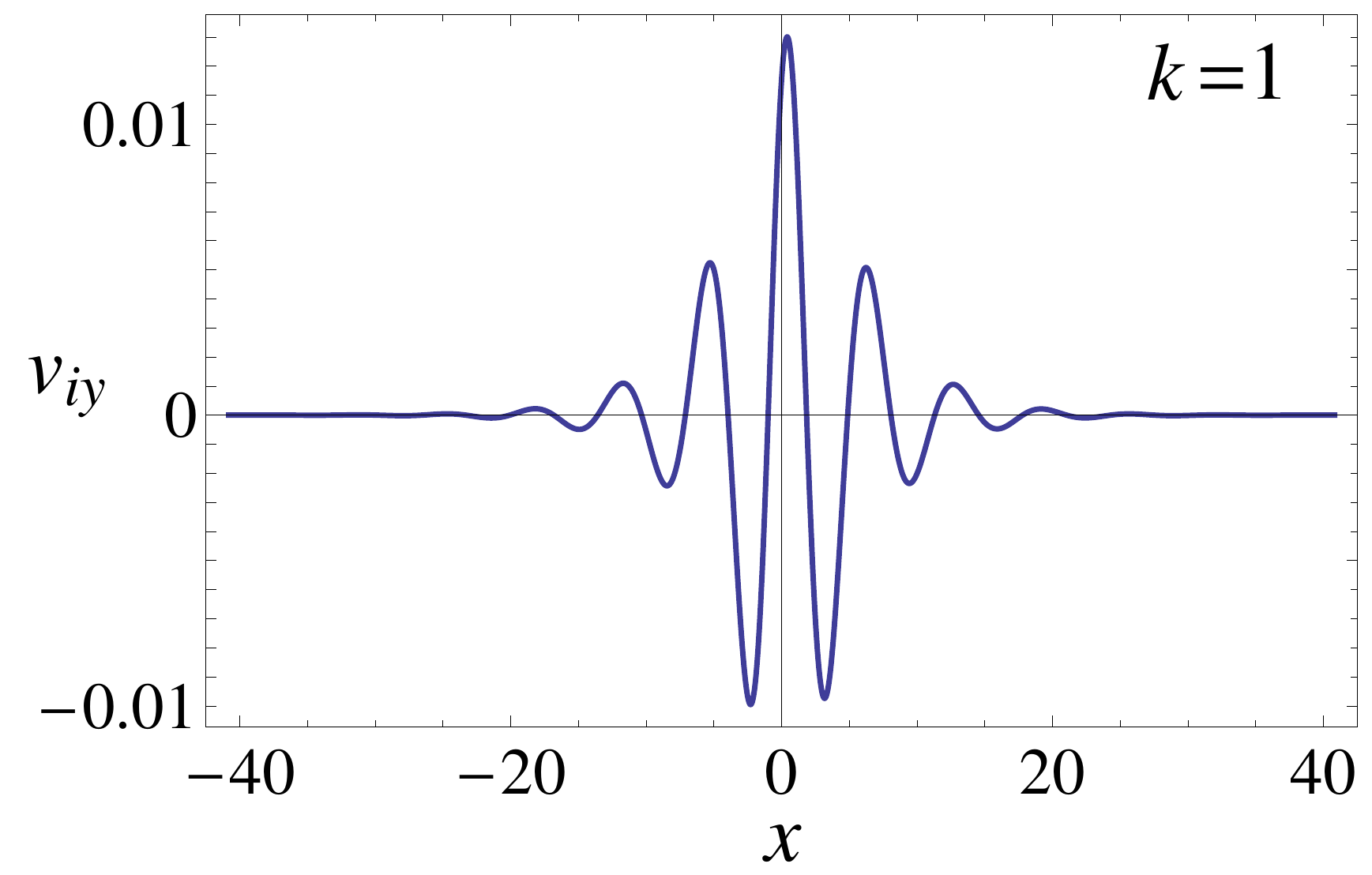} & %
\includegraphics[scale=0.3]{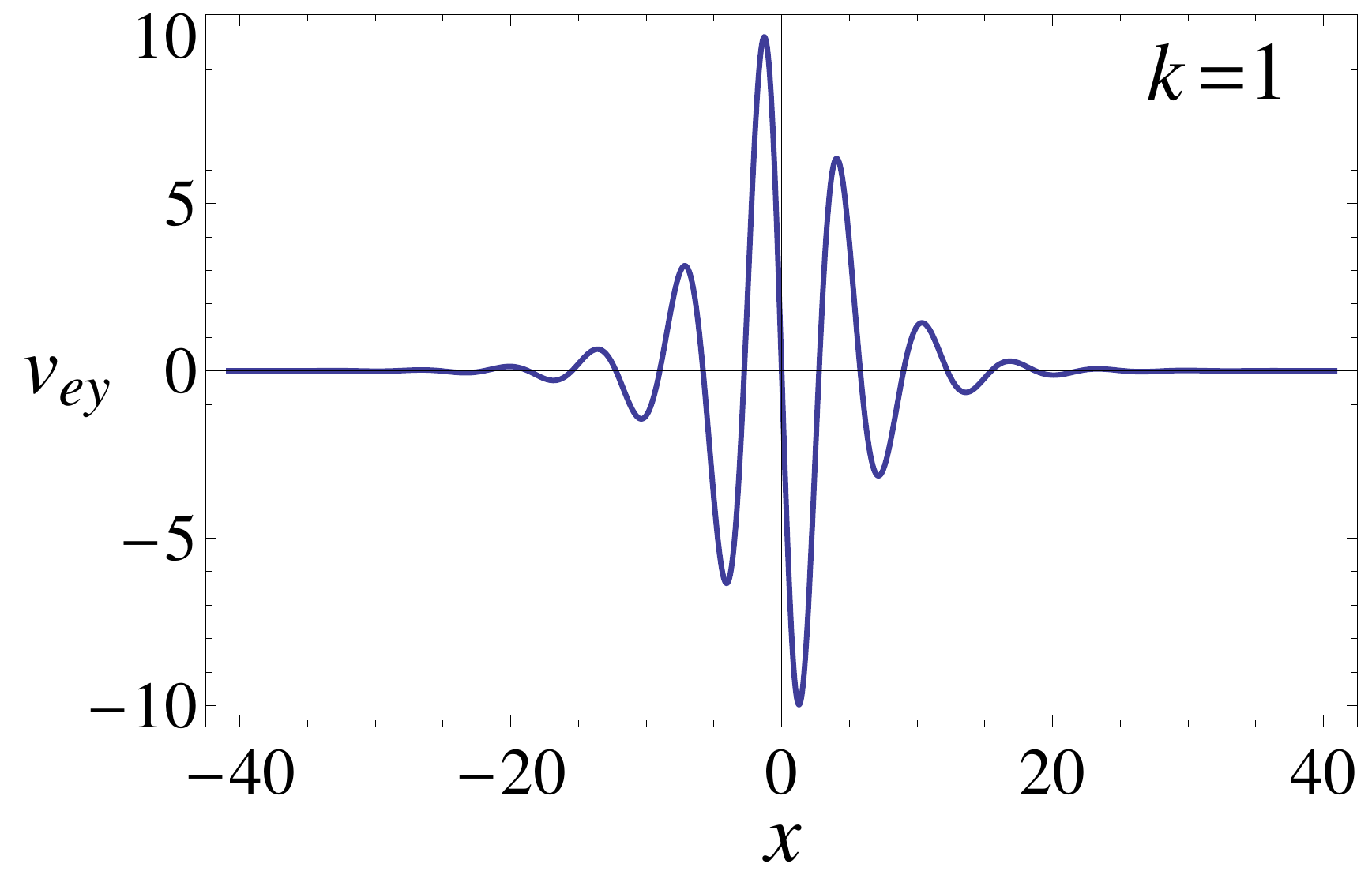} & %
\includegraphics[scale=0.3]{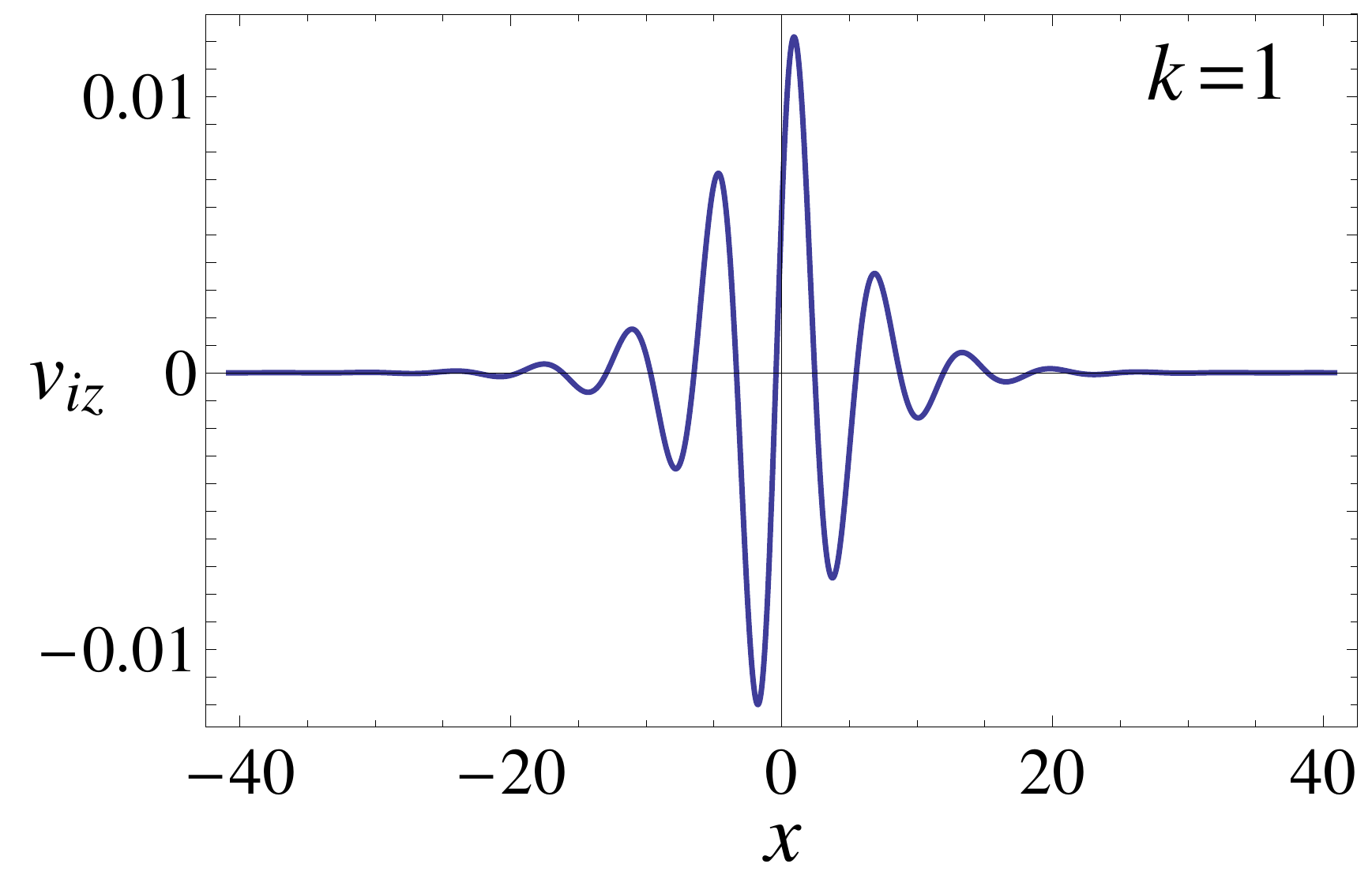} \\ 
\includegraphics[scale=0.3]{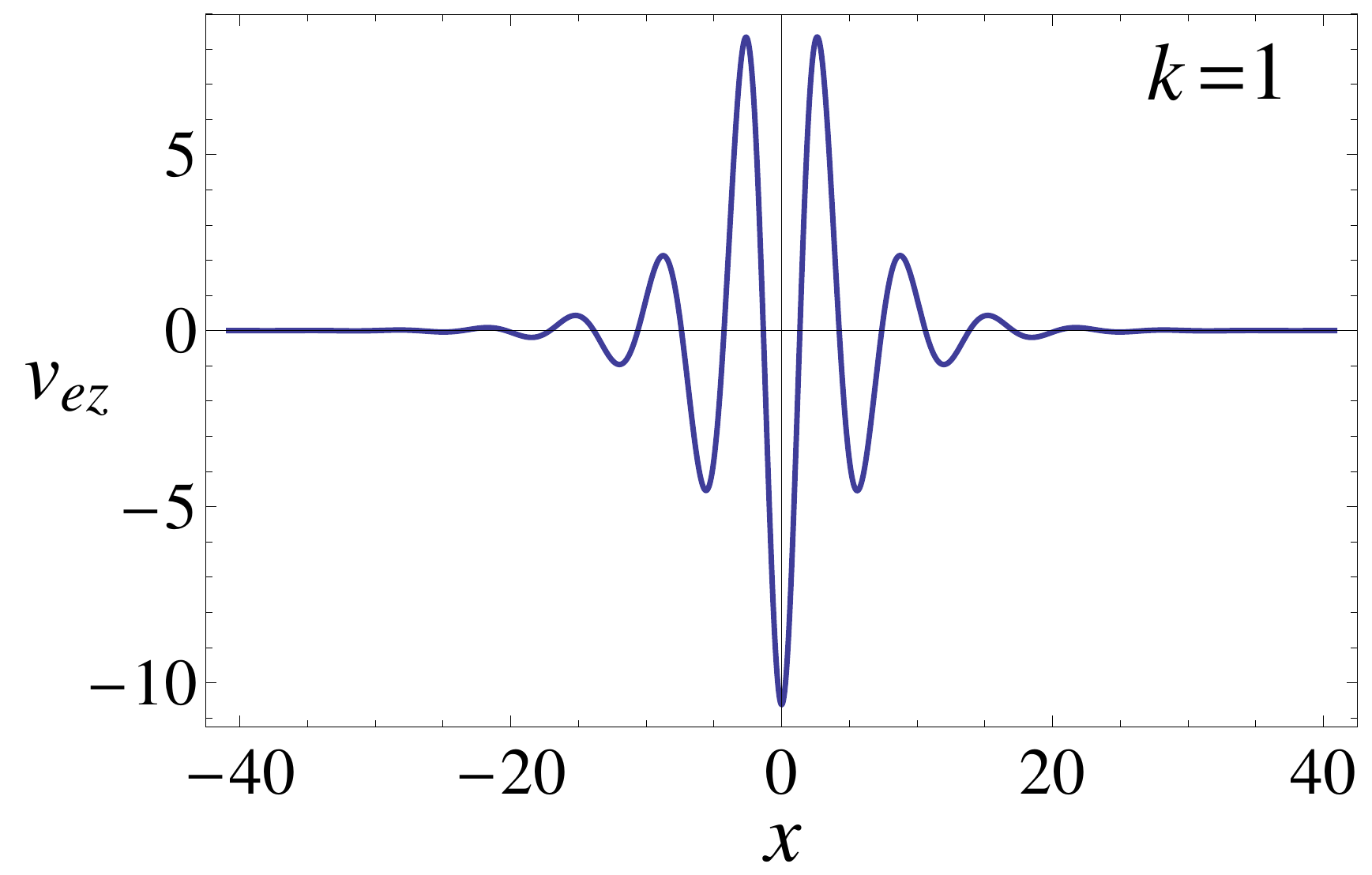} & %
\includegraphics[scale=0.3]{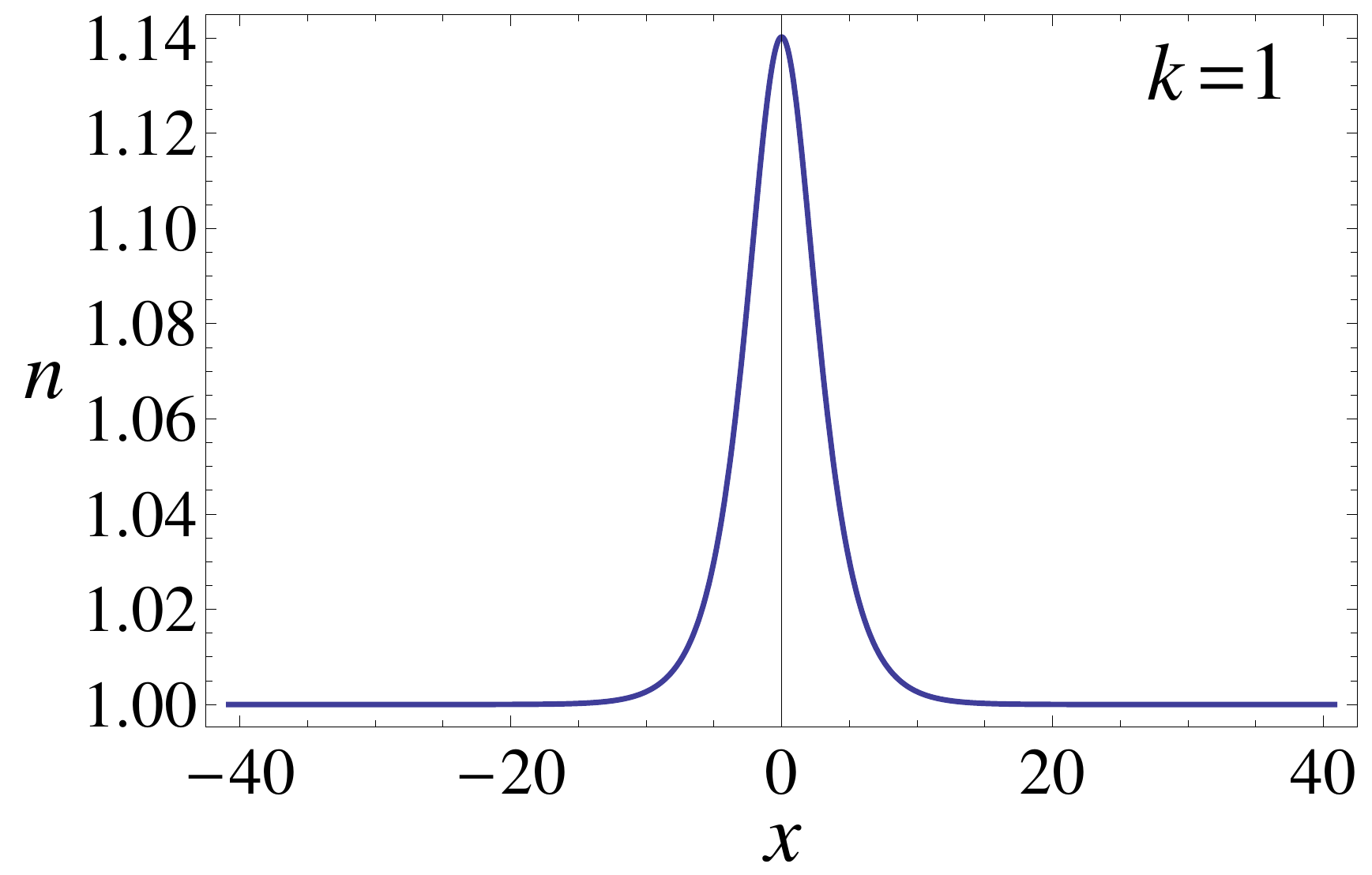} & %
\includegraphics[scale=0.3]{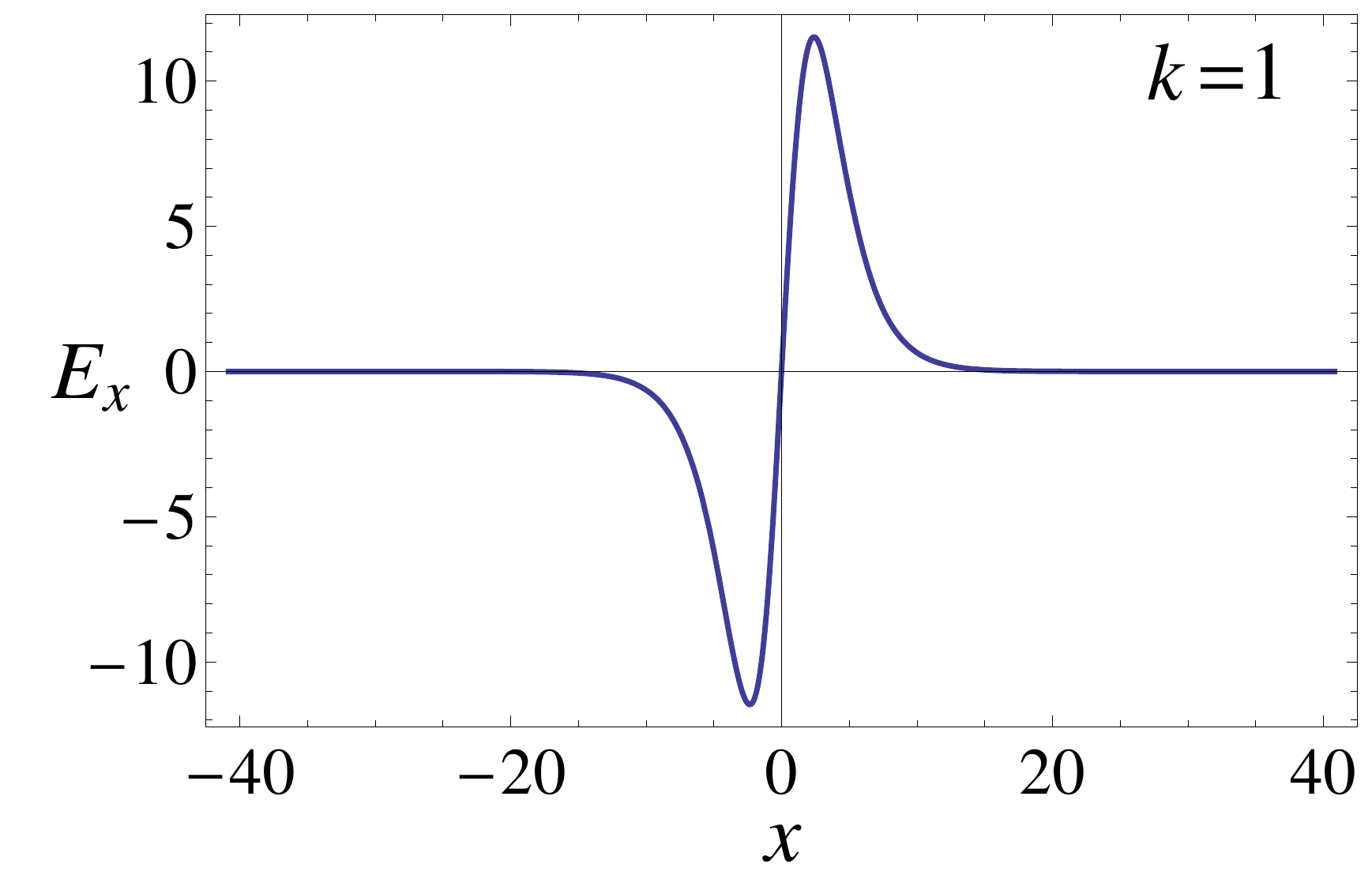} \\ 
&  & 
\end{tabular}%
\caption{The nine fields (5 velocity components, including a common one, for
ions and electrons, 2 transverse magnetic field components, a longitudinal
electric field and a charge density) which describe the system for $L=0$ and $k=1$.}
\label{fig:Fields_k_1}
\end{figure}

\subsubsection{Periodic solutions for $k=0.95$ and $k=0.4$ ($L=0$)}

\begin{figure}[tbp]
\includegraphics[height=5cm]{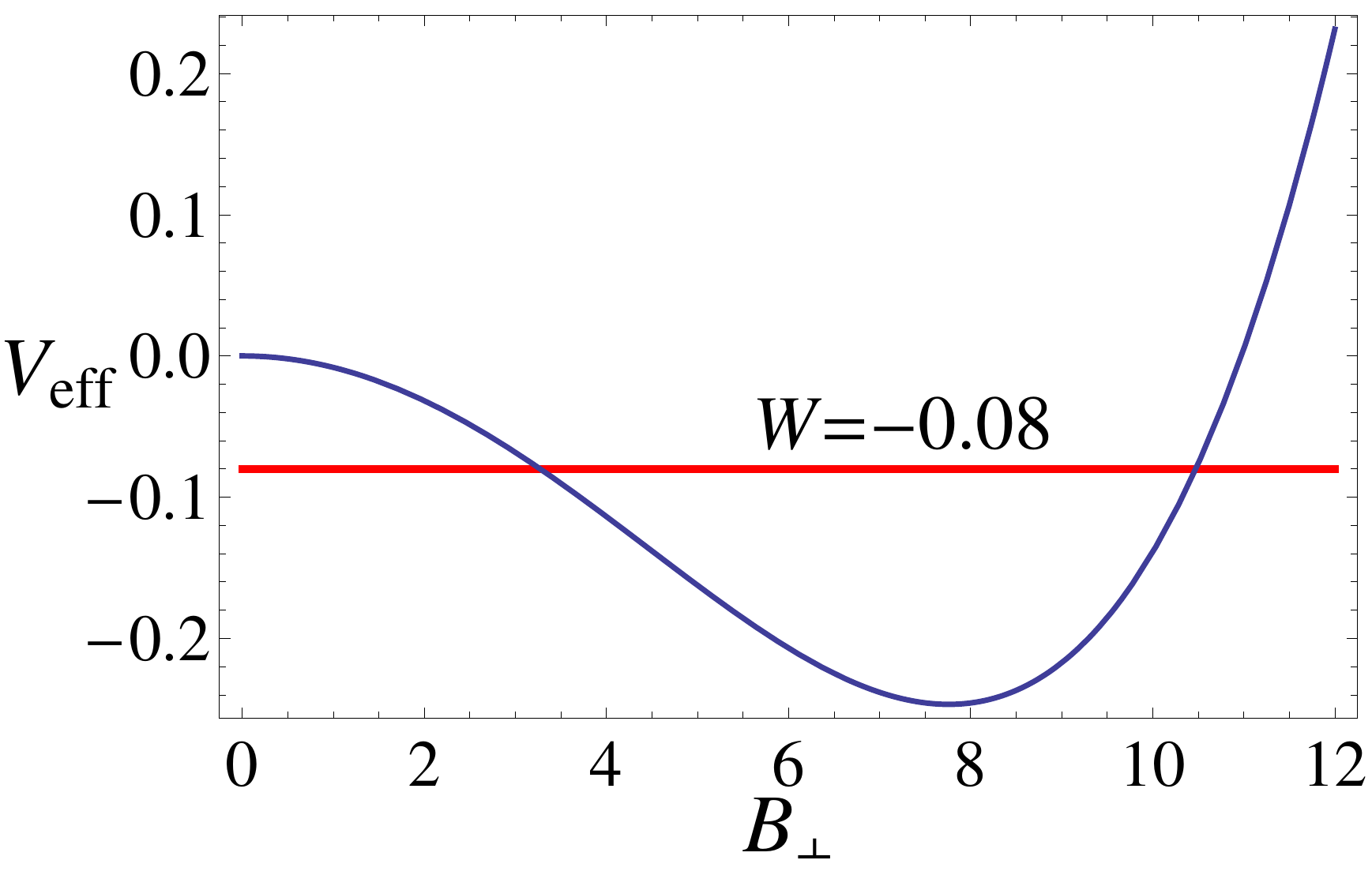}\hspace{0.5cm} %
\includegraphics[height=5cm]{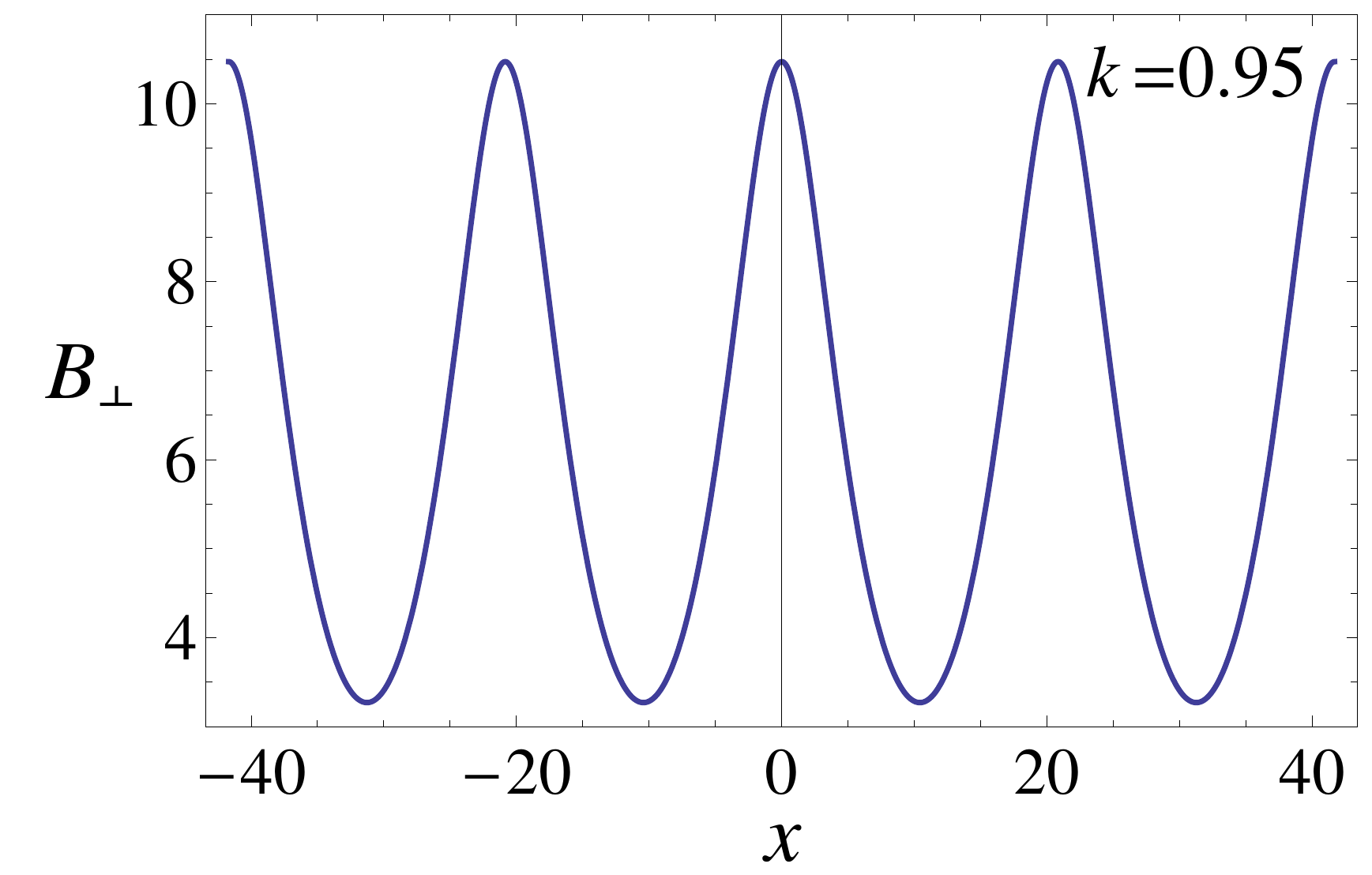}
\caption{Effective potential (left panel) and periodic solution 
of $B_\perp$ (right panel) for $L=0$ and $k=0.95$, as given analytically by Eqs.~(\protect\ref{eq:veff}) 
and (\protect\ref{eq:bperp}).}
\label{fig:V_k_0_95}
\end{figure}

We now turn to the generalization involving the genuinely periodic state
solutions. When $k\neq 1$, the total energy is $W<0$. In particular, for $%
k=0.95$ we get $W=-0.08$, as can be seen in the left panel of Fig.~\ref%
{fig:V_k_0_95}. The corresponding solution of $B_\bot$ is now indeed
periodic and not solitonic; see the right panel of Fig.~\ref{fig:V_k_0_95}.

\begin{figure}[tbp]
\begin{tabular}{ccc}
\includegraphics[scale=0.3]{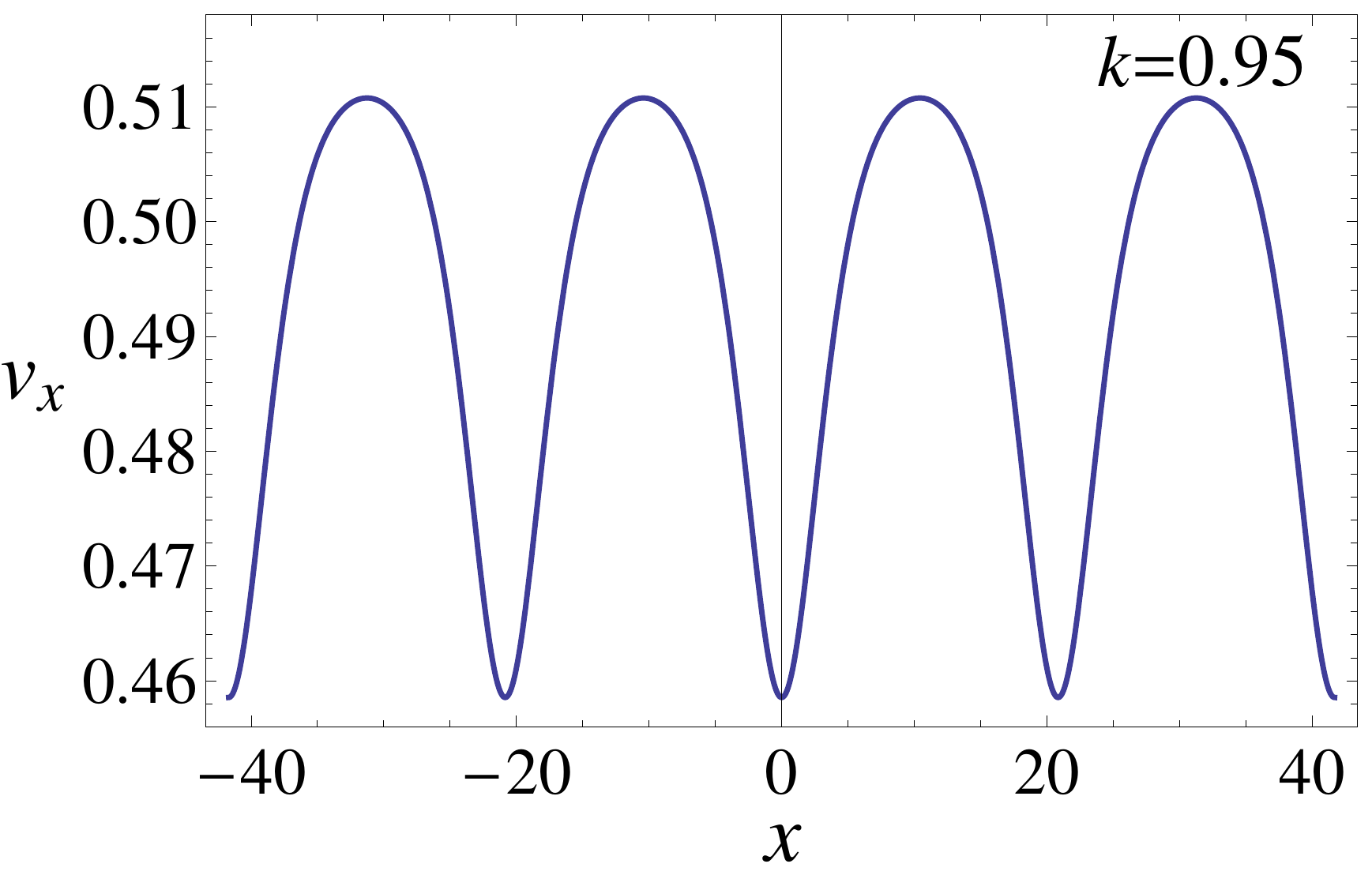} & %
\includegraphics[scale=0.3]{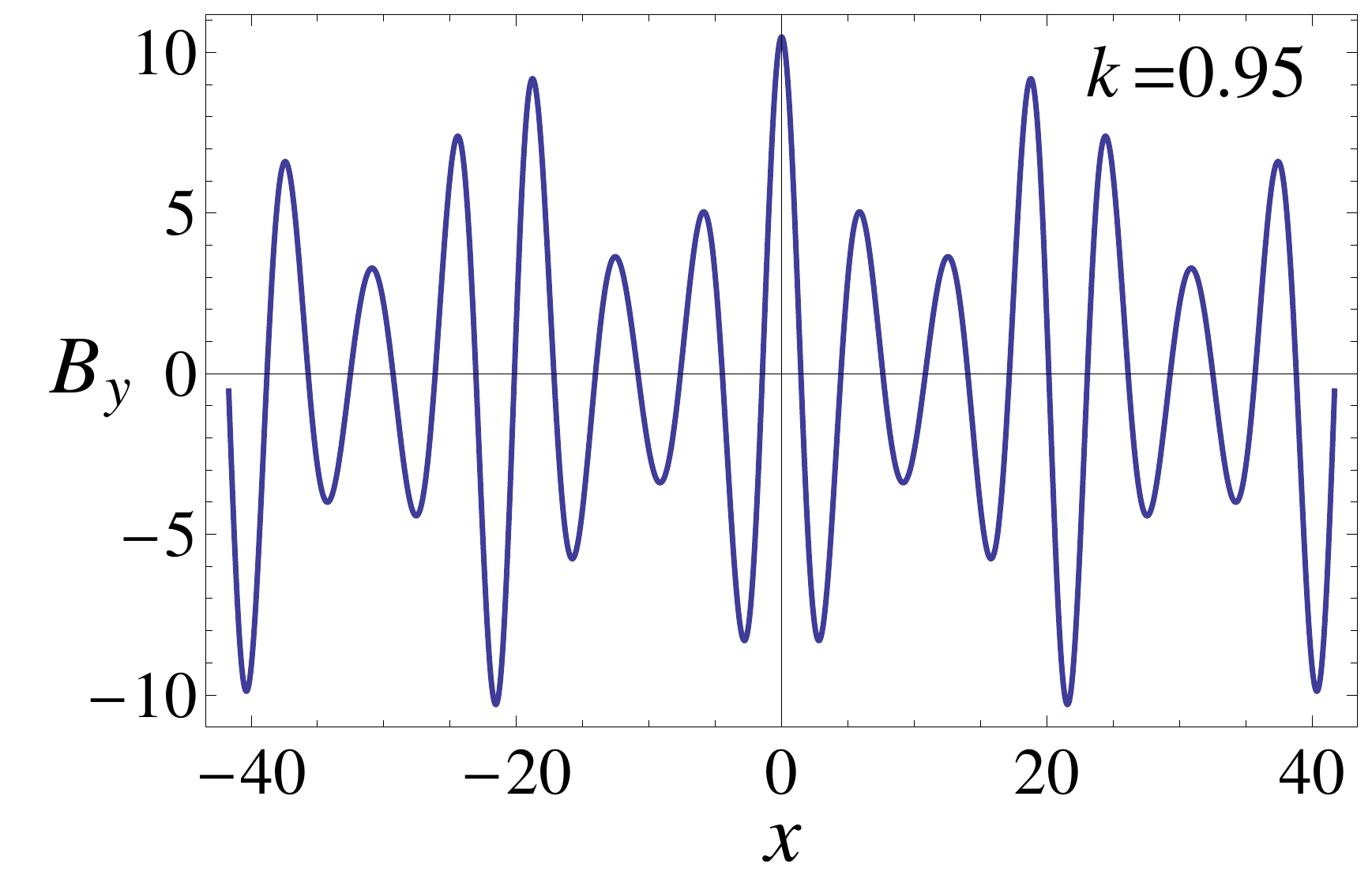} & %
\includegraphics[scale=0.3]{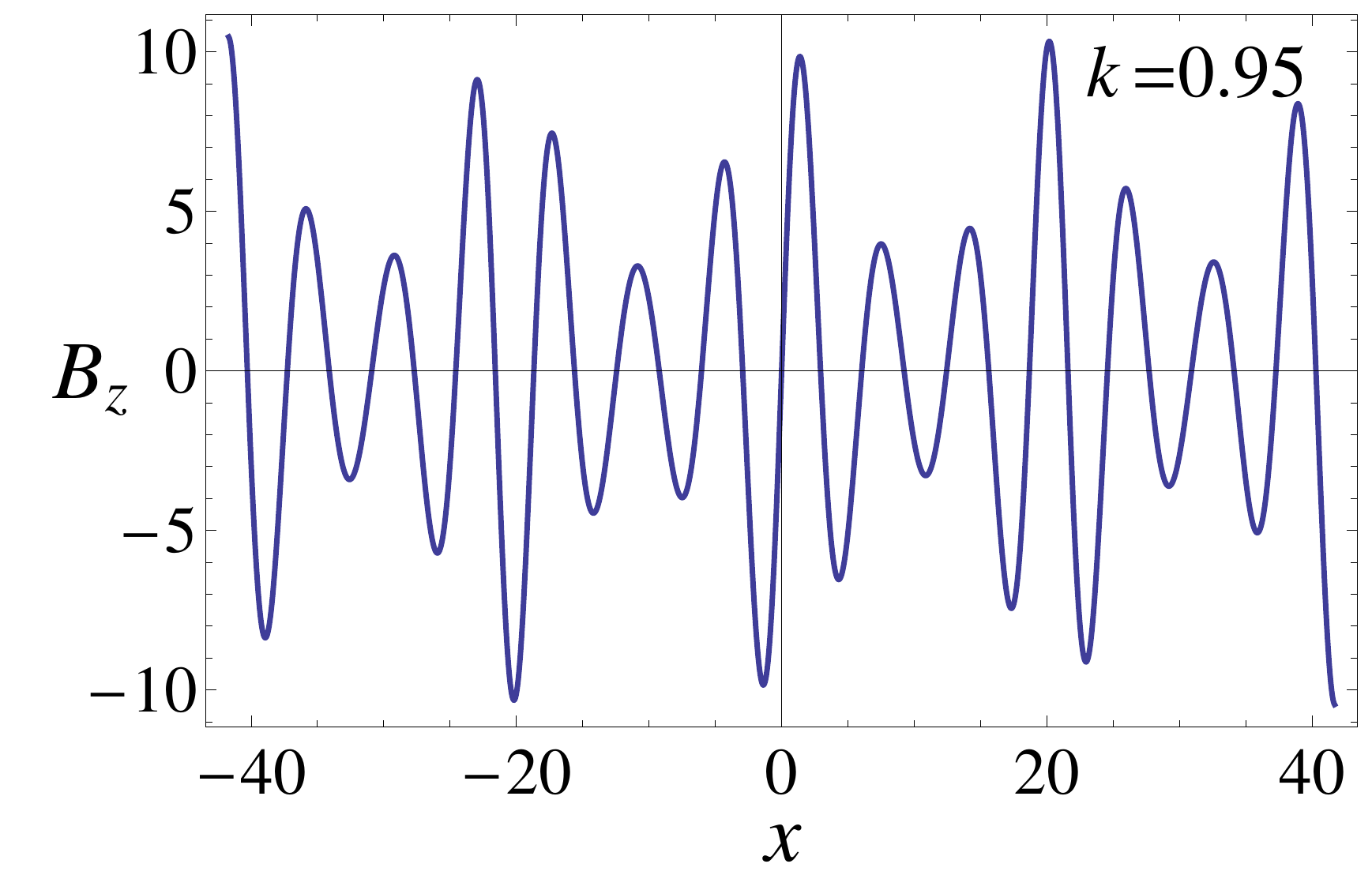} \\ 
\includegraphics[scale=0.3]{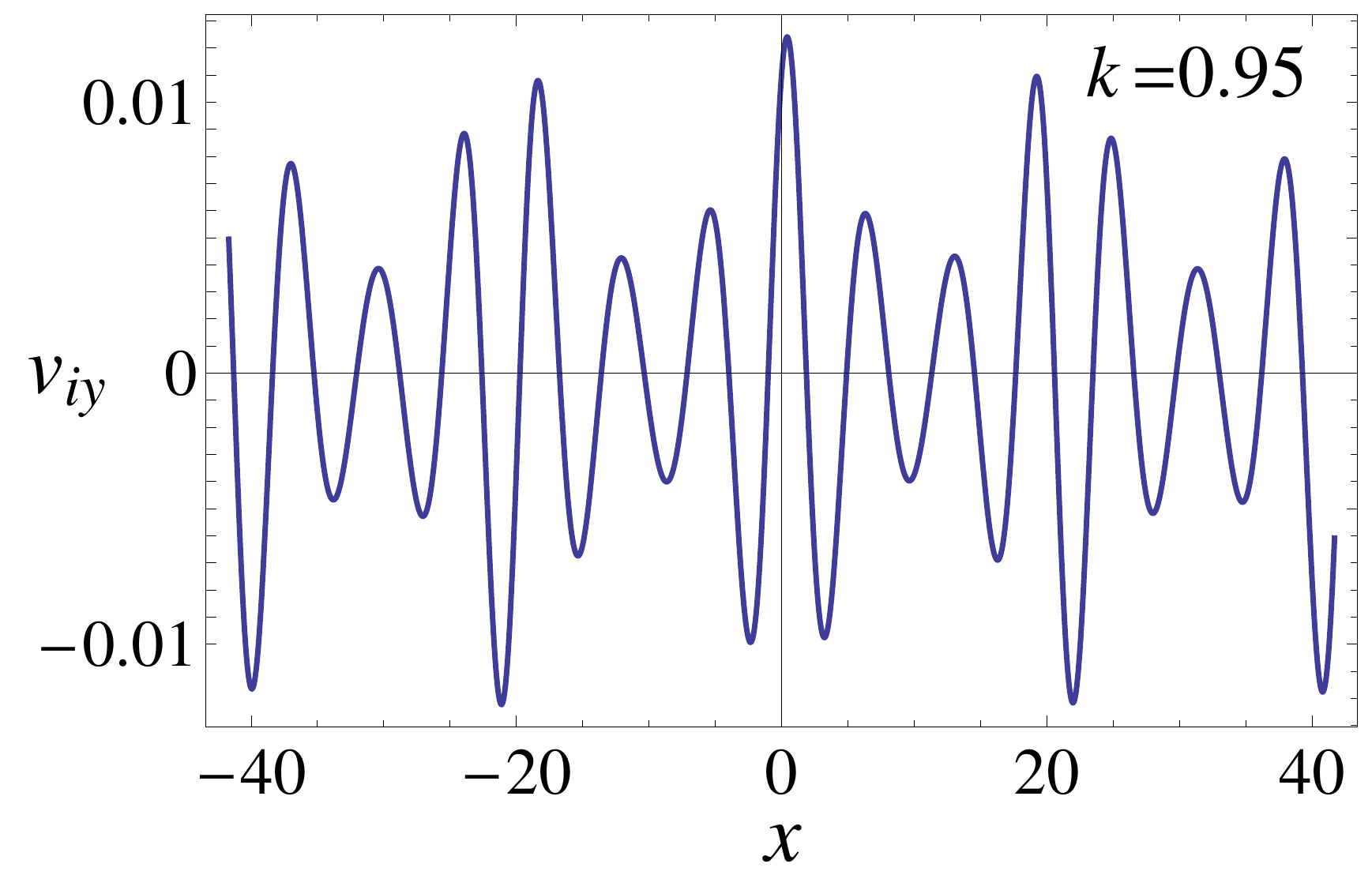} & %
\includegraphics[scale=0.3]{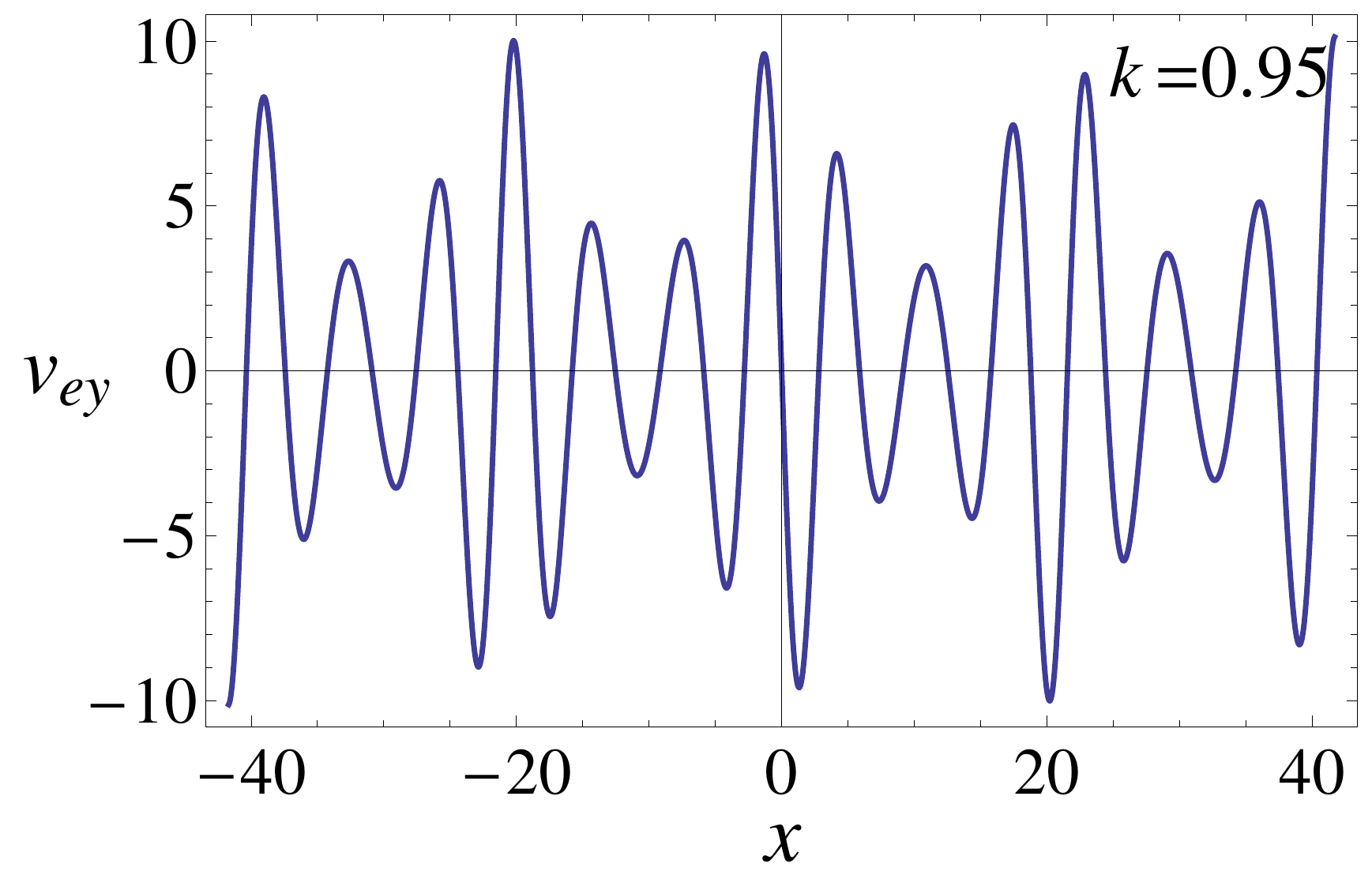} & %
\includegraphics[scale=0.3]{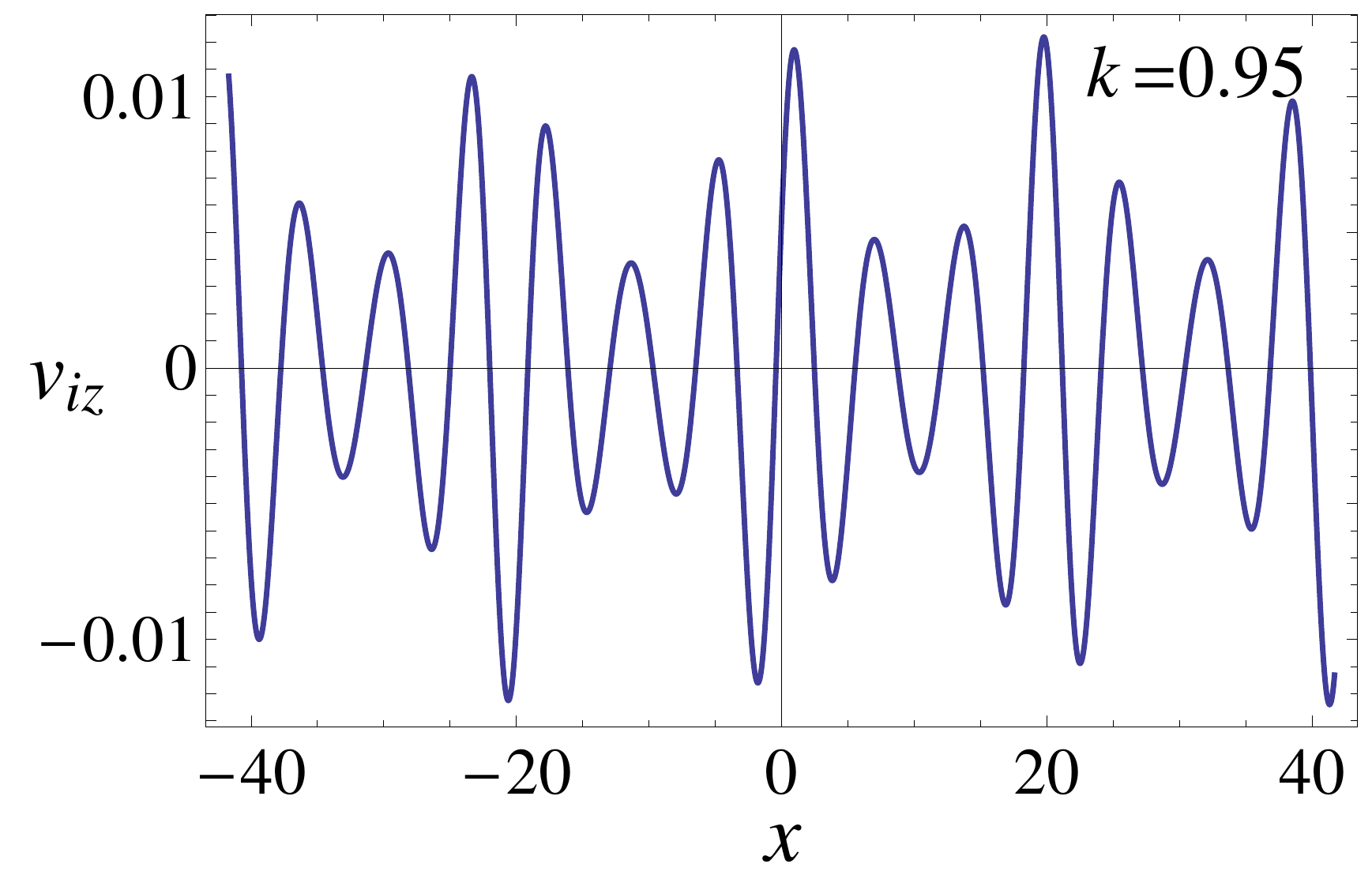} \\ 
\includegraphics[scale=0.3]{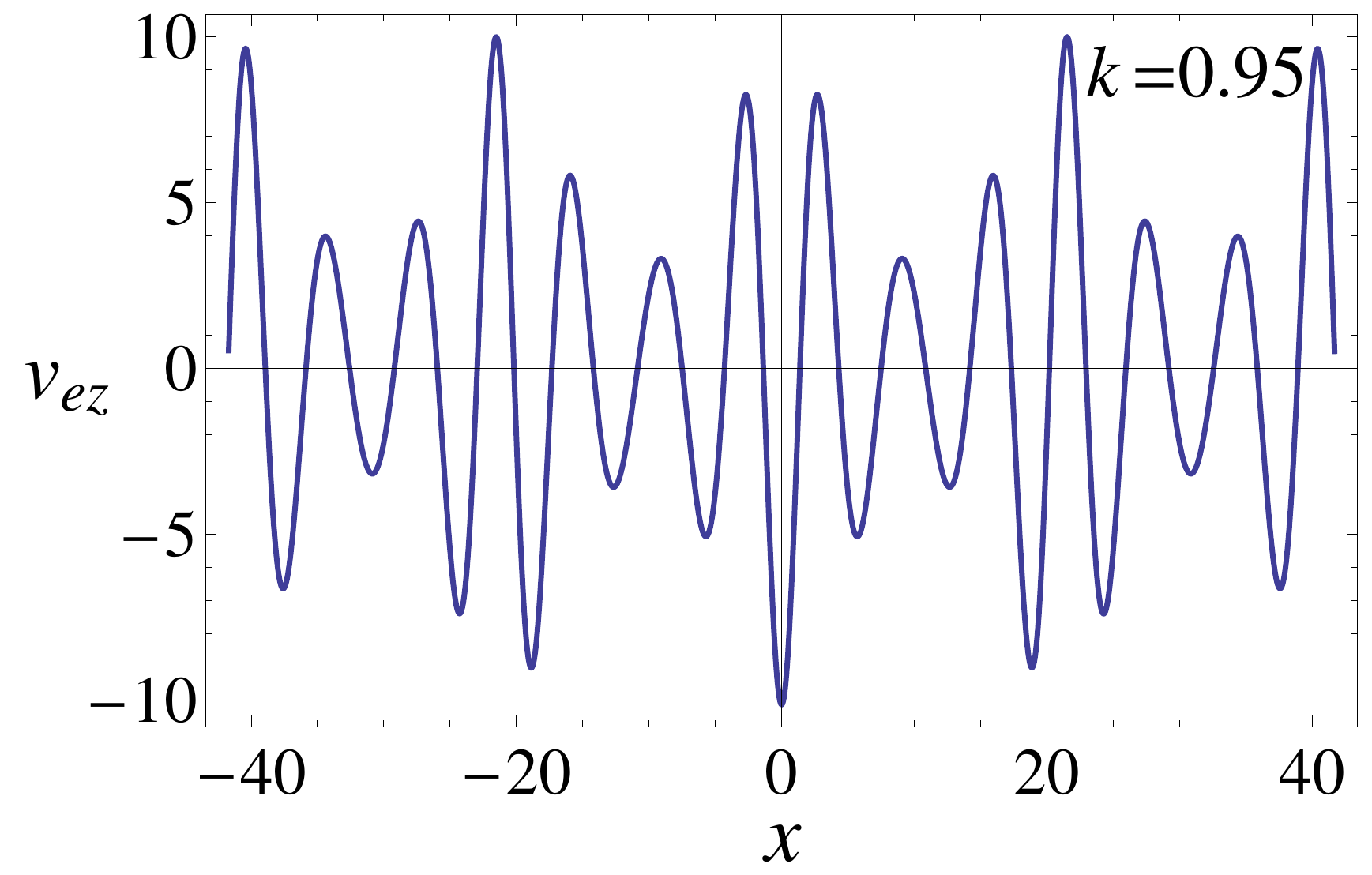} & %
\includegraphics[scale=0.3]{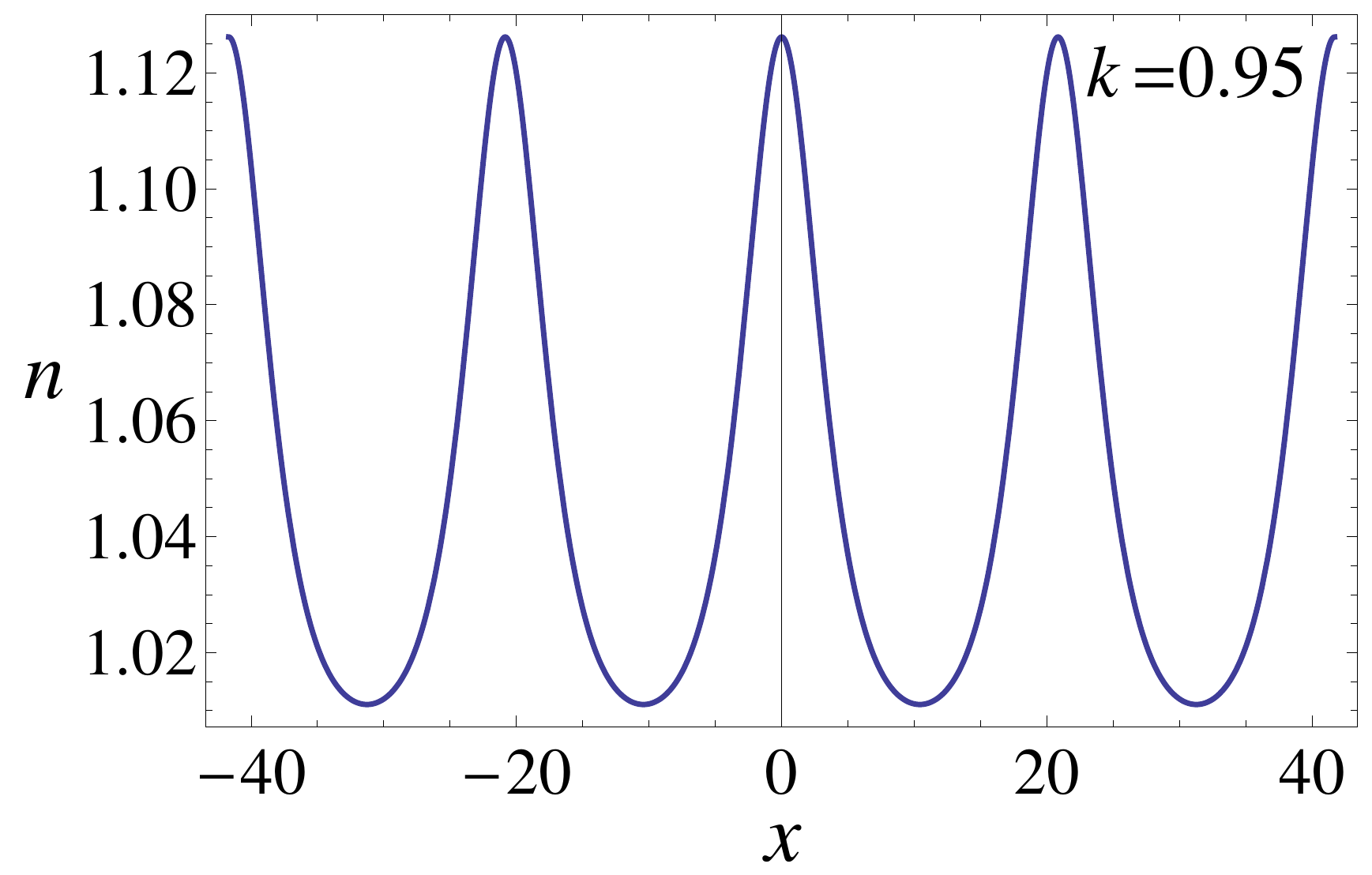} & %
\includegraphics[scale=0.3]{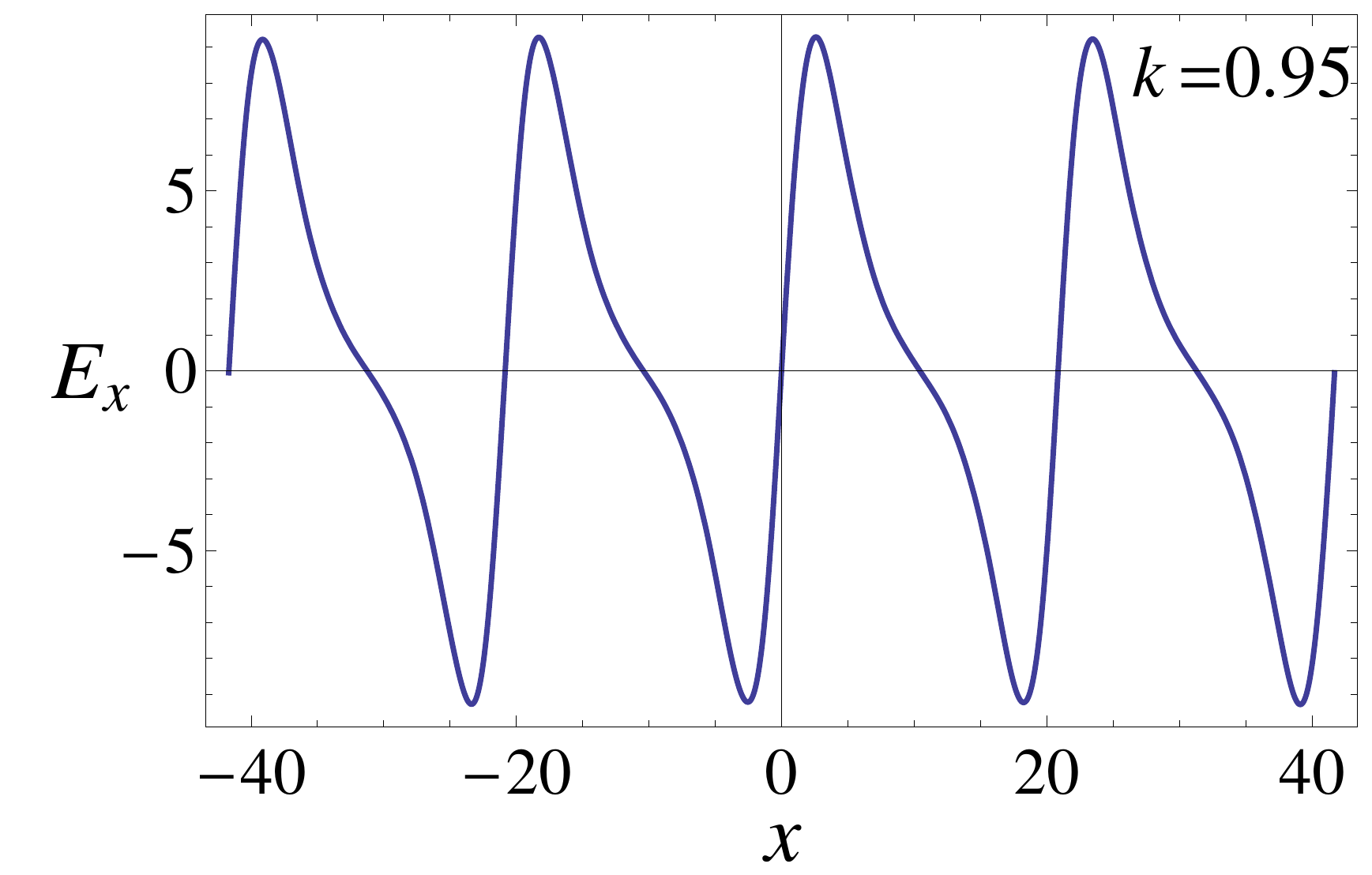} \\ 
&  & 
\end{tabular}%
\caption{Reconstruction of the nine fields describing the system, as was
done previously in Fig.~\protect\ref{fig:Fields_k_1}, but now the solutions
with $L=0$ and $k=0.95$.}
\label{fig:Fields_k_0_95}
\end{figure}

Importantly, the same reconstruction path of Eqs.~(\ref{eq:xxpk1})-(\ref%
{elec}) can be utilized to obtain \textit{all} 9 of the associated
fields. In Fig.~\ref{fig:Fields_k_0_95}, the corresponding quantities and
their spatial profiles (with respect to $x$) are depicted. In the panels of $%
B_\bot$ and $v_x$ a clear elliptic function behavior can be recognized. The
same periodic behavior is obvious for $n$ and $E_x$; notice the analogy of
all of these features with the limiting case of $k \rightarrow 1$, where
essentially the additional periods of the central wave are pushed to 
$\infty$. On the other hand, the behavior of the rest of the examined fields is
quasi-periodic, due to the simultaneous action of two periodic quantities
bearing different periodicities, namely $B_\bot$ and $\theta$. Nevertheless,
this quasi-periodic pattern can be analytically constructed through the
decompositions and building blocks presented herein.

\begin{figure}[tbp]
\includegraphics[height=5cm]{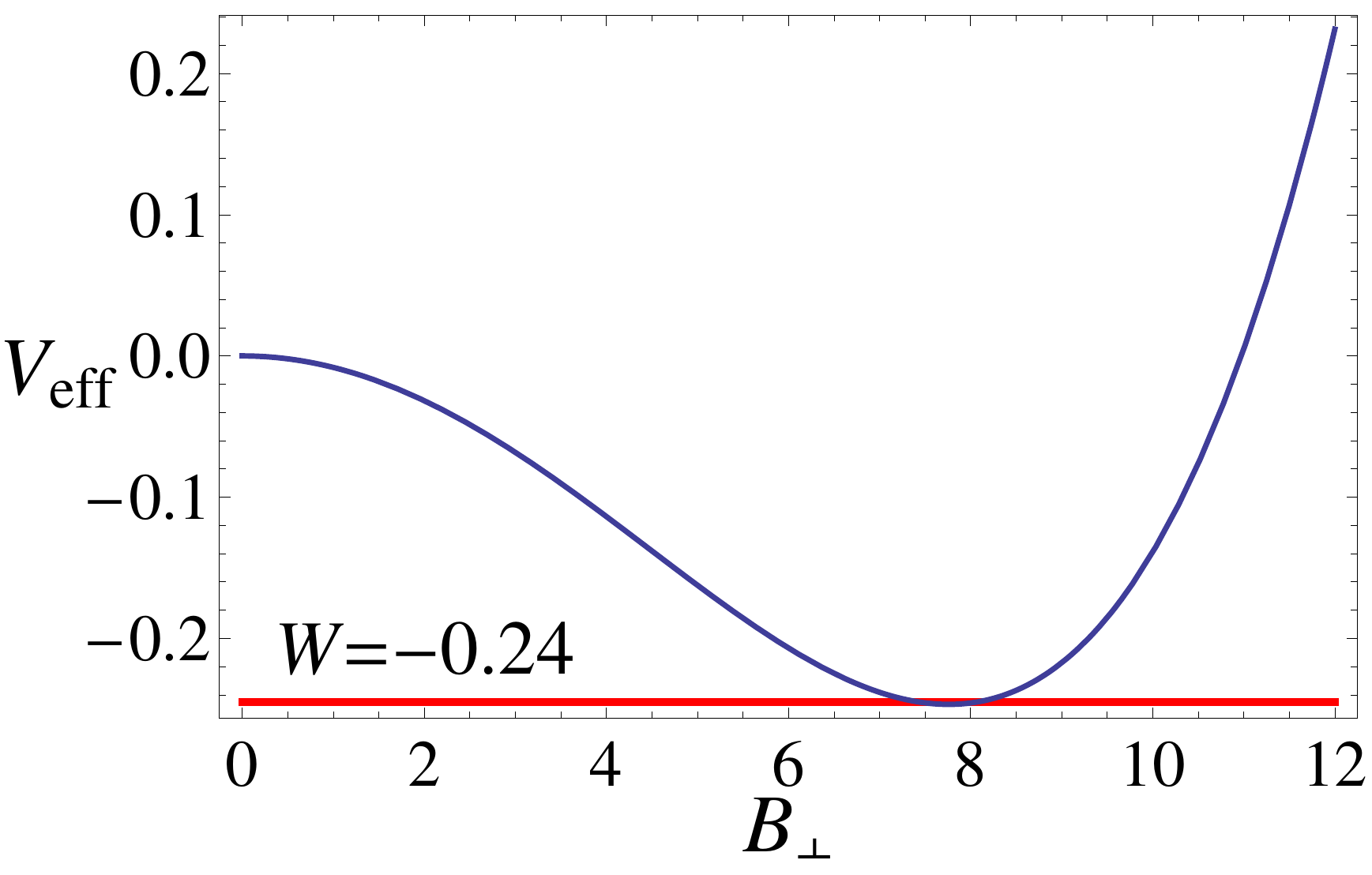}\hspace{0.5cm} %
\includegraphics[height=5cm]{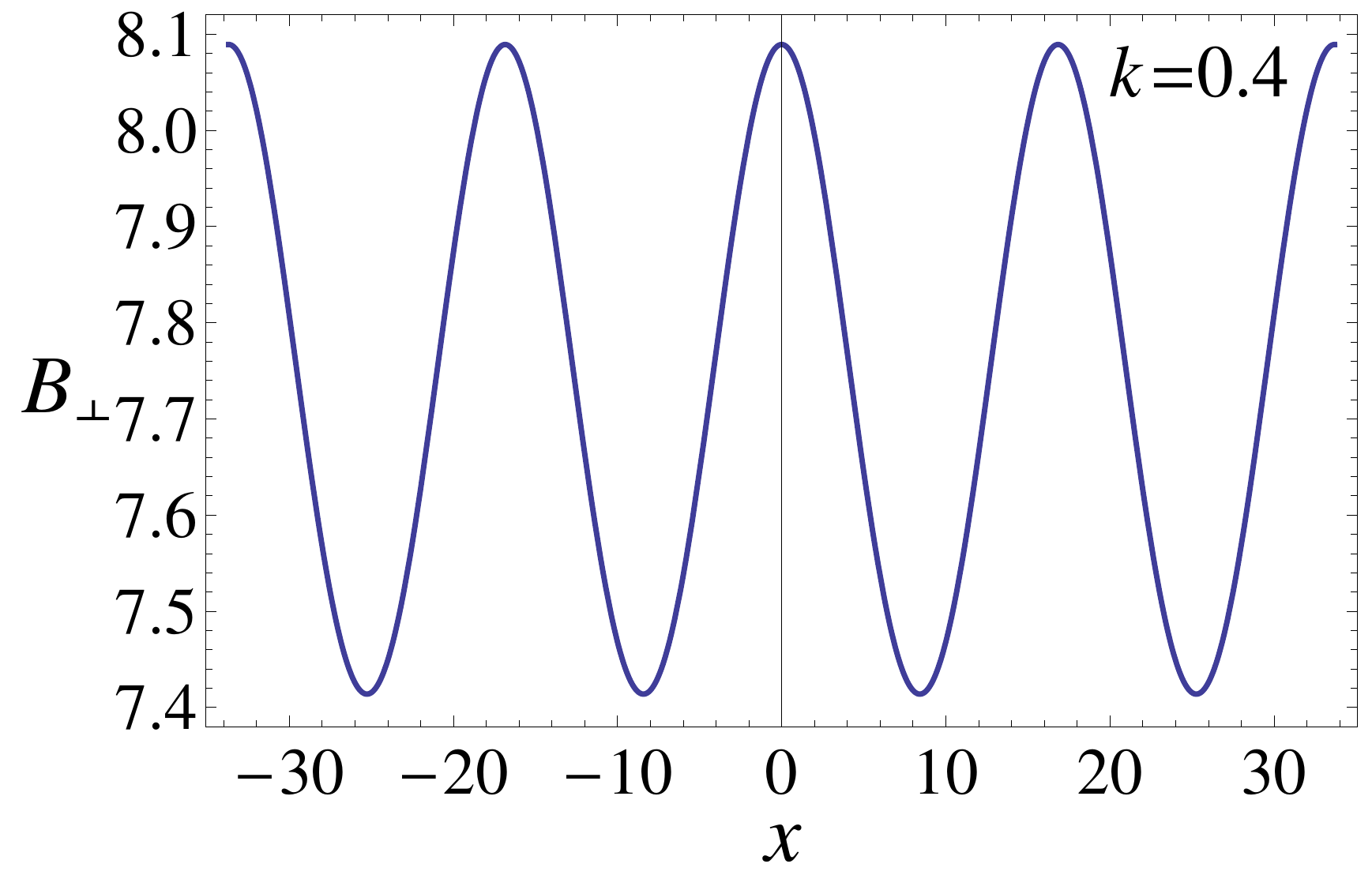}
\caption{Effective potential (left panel) and periodic solution (right panel) of $B_\perp$ for 
$L=0$ and $k=0.4$ as given analytically by Eqs.~(\protect\ref{eq:veff}) 
and (\protect\ref{eq:bperp}).}
\label{fig:V_k_0_4}
\end{figure}

\begin{figure}[tbp]
\begin{tabular}{ccc}
\includegraphics[scale=0.3]{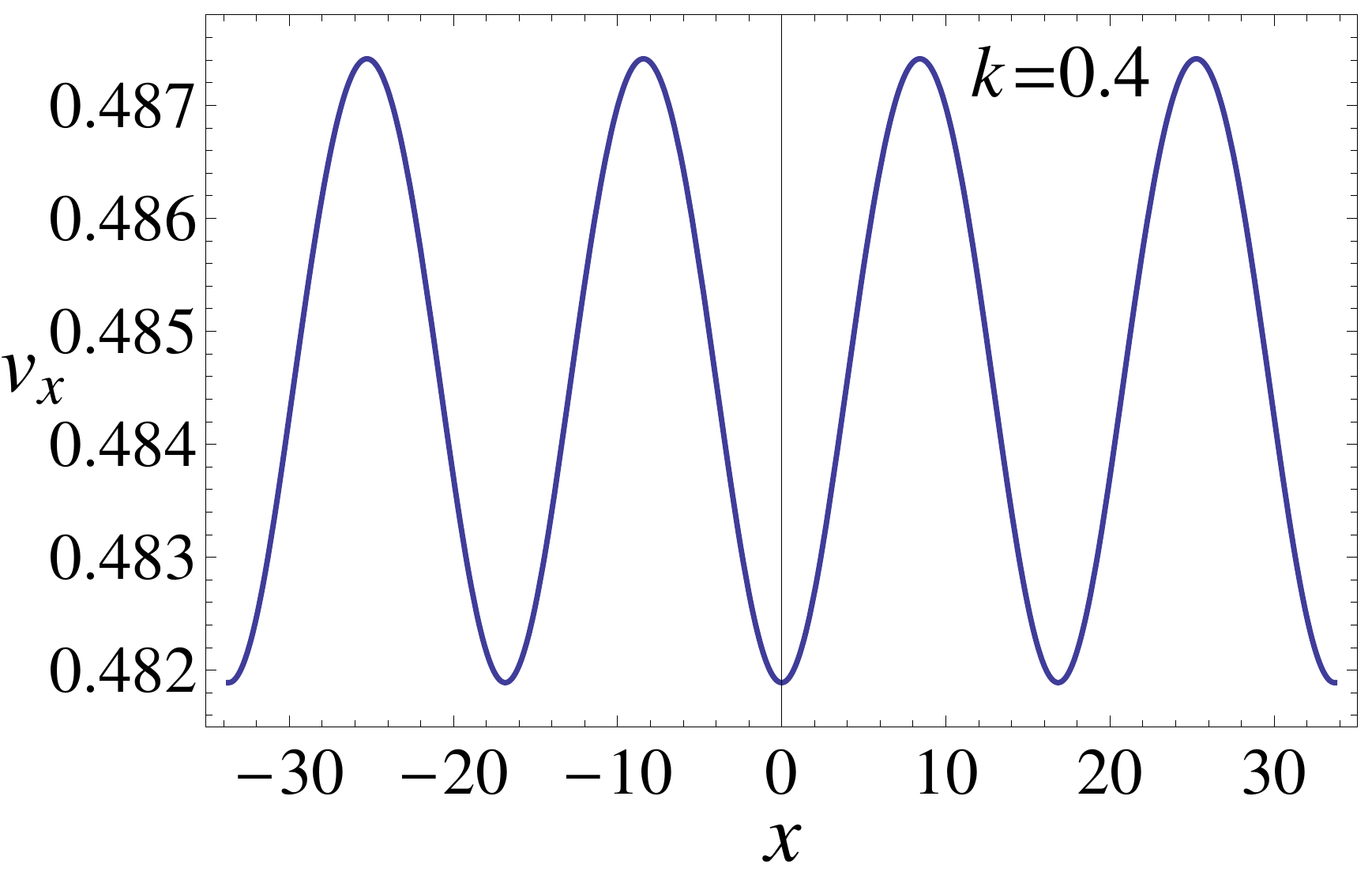} & %
\includegraphics[scale=0.3]{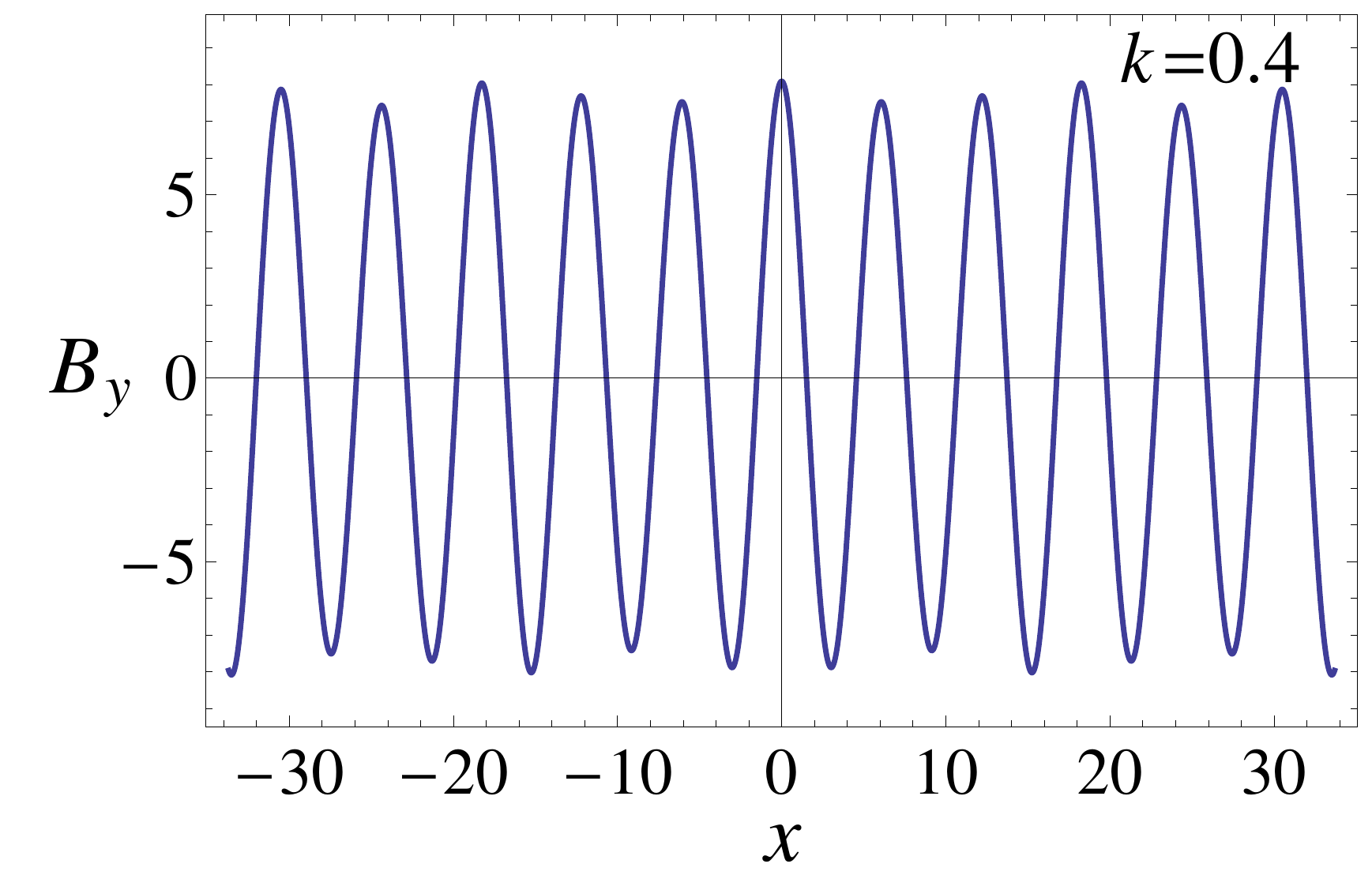} & %
\includegraphics[scale=0.3]{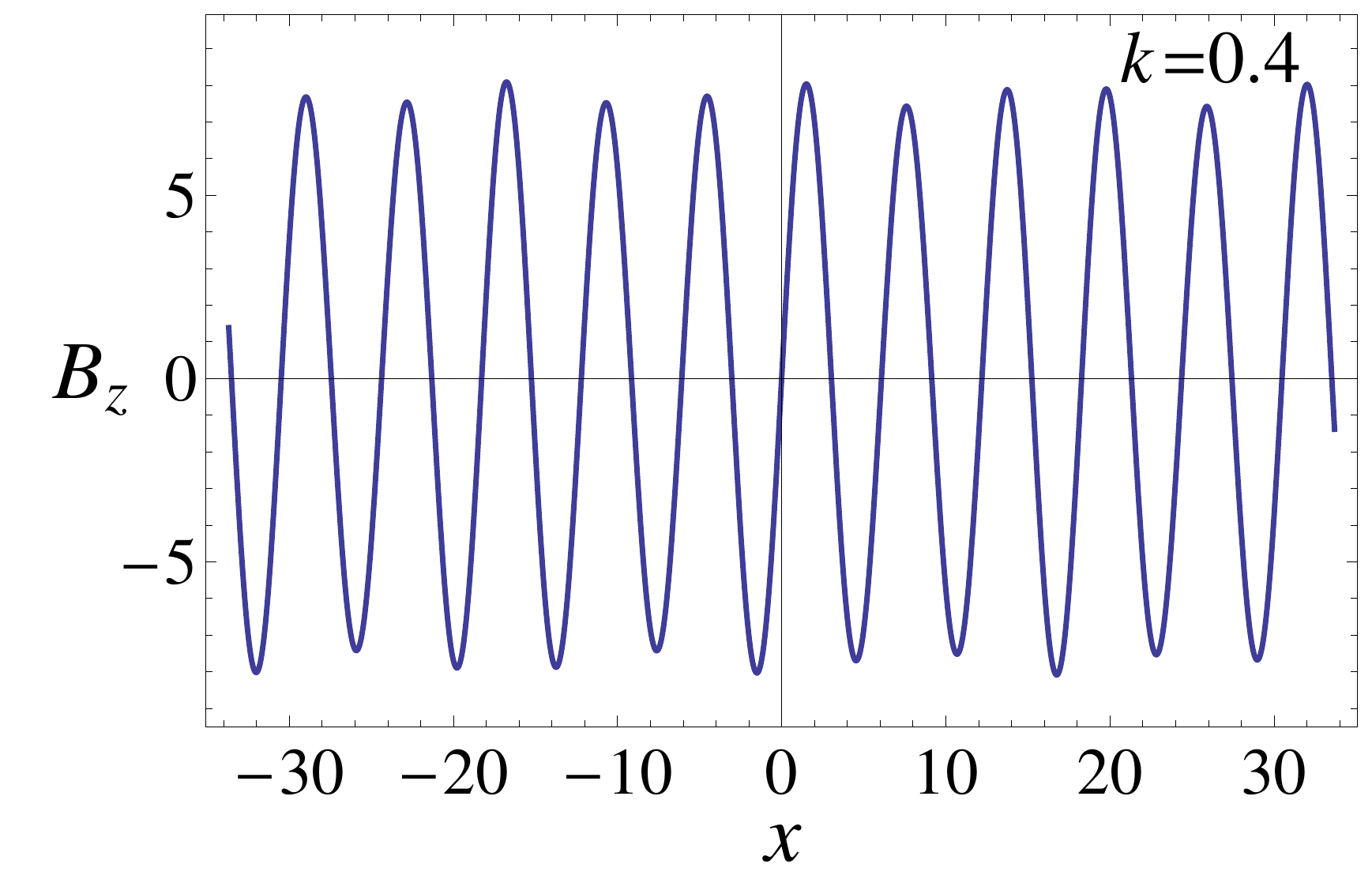} \\ 
\includegraphics[scale=0.3]{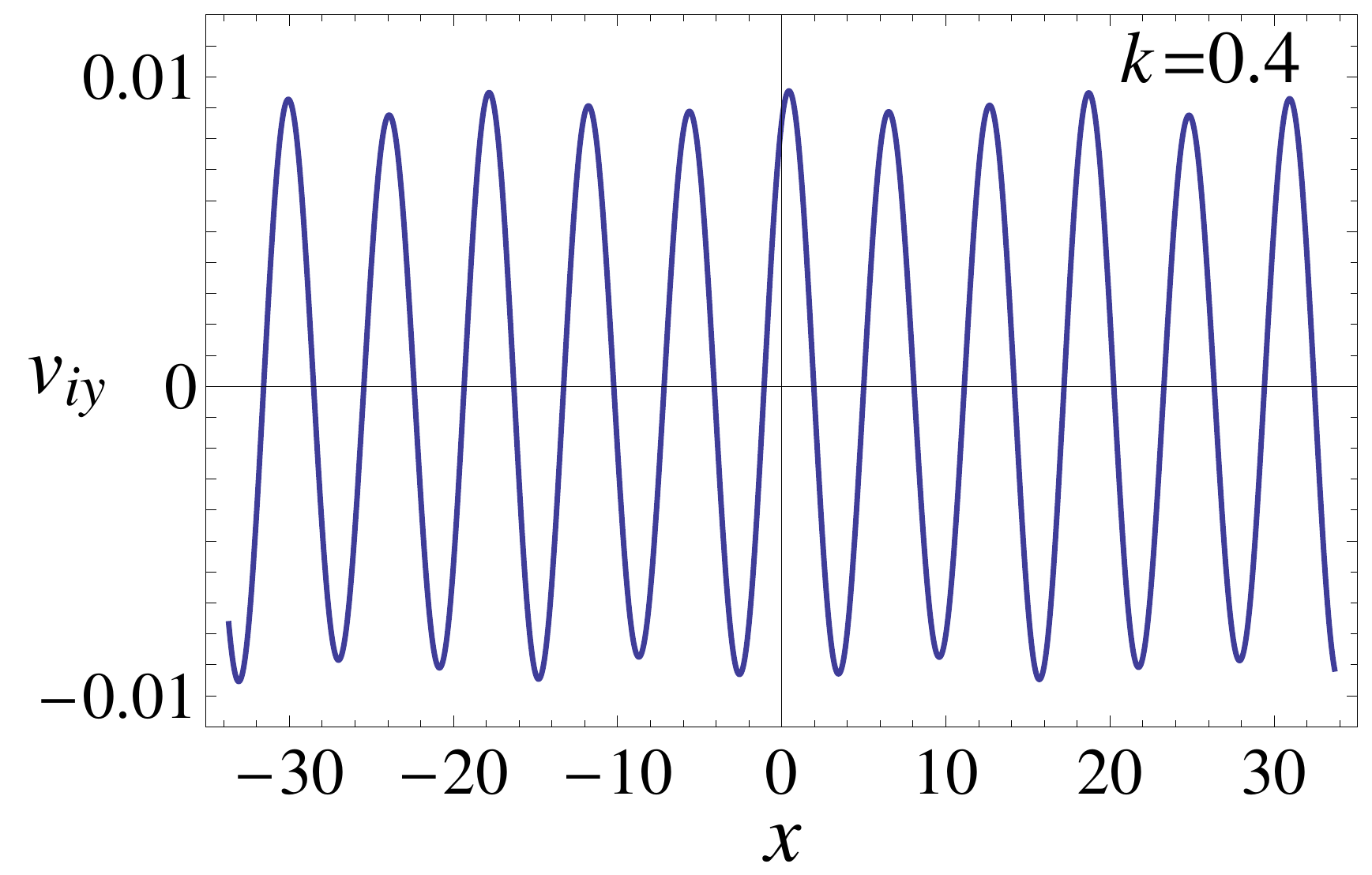} & %
\includegraphics[scale=0.3]{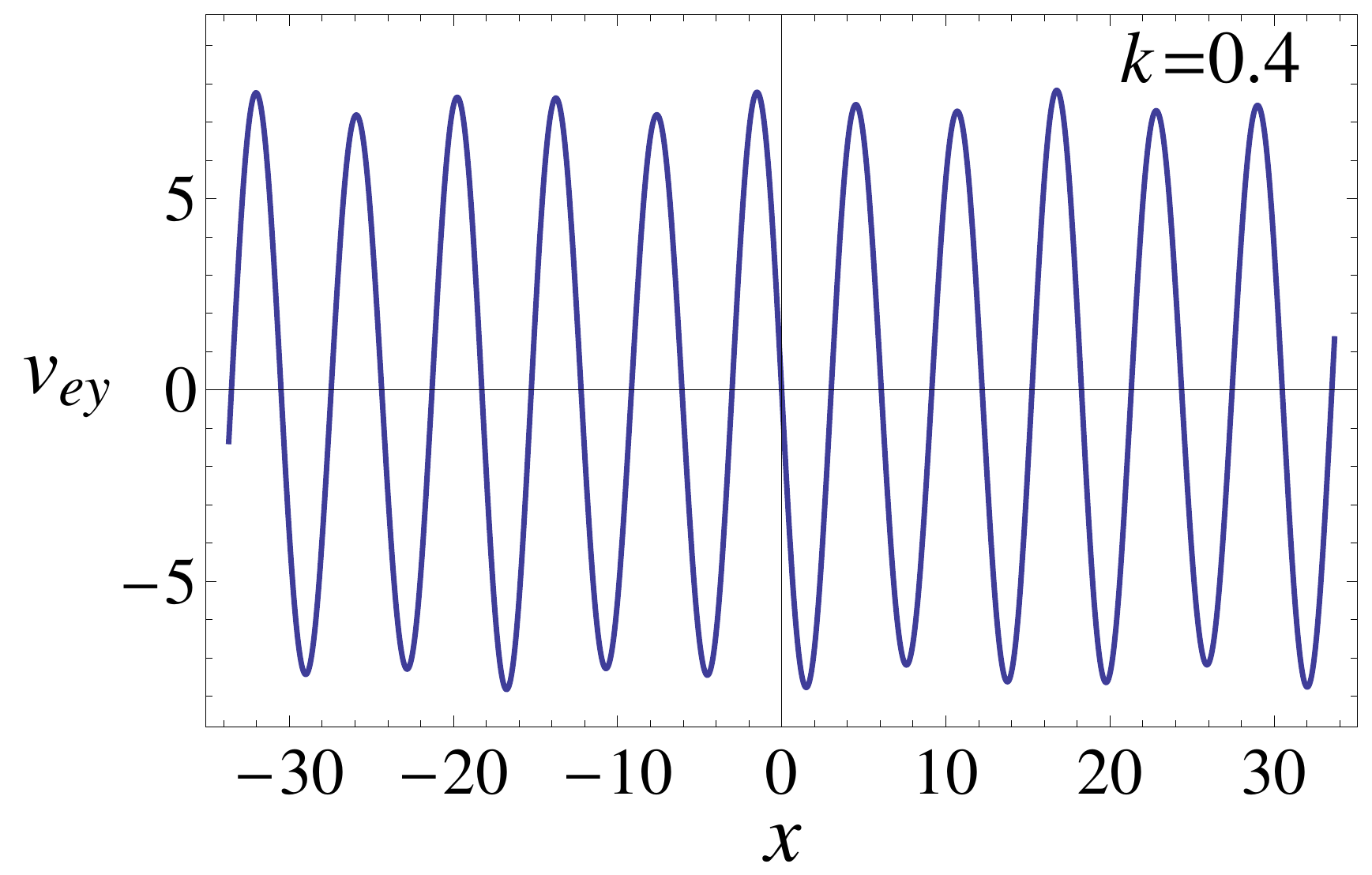} & %
\includegraphics[scale=0.3]{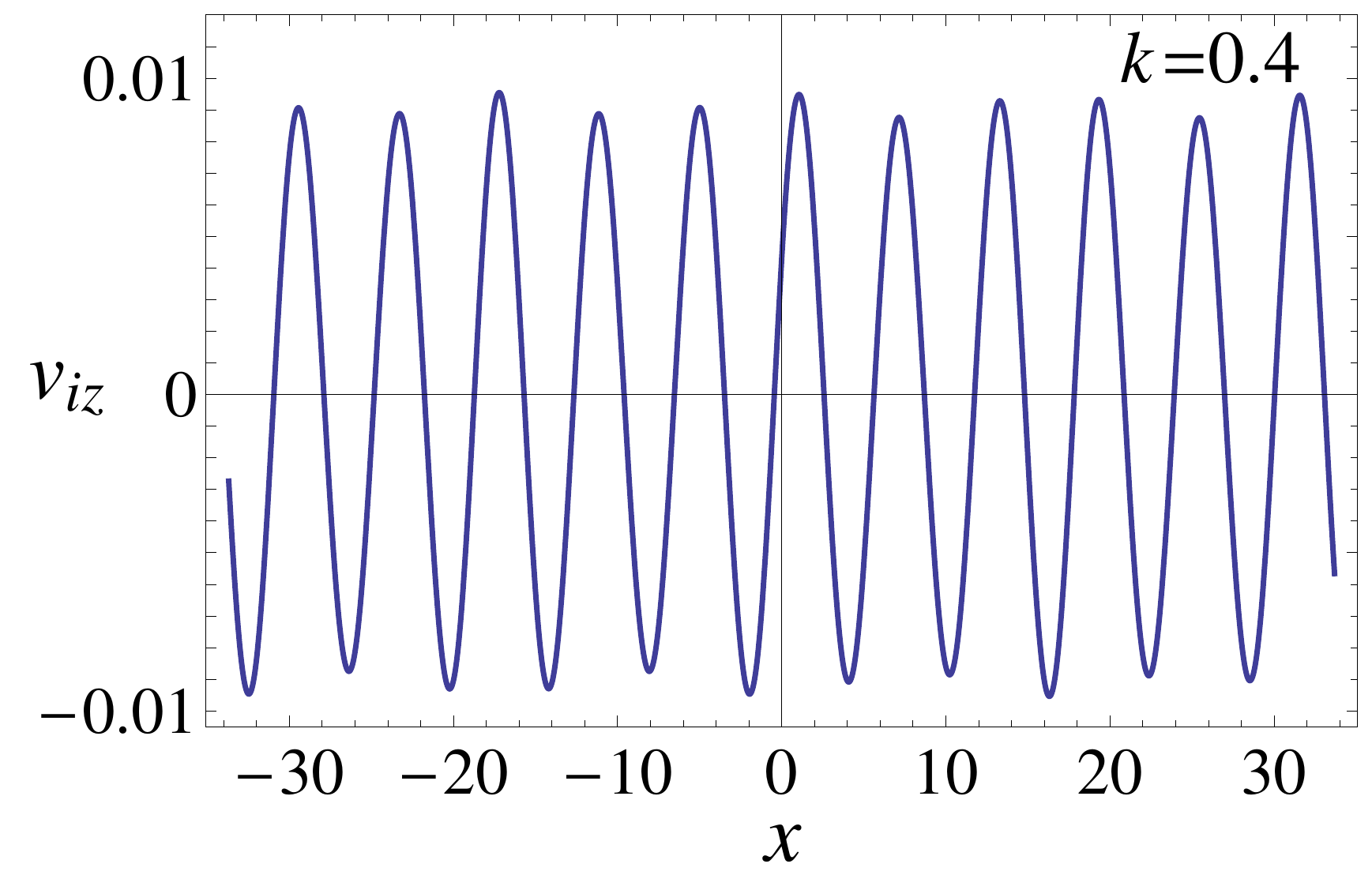} \\ 
\includegraphics[scale=0.3]{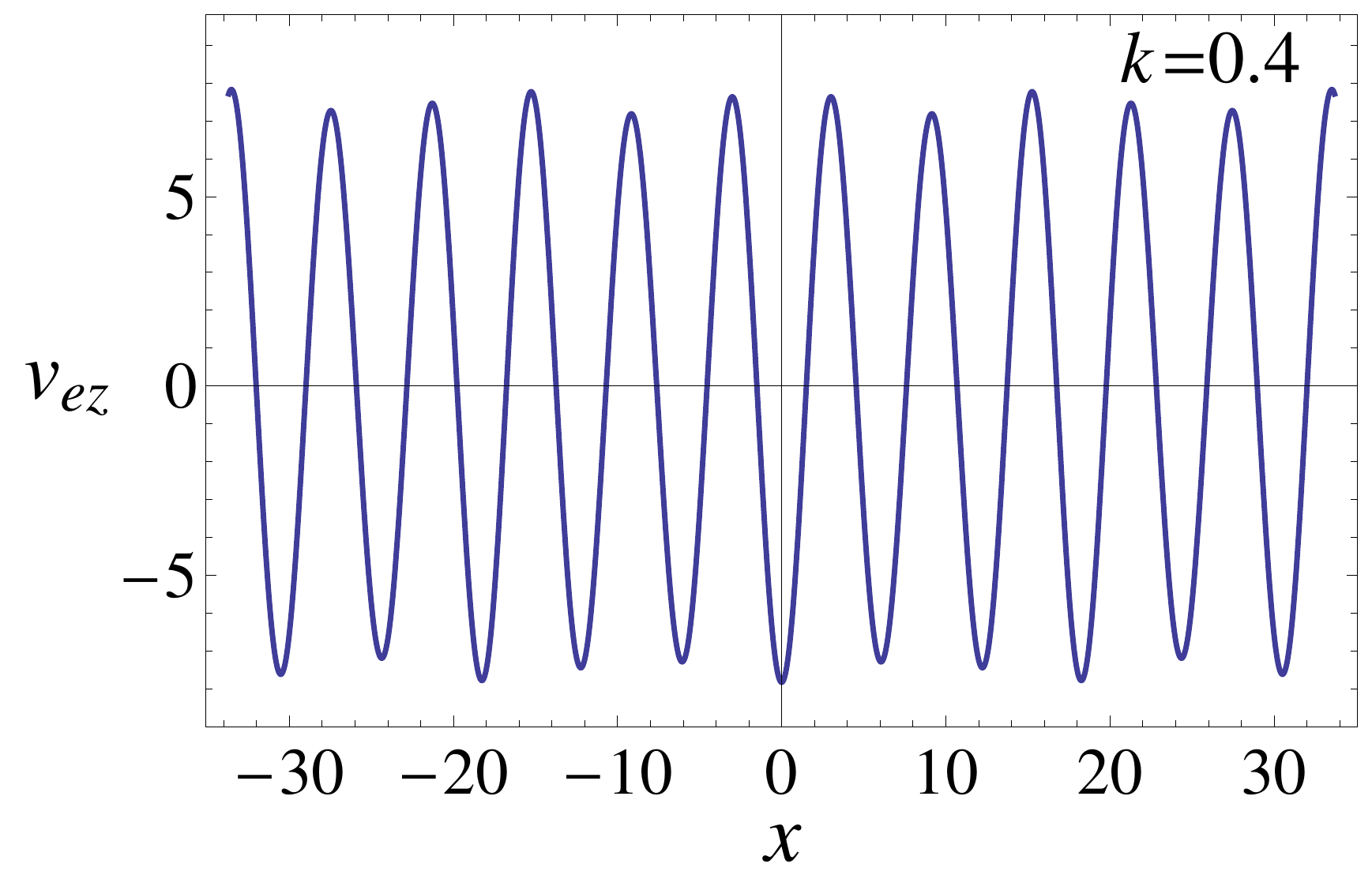} & %
\includegraphics[scale=0.3]{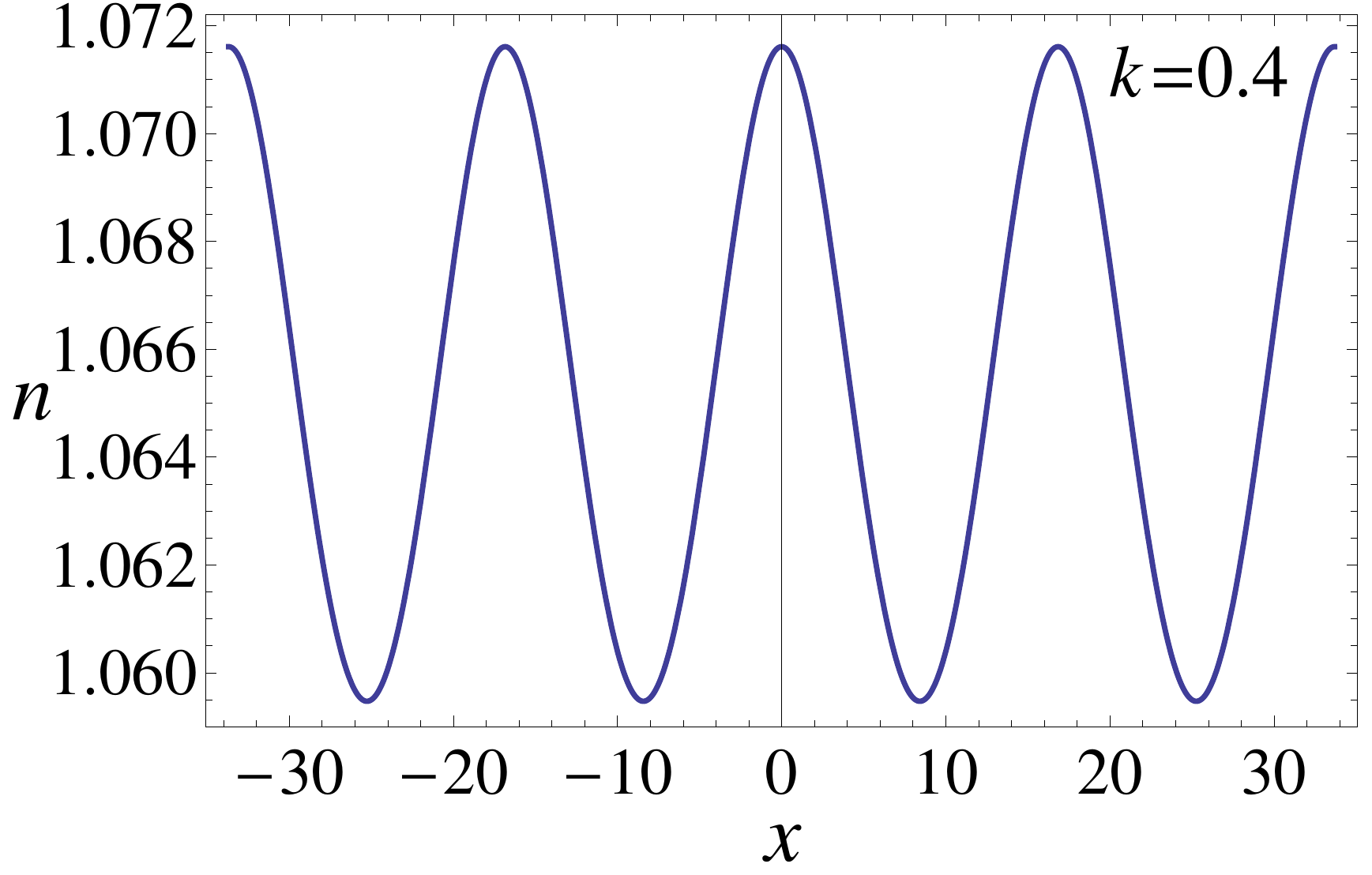} & %
\includegraphics[scale=0.3]{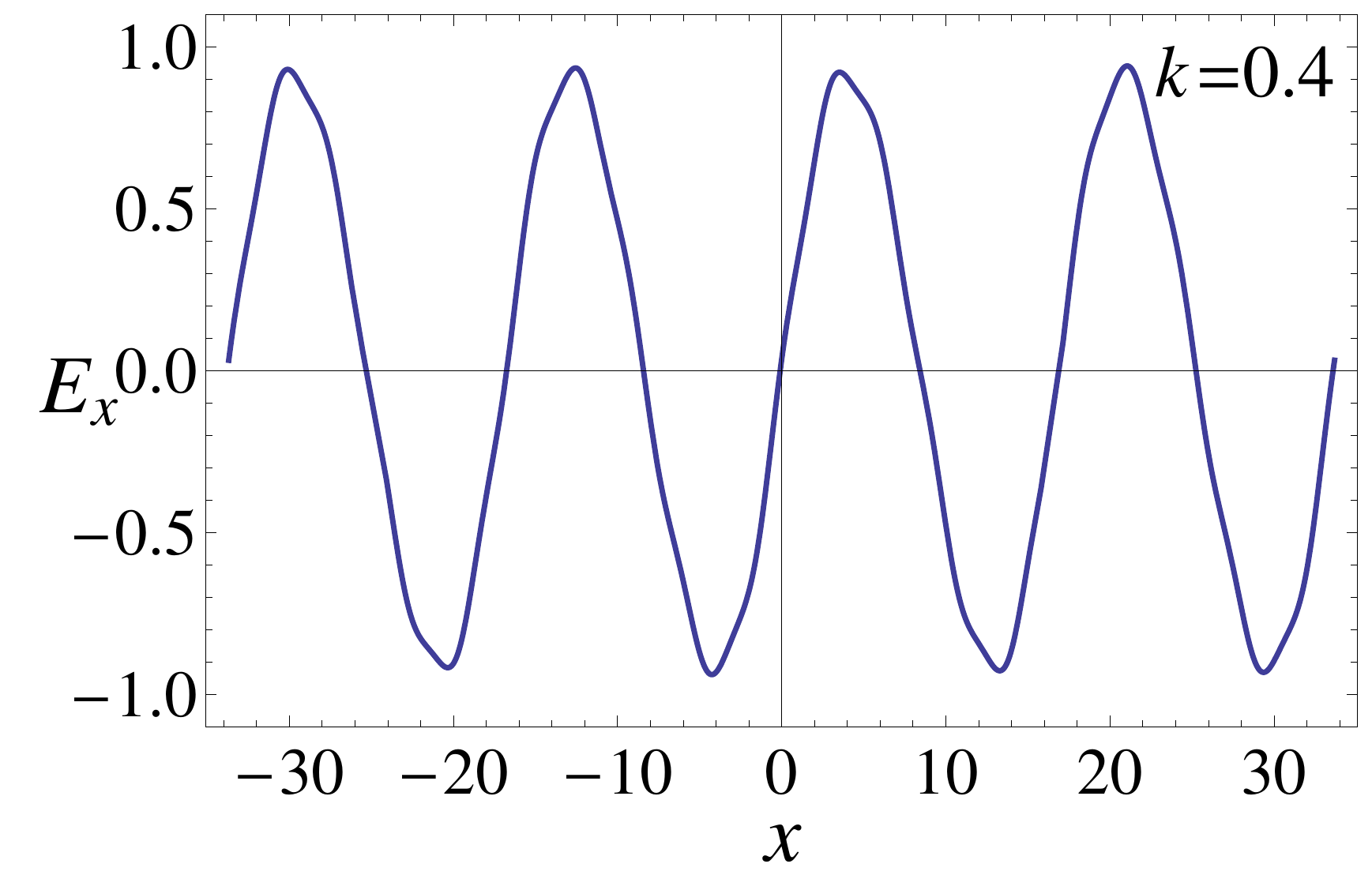} \\ 
&  & 
\end{tabular}%
\caption{Similar to Fig.~\protect\ref{fig:Fields_k_0_95}, but now for $L=0$ and $k=0.4$.}
\label{fig:Fields_k_0_4}
\end{figure}

In the case $k=0.4$, the value of the total energy is $W=-0.24$ (see left panel
of Fig.~\ref{fig:V_k_0_4}). Since the motion now occurs close to the 
local minimum of $V_\mathrm{eff}$, the corresponding
behavior is close to the harmonic one as it can be seen both in the
right panel of Fig.~\ref{fig:V_k_0_4} and in Fig.~\ref{fig:Fields_k_0_4}. This
will be progressively more so as we approach the minimum of the potential, which is $V_{\mathrm{eff}_\mathrm{min}%
}\simeq-0.2467$ and occurs for $B_\bot\simeq 0.759$
. Nevertheless, the transverse velocity and magnetic
fields retain their quasi-periodic functional form in this case too, as $k \rightarrow 0$ and we tend to the near-linear, small
  amplitude (trigonometric) limit of the theory.


\section{Conclusions and Future Work}

In the present work we have provided a full formulation, in the form of
partial differential equations for nine fields, of the problem of
longitudinal wave propagation in a cold, quasi-neutral, collision-free
plasma. The guiding principles involved the Newtonian dynamics at the level
of the ions and the electrons, the equation of continuity and the 
Amp{\'e}re law's components. This enabled the development of a system of nine
equations for the five components of the velocity $v_{x},v_{iy,ey},v_{iz,ez}$, 
the two components of the transverse magnetic field $(B_{y},B_{z})$, the
density $n$ and the electric field $E_{x}$. While considering the full set
of nine PDEs at the numerical level remains a particularly challenging task,
we have been able to achieve a substantial simplification by considering the
co-traveling frame. We have considered the reduction of the resulting nine
ODEs into, arguably, the simplest formulation of a $2\times 2$ system for
the transverse magnetic field components. This was eventually converted to a
single degree of freedom setting via $L$ considerations (with $L$ being the 
total angular momentum flux). This allowed us to remarkably not only 
retrieve the solitary wave solutions of Ref.~\cite{abbas} for \textit{all} 
the relevant fields, but  also provided a platform for generalizing these 
considerations to a mono-parametric family of periodic function solutions. 
We explained how/why some of the fields
(like $B_{\bot }$, $v_{x}$, $n$ or $E_{x}$) feature solitonic or periodic
character, while others (the transverse components of the velocity or the ones of the
magnetic field) have, respectively, an envelope-soliton- or periodic- nature.
%%%%%%%%%%% PGK: Change re: L \neq 0
%We also distinguished the case where the transverse magnetic field has a
In Appendix \ref{appA}, we also consider the case of mathematical
interest with
non-vanishing $L$. There, the additional centrifugal
contribution to the effective potential energy landscape precludes
analytical solutions (and especially so solitary waves), yet still we can
numerically identify periodic orbits in the system.

Naturally, there are numerous directions that are worthwhile to consider for
future study in this vein. From a numerical point of view, it would be
particularly interesting to explore the full set of partial differential
equations describing this system, namely Eqs.~(\ref{ElecEqM58})-(\ref%
{MFlux62}). In that framework, our solutions are traveling ones, or
equivalently stationary ones in the co-traveling frame and hence it would
also be natural to explore their dynamical stability. From the point of view
of dynamical reductions, it would be particularly interesting (although
quite challenging in its own right) to reduce the solutions via a reductive
perturbation method to solutions of the Korteweg-de Vries or similar
equations as was recently done for the transverse case in Ref.~\cite{allen2020}. 
In a similar spirit as the latter work, exploring numerically at first the
interactions between different solitary waves would be a topic of interest
in its own right, especially given the fundamentally more complex form
(reminiscent of an envelope soliton in the transverse components) of the
solitary waves herein. Lastly, generalizing corresponding considerations to
more complex scenarios, where the traveling wave may not be genuinely one
dimensional, or where a dust granule may interact with the wave, are also of
interest. This is essential because the dusty solitary
  currents \cite{Trukh2019} can influence the phenomena considered in
  the present analysis. Work along some of these directions is currently in progress and
will be reported in future publications.

\begin{acknowledgments}
 This material is based upon work supported by the US National Science
Foundation under Grants No. PHY-1602994 and DMS-1809074
(PGK). PGK also acknowledges support from the Leverhulme Trust via a
Visiting Fellowship and thanks the Mathematical Institute of the University
of Oxford for its hospitality during part of this work.
\end{acknowledgments}

\appendix 
\section{The $L\neq0$ Case}
\label{appA}
\begin{figure}[tbp]
\includegraphics[height=5cm]{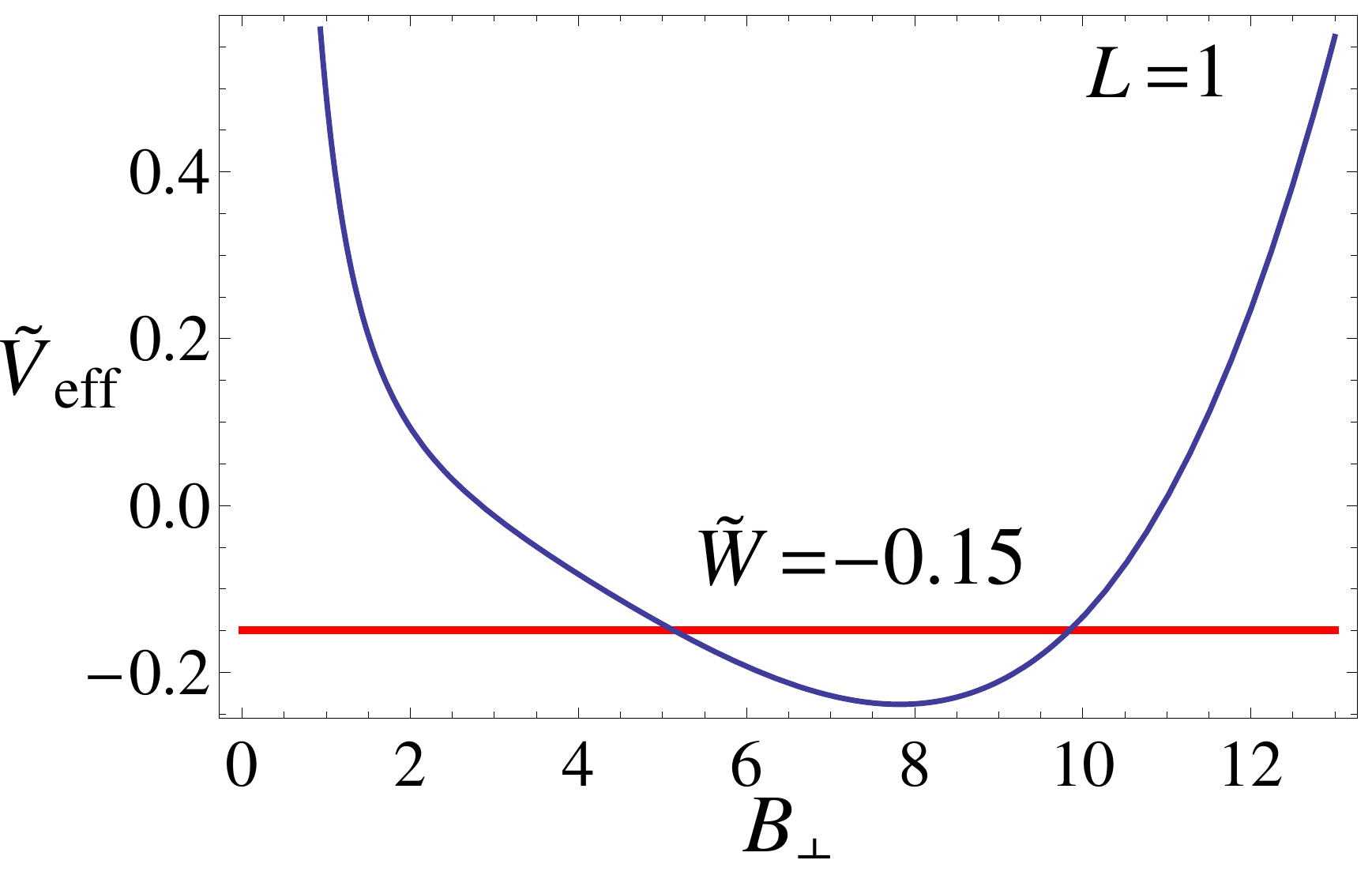}%
\hspace{0.5cm} %
\includegraphics[height=5cm]{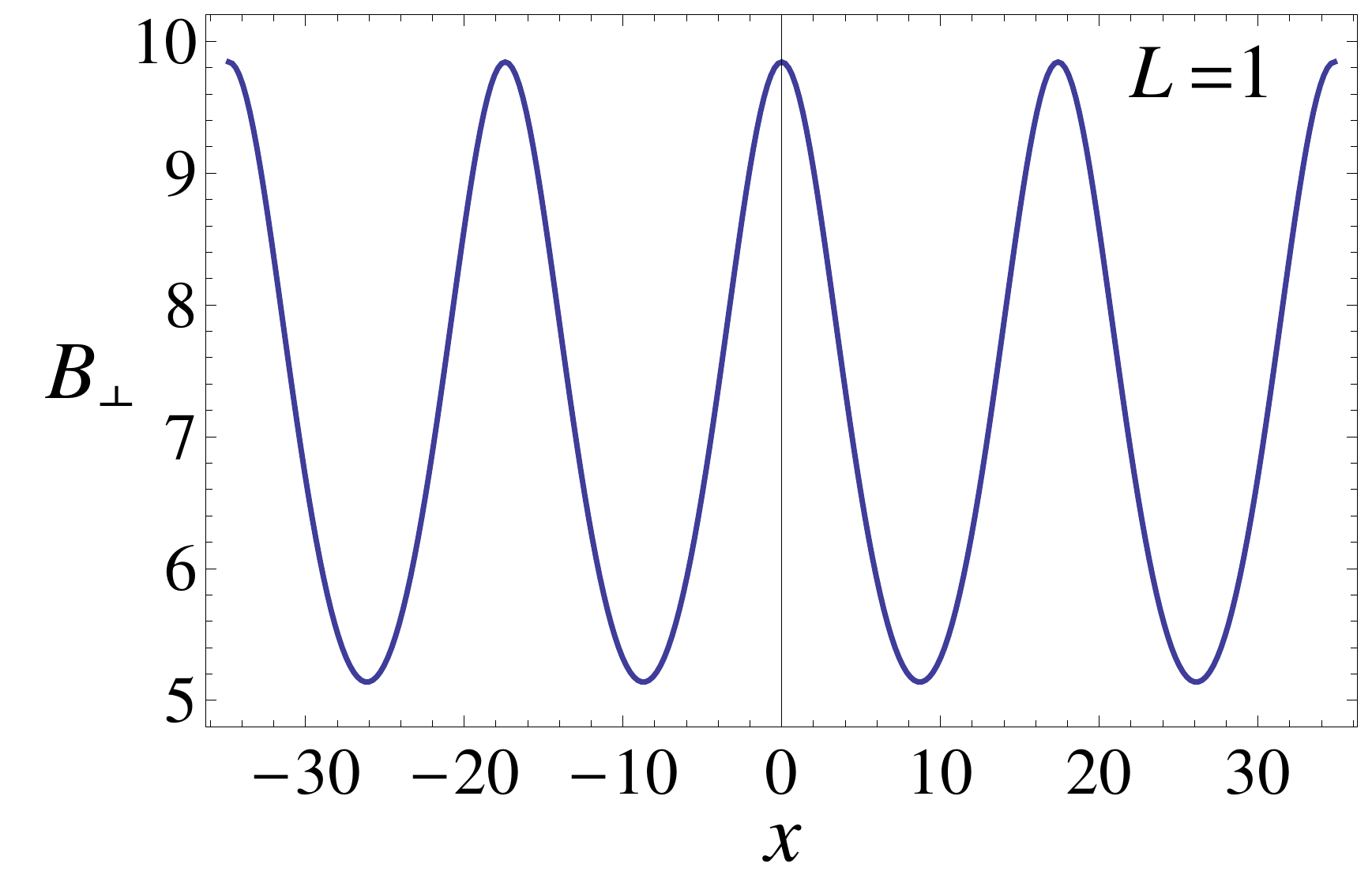}
\caption{Effective potential (left panel) and periodic solution 
of $B_\perp$ (right panel) for $L=1$ and $\tilde{W}=-0.15$. $\tilde{V}_\mathrm{eff}$ is given analytically by \eqref{eff2} while $B_\bot$ is calculated numerically.}
\label{fig:V_L_1_W_m0_15}
\end{figure}

In the case where $L\neq0$, we still select $E_2=E_3=0$, so as to have a
tractable, effectively one degree-of-freedom scenario as concerns $B_\bot$.
Then, the equation of energy, \eqref{Eq82}, becomes:
%\begin{equation}
%\frac{1}{2}\left(\frac{\partial B_{\bot }}{\partial x^{\prime}}\right) ^{2} +%
%\frac{B_\bot^2}{2}\left(\frac{1-\alpha}{2}-\frac{L}{B_\bot^2}\right)^2+\frac{%
%\alphaB_\bot^4}{8}-\alpha(E_1-1)\frac{B_\bot^2}{2}=W,
%\end{equation}
%or 
\begin{equation}
\frac{1}{2}\left(\frac{\partial B_{\bot }}{\partial x^{\prime}}\right) ^{2} +%
\tilde{V}_\mathrm{eff}=\tilde{W},  \label{eq:energy_L_neq_0}
\end{equation}
where 
\begin{equation}
\tilde{V}_\mathrm{eff}=V_\mathrm{eff}+\frac{L^2}{2B_\bot^2}\quad\mathrm{and}%
\quad \tilde{W}=W+\frac{(1-a)L}{2}.  \label{eff2}
\end{equation}
Since the effective potential $\tilde{V}_\mathrm{eff}$ does not possess a
local maximum at $B_\bot=0$ as in the $L=0$ case, but it tends to infinity
as $B_\bot\rightarrow0$ (e.g.~left panel of Fig.~\ref{fig:V_L_1_W_m0_15}), we \textit{cannot} achieve solitonic solutions with
respect to $B_\bot$. That is to say, solutions with nontrivial $L$ of the transverse 
magnetic field can only be periodic and not solitary wave states; see e.g.~right panel of Fig.~\ref{fig:V_L_1_W_m0_15}. On the other hand, 
since the potential is convex, there
exists an infinite number of periodic solutions. We cannot acquire these
solutions in closed form as before, due to the presence of the centrifugal
potential term in Eq.~(\ref{eff2}), but can study them numerically.

The inverse transformation $x^{\prime}\mapsto x$ is calculated as in the $%
L=0 $ case. In addition, since it is now true (per Eq.~\eqref{Eq81}) that
\begin{equation*}
\frac{\partial\theta}{\partial x^{\prime}}=\frac{1-\alpha}{2}-\frac{L}{%
B_\bot^2},
\end{equation*}
the angle $\theta$ is calculated as 
\begin{equation}
\theta=\frac{1-\alpha}{2}x^{\prime}-\int\frac{L}{B_\bot^2(x^{\prime})}{%
\mathrm{d}} x^{\prime}.
\end{equation}
The various fields which describe our system are calculated as in the $L=0$
case.

\begin{figure}[tbp]
\begin{tabular}{ccc}
\includegraphics[scale=0.3]{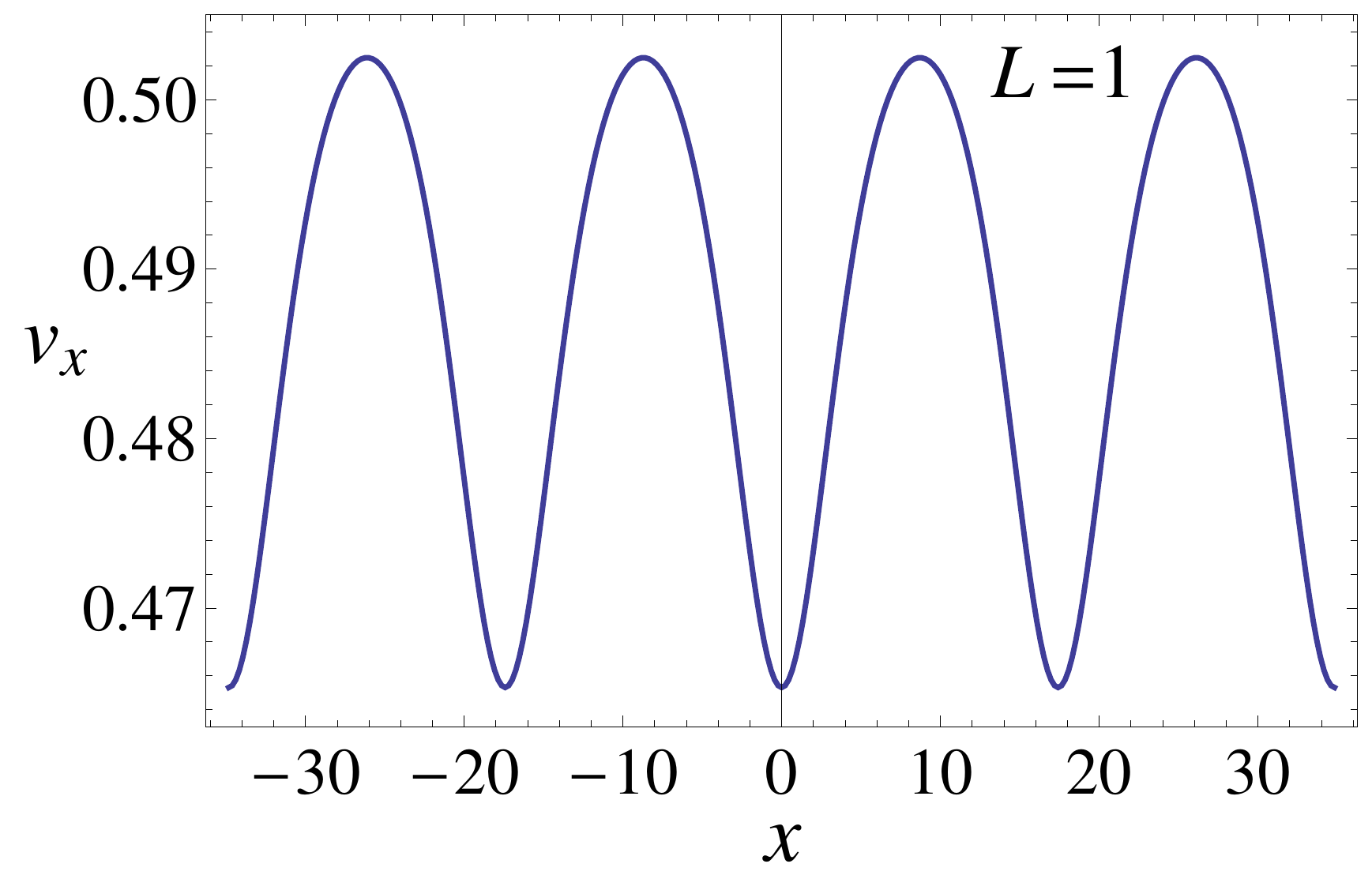} & %
\includegraphics[scale=0.3]{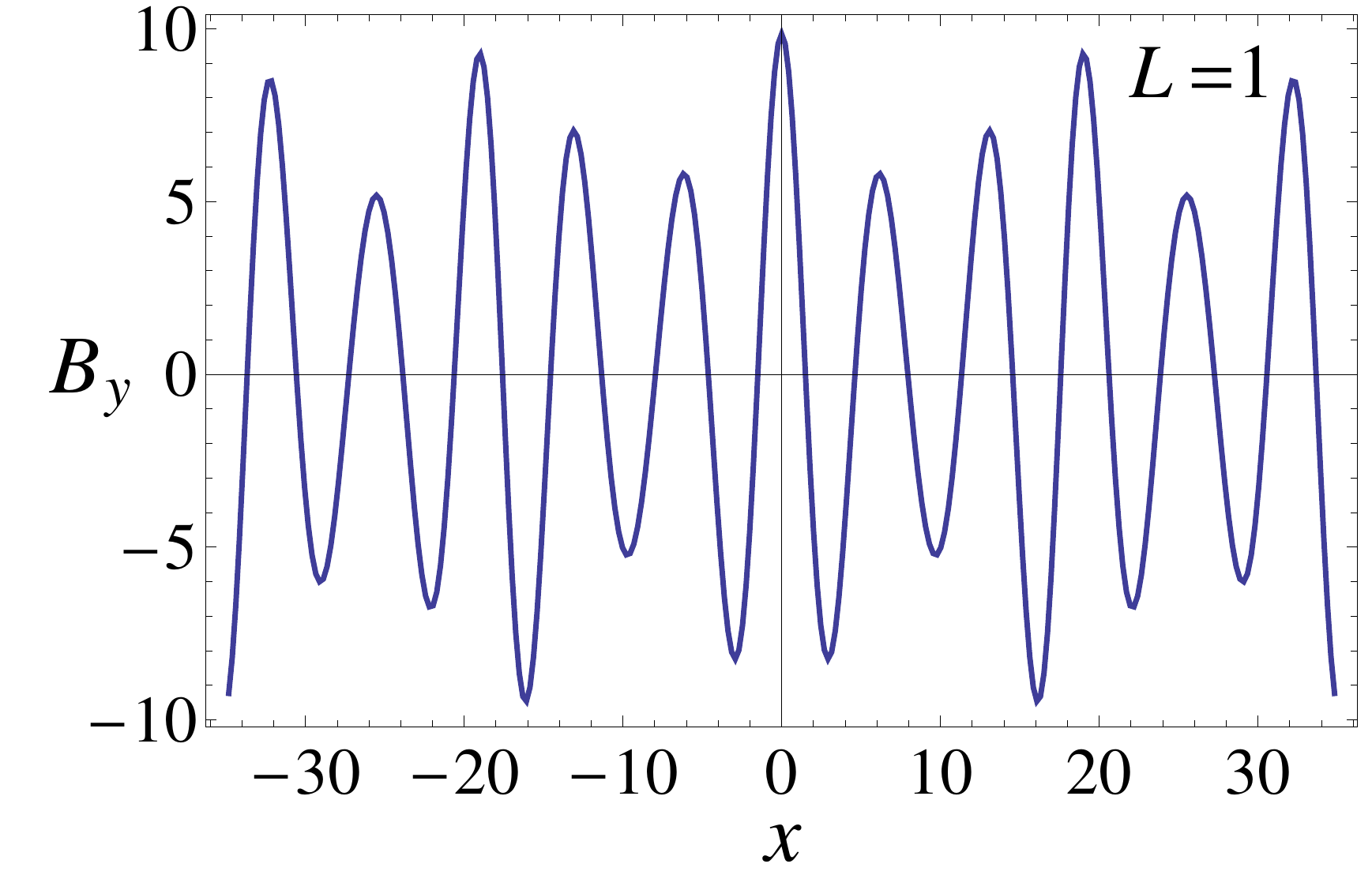} & %
\includegraphics[scale=0.3]{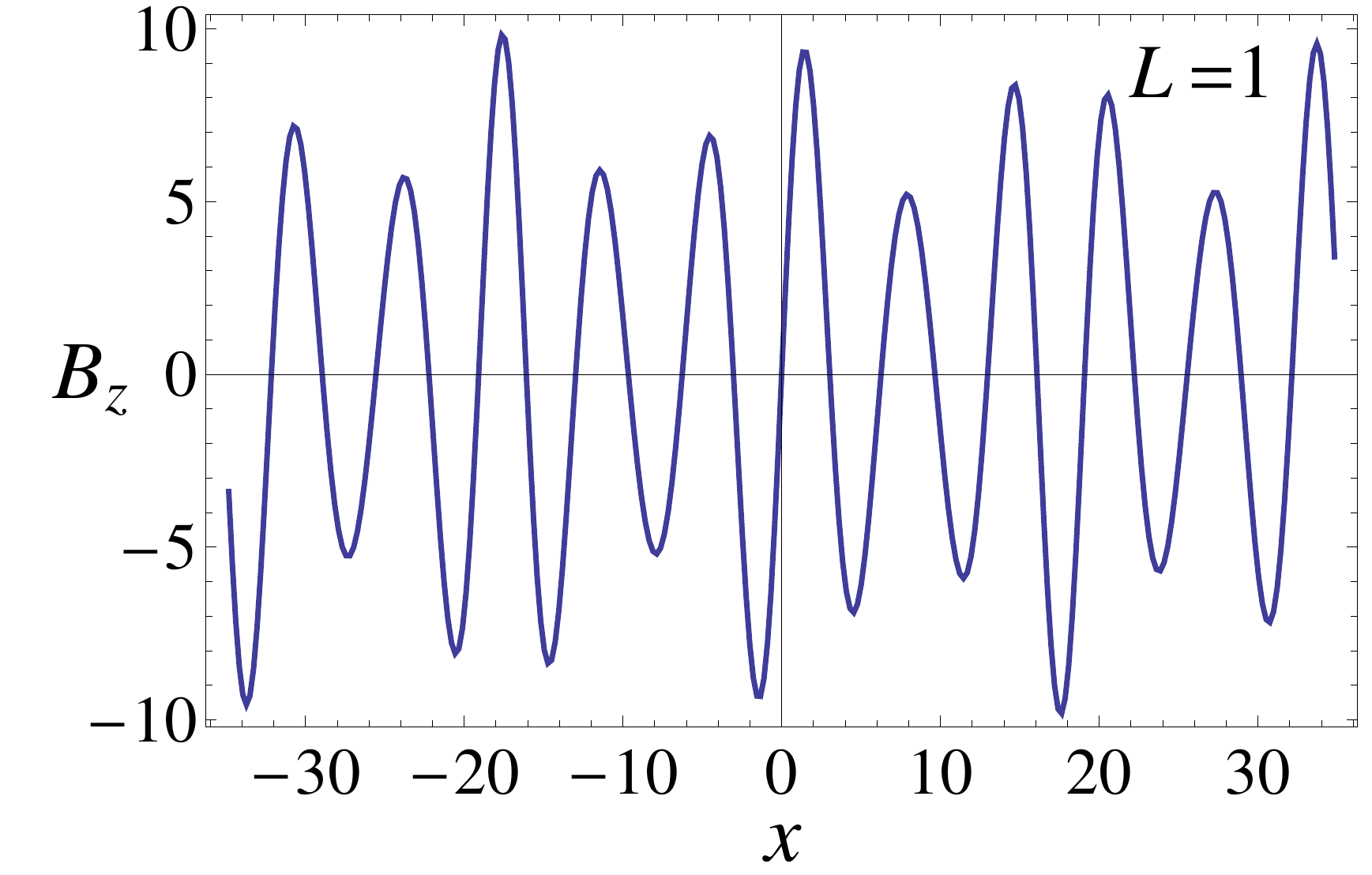} \\ 
\includegraphics[scale=0.3]{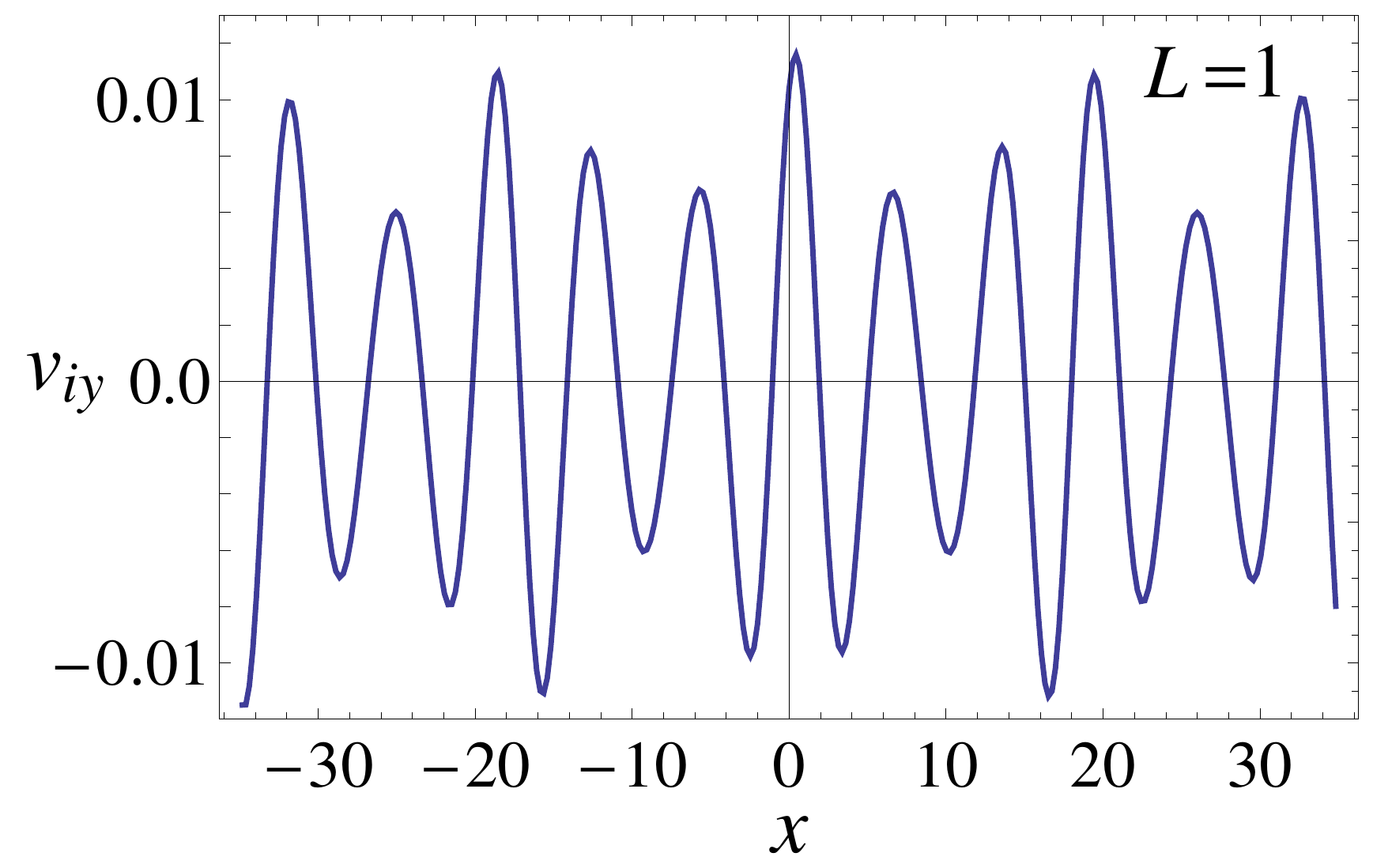} & %
\includegraphics[scale=0.3]{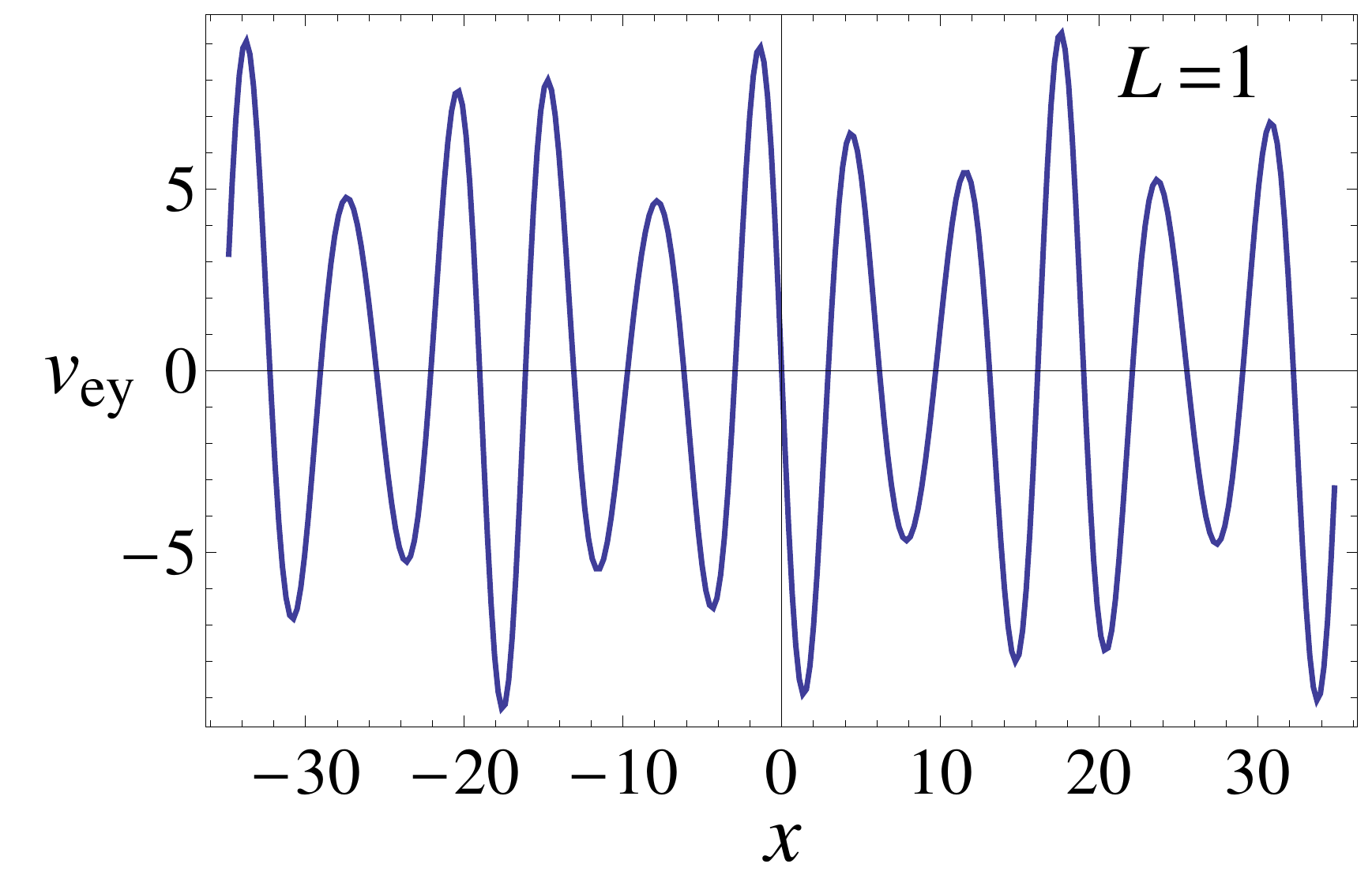} & %
\includegraphics[scale=0.3]{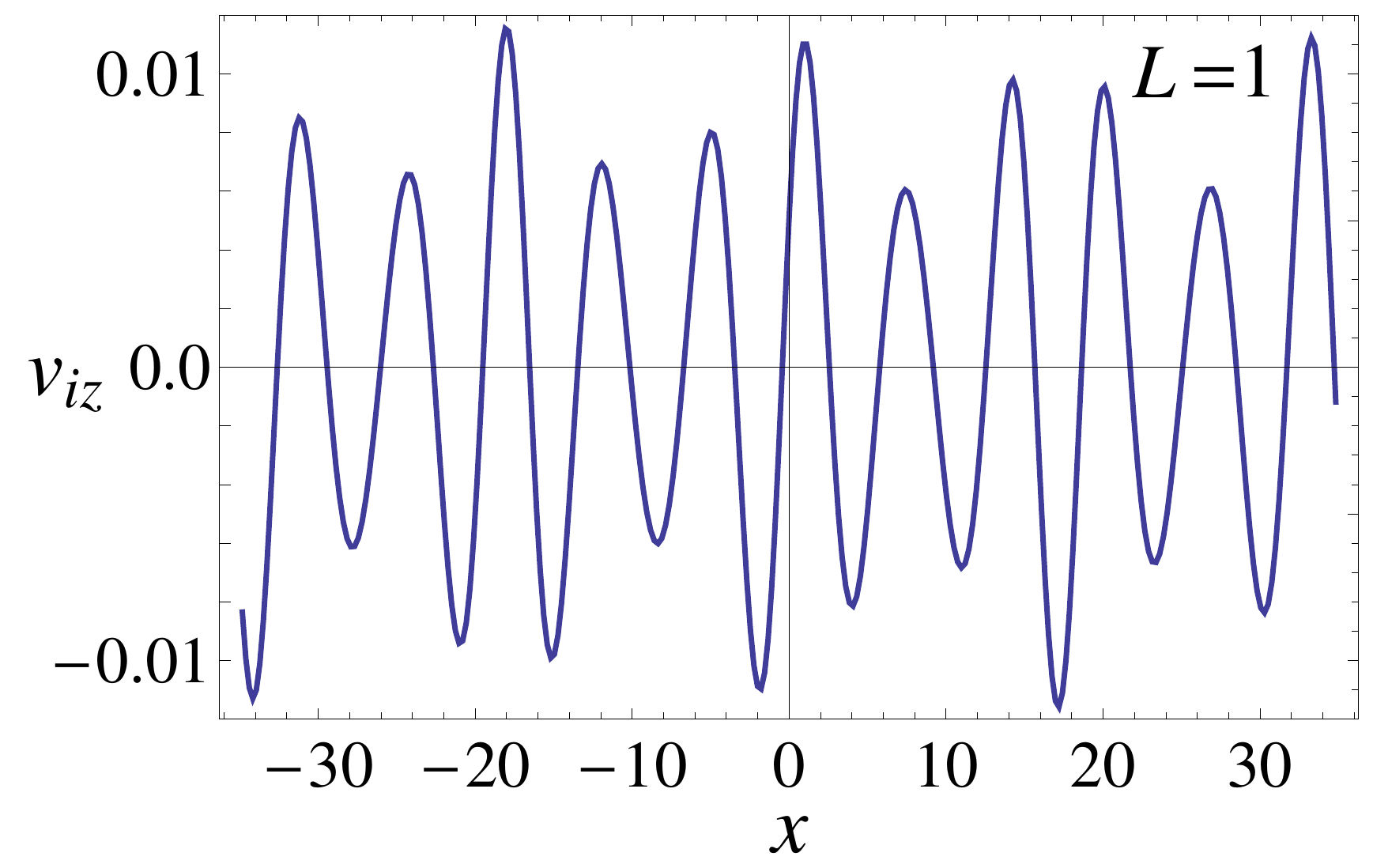} \\ 
\includegraphics[scale=0.3]{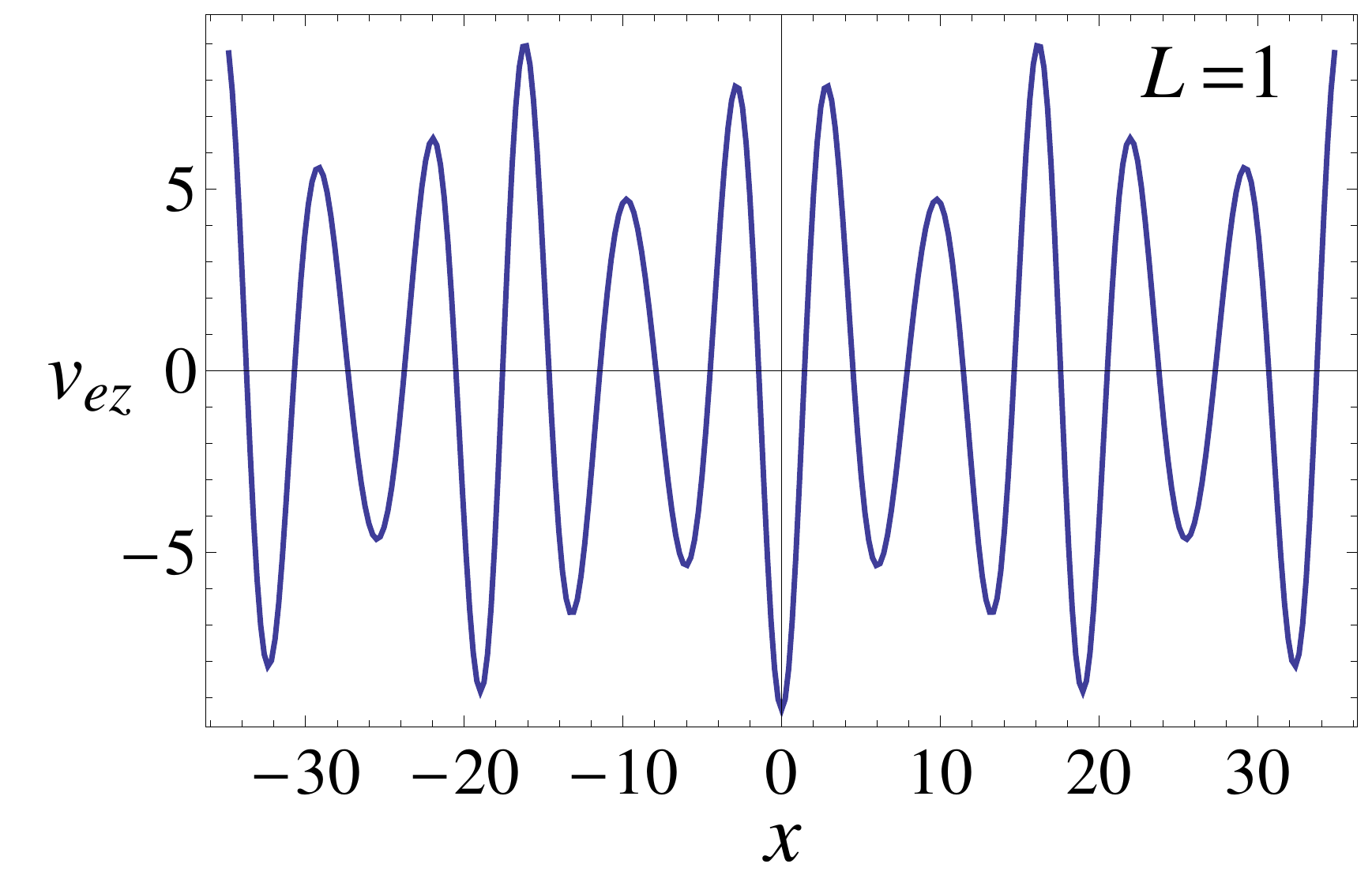} & %
\includegraphics[scale=0.3]{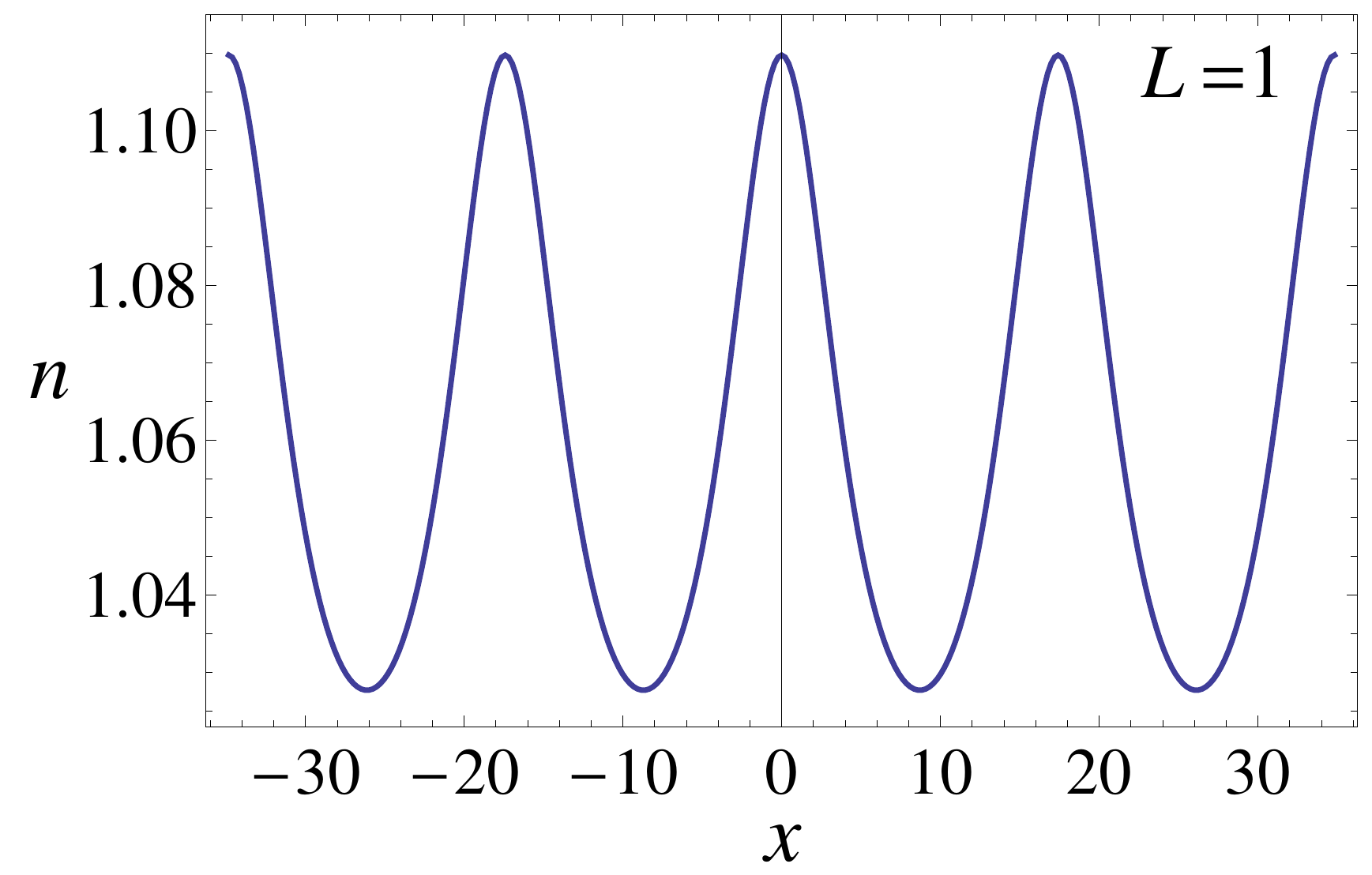} & %
\includegraphics[scale=0.3]{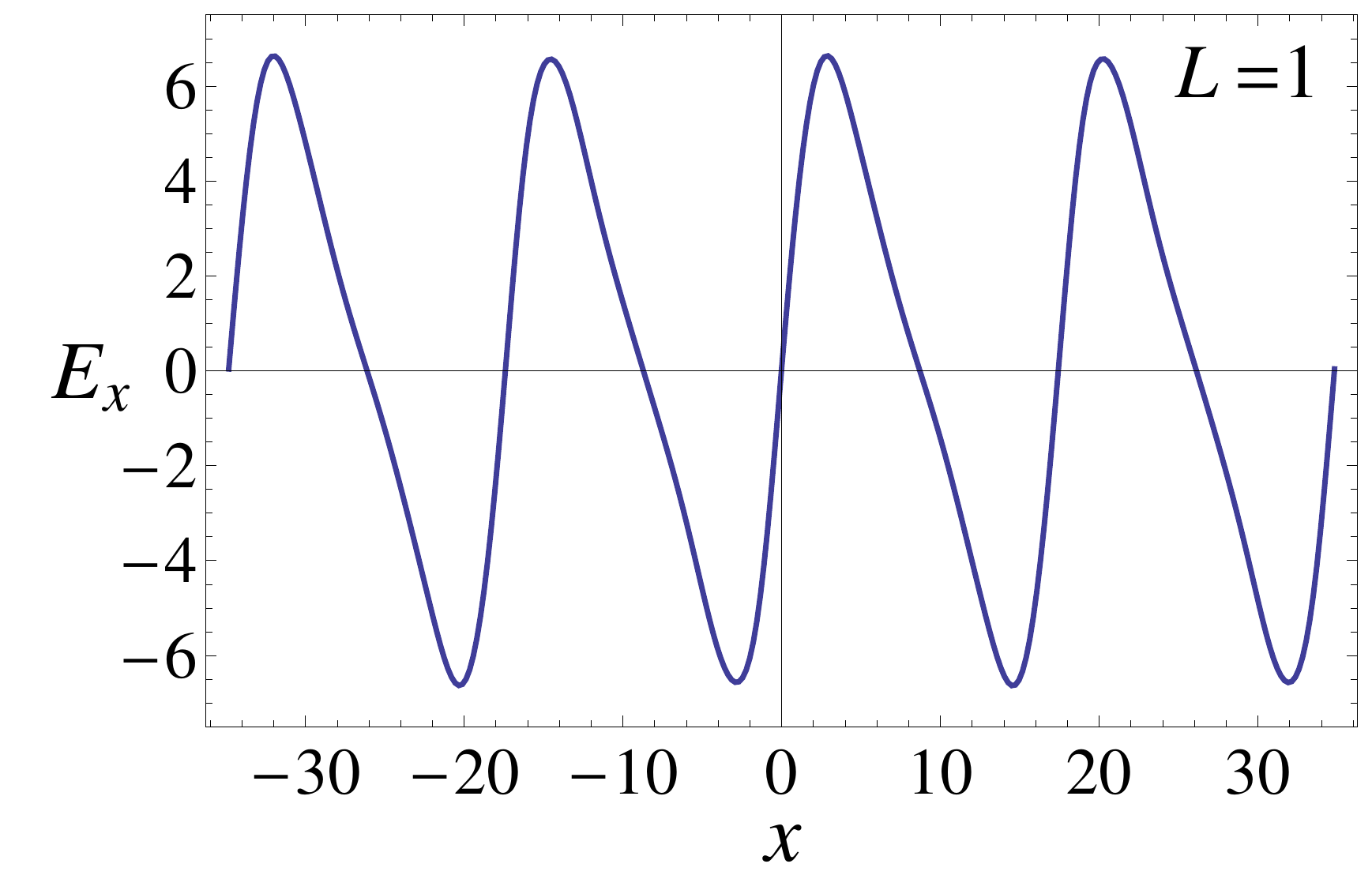} \\ 
&  & 
\end{tabular}%
\caption{Reconstruction of the nine fields describing the system, in this case for $L=1$ and effective energy $\tilde{W}=-0.15$}
\label{fig:Fields_L_1_W_m0_15}
\end{figure}

%\subsection{Periodic solution for $L=1$ and $\tilde{W}=-0.15$}
Let us consider a situation where the total energy of the system is negative $\tilde{W%
}=-0.15$ and $L$ is non-zero, i.e.,~$L=1$. This is the case actually depicted in Fig.~\ref{fig:V_L_1_W_m0_15}, where the form of the
effective potential, as well as the corresponding periodic orbit are shown. We can clearly observe how the divergence
of the centrifugal potential as $B_\bot$ tends to smaller values leads to
the sole existence of periodic orbits for the allowable range of $B_\bot$,
such that $V_\mathrm{eff}(B_\bot)< \tilde{W}$. The various fields are shown
in Fig.~\ref{fig:Fields_L_1_W_m0_15}. The central core of the solution
retains a form reminscent from before, yet the periodic character of the
solution is also evident. We have also explored a variety of other cases,
including ones with the same $L$ but higher $\tilde{W}$, as well as ones
with larger $L$ and the same $\tilde{W}$. While the specifics of the
solution (e.g., its period or the specifics of the quasi-periodicity of the
transverse velocity or magnetic components) may change, the over-arching
character of the waveforms does not. Hence, we do not show further such
examples here.



\begin{thebibliography}{99}
\bibitem{fpu} E. Fermi, J. Pasta, and S. Ulam, 
% Studies of Nonlinear Problems,
Tech. Rep. Los Alamos Nat. Lab. LA1940 (1955).

\bibitem{kz} N. J. Zabusky and M. D. Kruskal, Phys. Rev. Lett. \textbf{15},
240 (1965).

\bibitem{tan} H. Washimi and T. Taniuti, Phys. Rev. Lett. \textbf{17}, 996
(1966)

%\bibitem{aa1} J. H. Adlam and J. E. Allen, Philosophical Magazine {\bf
%    3}, 448--455 (1958).

\bibitem{Allen1958} J. H. Adlam and J. E. Allen Phil. Mag. \textbf{3}, 448
(1958).

\bibitem{aa2} J. H. Adlam and J. E. Allen, Proc. Phys. Soc. \textbf{75}, 640
(1960).

\bibitem{Allen2017} J. E. Allen and J. Gibson, Phys. Plasmas \textbf{24},
042106 (2017).

\bibitem{Montgomery1959} D. C. Montgomery, Phys. Fluids, 2, 585 (1959).

\bibitem{Saffman1961} P. G. Saffman, J. Fluid Mech. \textbf{11,} 16 (1961).

\bibitem{Tidman-Book} D. A. Tidman and N. A. Krall, \textit{Shock Waves in
Collisionless Plasmas}, John Wiley \& sons, Inc. (1971).

\bibitem{Sauer2002} K. Sauer, E. Dubinin, and J. F. McKenzie, Geophys. Res.
Lett. \textbf{29}, 2226 (2002).

\bibitem{Dubinin2003} E. Dubinin, K. Sauer, and J. F. McKenzie, J. Plasma
Phys. \textbf{69}, 305 (2003).

\bibitem{Cattaert2005} T. Cattaert and F. Verheest, Phys. Plasmas \textbf{12}%
, 012307 (2005).

\bibitem{Gibson2017} J. Gibson, \textquotedblleft\ Theoretical and
computational studies of magnetized dusty plasmas,\textquotedblright\ Ph.D.
thesis ( Imperial College London, 2018).

\bibitem{Cramer2018} N. F. Cramer and F. Verheest, Phys. Plasmas \textbf{25}%
, 123702 (2018).

\bibitem{abbas} G. Abbas, J. E. Allen, M. Coppins, L. Simons, and L. James, Phys.
Plasmas \textbf{27}, 042102 (2020).

\bibitem{allen2020} J. E. Allen, D. J. Frantzeskakis, N. I. Karachalios, P. G.
Kevrekidis, and V. Koukouloyannis, Phys. Rev. E \textbf{102}, 013209 (2020).

\bibitem{abra} M. Abramowitz and I. A. Stegun, {\it Handbook of Mathematical Functions} 
(New York, Dover, 1970).

\bibitem{rem} M. Remoissenet, \textit{Waves Called Solitons} (Springer,
Berlin, 1999).

\bibitem{Trukh2016} F. M. Trukhachev, A.V. Tomov, Cosmic Res. {\bf 54},
  351 (2016).
  %https://doi.org/10.1134/S0010952516050075.

\bibitem{Trukh2018} F. M. Trukhachev, A.V. Tomov, M.M. Mogilevsky and
  D.V. Chugunin,
  %Electric Currents Induced in Plasma by Ion-Acoustic Solitons:
  %Allowance for Trapped Electrons.
  Tech. Phys. Lett. {\bf 44}, 494 (2018).
  %https://doi.org/10.1134/S1063785018060123.

\bibitem{Trukh2019} F. M. Trukhachev, M.M. Vasiliev, O.F. Petrov,
  E.V. Vasilieva,
  Phys. Rev. E {\bf 100}, 063202 (2019).

\bibitem{Sauer2001}K. Sauer, E. Dubinin and J.E McKenzie,  Geophys. Res. Lett. \textbf{28}, 3589 (2001).
%\bibitem{Allen2005} C. M. C. Nairn, R. Bingham and, J. E. Allen, J. Plasma
%Phys. \textbf{71}, 631 (2005).
\end{thebibliography}
\end{document}